\documentclass[aps,prd,amsmath,amssymb,superscriptaddress,twocolumn,nofootinbib,showpacs,preprintnumbers]{revtex4-1}
\usepackage{graphics,hyperref,color,subfigure}

\hypersetup{ 
    pdfnewwindow=true,      
    colorlinks=true,       
    linkcolor=blue,          
    citecolor=blue,        
    filecolor=blue,      
    urlcolor=blue        
}  

\begin{document}
\title{Coupled-Channel Model for $\bar{K}N$ Scattering in the Resonant Region}

\author{C.~Fern\'andez-Ram\'{\i}rez}\email{cesar@jlab.org}
\affiliation{Theory Center, Thomas Jefferson National Accelerator Facility,
12000 Jefferson Avenue, Newport News, VA 23606, USA}
\affiliation{Instituto de Ciencias Nucleares, Universidad Nacional Aut\'onoma de M\'exico, 
Circuito Exterior C.U., A.P. 70-543, M\'exico D.F. 04510, M\'exico}

\author{I.~V.~Danilkin}
\affiliation{Theory Center, Thomas Jefferson National Accelerator Facility,
12000 Jefferson Avenue, Newport News, VA 23606, USA}
\affiliation{Institut f\"ur Kernphysik and PRISMA Cluster of Excellence, Johannes Gutenberg Universit\"at, D-55099 Mainz, Germany}

\author{D.~M.~Manley}
\affiliation{Department of Physics, Kent State University, Kent, OH 44242, USA}

\author{V.~Mathieu}
\affiliation{Center for Exploration of Energy and Matter, Indiana University, Bloomington, IN 47403, USA}
\affiliation{Physics Department, Indiana University, Bloomington, IN 47405, USA}

\author{A.~P.~Szczepaniak}
\affiliation{Theory Center, Thomas Jefferson National Accelerator Facility,
12000 Jefferson Avenue, Newport News, VA 23606, USA}
\affiliation{Center for Exploration of Energy and Matter, Indiana University, Bloomington, IN 47403, USA}
\affiliation{Physics Department, Indiana University, Bloomington, IN 47405, USA}

\preprint{JLAB-THY-15-2152}
\collaboration{Joint Physics Analysis Center}

\begin{abstract}
We present a unitary multichannel model for $\bar{K}N$ scattering in the resonance region
that fulfills unitarity.  It has the correct analytical properties for the amplitudes once they are extended 
to the complex-$s$ plane and the partial waves have the right threshold behavior. 
To determine the parameters of the model, we have fitted 
single-energy partial waves up to $J=7/2$ and  up to 2.15 GeV of energy 
in the center-of-mass reference frame obtaining the poles of the $\Lambda^*$ and $\Sigma^*$ resonances,
which are compared to previous analyses.
We provide the most comprehensive picture of the $S=-1$ hyperon spectrum to date.
Important differences are found between the available analyses making the
gathering of further experimental information on $\bar{K}N$ scattering
mandatory to make progress in the assessment of the 
hyperon spectrum.
\end{abstract}
\pacs{13.75.Jz,14.20.Jn}
\date{\today}
\maketitle

\section{Introduction}\label{sec:introduction}
The comprehensive understanding of strong interactions 
in the resonance region is an important unresolved issue in particle and nuclear physics.
 Non-perturbative aspects of QCD related to the question of 
how quarks and gluons aggregate to build hadrons, 
can be investigated by analyzing the excited baryon spectrum. 
Several  experiments were devoted in the past to the measurement of  
$\pi N$ and $\bar{K}N$ scattering as well as meson photoproduction to garner information on the baryon spectrum.
 The amount of experimental data on hyperon resonances with $S=-1$ ($Y^*=\Lambda^*, \Sigma^*$) 
is not as large as in the case of strangeness zero  ($S=0$) nucleon excitations
and, as a consequence, the hyperon spectrum is somewhat less understood.  
For example only recently,  following developments in models for $\bar{K}N$ 
scattering \cite{Manley13b,Kamano14,Kamano15} 
and kaon electroproduction
\cite{Qiang10},  the \textit{Review of Particle Physics} (RPP) 
\cite{PDG2014} began to report  $Y^*$  resonance  pole positions. 
The $\bar{K}N \to \bar{K}N$ 
reaction amplitudes,  besides their importance for studies of the $Y^*$ spectrum, 
play a role in amplitude analysis of more complicated reactions, which 
include, for example three-body decays, pentaquark searches \cite{LHCbpentaquark}, 
or  $K \bar{K}$ pair photoproduction \cite{ATHOS,JLAB}.
For example, recent observation of two pentaquark states in  
$\Lambda_b^0 \to J \slash \psi \: K^- p$ decay \cite{LHCbpentaquark}  uses 
a specific model to incorporate  $Y^*$ resonances in the $K^- p$ channel. 
Studies of systematic uncertainties should, however,  involve comparison with other models of 
 $\bar{K}N$ interactions. 
Real and quasi-real diffractive photoproduction of $K \bar{K}$ pairs can produce 
the poorly known \textit{strangeonia}, \textit{i.e.}, mesons containing $s\bar{s}$ pairs that also include
exotic mesons with hidden strangeness. 
To factorize the $K\bar{K}$ photoproduction vertex   
requires, however,  separation of target fragmentation at the amplitude level.    
Hence,  the provision of amplitudes describing, the $\bar K N$ interactions in  target fragmentation 
is relevant to  future  partial-wave analyses of the $\gamma p \to K \bar{K} p$ process.   
At Jefferson Lab \cite{JLAB}, both CLAS12 (Hall B) and GluEx (Hall D) experiments will devote
part of their effort to study this reaction.

In this article we present a coupled-channel model for $\bar{K}N$ partial waves 
that  incorporates a number of relevant channels,  including, for example,  $\pi \Sigma$ and $\pi \Lambda$.
The approach  is based on the $K$-matrix formalism and   
we pay special attention to the analytical properties of the amplitudes determined 
by the square-root unitary branch points. 
This enables continuation of partial waves to the complex $s$-plane 
and permits a search for amplitude poles (resonances). 

The poles of the amplitude that are close to the physical axis in unphysical Riemann sheets determine 
the behavior of  partial waves in the physical region. Identification of baryon resonance  poles 
is one of the goals of meson-baryon amplitude analysis. 
In recent years, poles of  $\bar{K}N$ scattering amplitudes have been reported, 
initially for the narrow-width $\Lambda (1520)$ \cite{Qiang10}, 
and subsequently from a comprehensive analysis 
by Zhang \textit{et al.}  \cite{Manley13b}.
Other recent results come from a  dynamical coupled-channel 
model by Kamano \textit{et al.} \cite{Kamano14,Kamano15}. Both  \cite{Manley13b} and \cite{Kamano15} 
analyses are in a fair 
agreement for most of the resonances with a four-star status assigned in the RPP. 
  
This article is organized as follows. In Section \ref{sec:model} we describe the details 
of the theoretical model for the partial waves based on the analytical, coupled-channel $K$-matrix representation. 
In  Section \ref{sec:results} we discuss the fits to the single-energy partial waves, 
extraction of resonance parameters,  and comparison with experimentally measured observables. 
We also compare with other extractions of $\Lambda^*$ and $\Sigma^*$ resonance parameters. 
Finally, in Section \ref{sec:conclusions} we present our conclusions and outlook. 
  
\section{Model} \label{sec:model}
We construct an analytical model that relies on unitarity that enforces square-root singularities at thresholds. 
Amplitudes are constructed by means of an analytical  $K$-matrix representation. 
Summary of the construction is given below with more details given in Appendix \ref{sec:appendix}. 

\subsection{Observables and Definition of Partial Waves} \label{sec:partialwaves}
The differential cross section and polarization for the 
processes involving $S=-1$ meson-baryon states, 
which include $\bar{K}N, \pi \Sigma,  \ldots \to \bar{K}N,\pi \Sigma,  \ldots$
are given by \cite{Hoehler1983}
\begin{alignat}{2}
\frac{d\sigma}{d\Omega} (s,\theta)=& \frac{1}{q^2} \left[ |f(s,\theta)|^2 + |g(s,\theta)|^2 \right]  , \\
P(s,\theta) =& \frac{2 \: \text{Im} \left[ f (s,\theta)\: g^*(s,\theta)  \right] }{|f(s,\theta)|^2 + |g(s,\theta)|^2} ,
\end{alignat}
where $q$ is the magnitude of the relative momentum in the center-of-momentum frame and 
 $\theta$ is the scattering angle. The amplitudes 
 $f(s,\theta)$  and   $g(s,\theta)$ correspond to no spin-flip and spin-flip contributions, respectively.  
 These amplitudes are related to the $s$-channel isospin $I=0$ and $I=1$ amplitudes through a general relation 
\begin{alignat}{1}
f(s,\theta) = \alpha^0\: f^0_{kj}(s,\theta) + \alpha^1\: f^1_{kj}(s,\theta), \\
g(s,\theta) = \alpha^0\:  g^0_{kj}(s,\theta) + \alpha^1\: g^1_{kj}(s,\theta),
\end{alignat}
where $f^I_{kj}(s,\theta)$ and $g^I_{kj}(s,\theta)$ are the isospin amplitudes. Here 
$\alpha^0$ and $\alpha^1$ are the corresponding Clebsch-Gordan 
coefficients for isospin zero and one  and $kj$ label the initial ($k$) and final ($j$) state, respectively. Specifically, 
in this work we consider the following cases, for which data are available 
 \begin{alignat}{5}
f^{K^-p\to K^-p} &=& &\frac{1}{2}f_{\bar{K}N\to\bar{K}N}^{1} +\frac{1}{2} f_{\bar{K}N\to\bar{K}N}^{0} ,&\\
f^{K^-p\to\bar{K}^0n} &=& &\frac{1}{2}f_{\bar{K}N\to\bar{K}N}^{1} - \frac{1}{2} f_{\bar{K}N\to\bar{K}N}^{0} ,&\\
f^{K^-p\to\pi^- \Sigma^+} &=&-&\frac{1}{2}f_{\bar{K}N\to\pi\Sigma}^{1}-  
\frac{1}{\sqrt{6}}f_{\bar{K}N\to\pi\Sigma}^{0} ,&\label{eq:piminussugmaplus} \\
f^{K^-p\to\pi^+ \Sigma^-} &=& &\frac{1}{2}f_{\bar{K}N\to\pi\Sigma}^{1} - 
\frac{1}{\sqrt{6}}f_{\bar{K}N\to\pi\Sigma}^{0}, & \label{eq:piplussigmaminus} \\
f^{K^-p\to\pi^0 \Sigma^0} &=& &\frac{1}{\sqrt{6}}f_{\bar{K}N\to\pi\Sigma}^{0},&\label{eq:pi0sigma0} \\
f^{K^-p\to\pi^0 \Lambda} &=& &\frac{1}{\sqrt{2}}f_{\bar{K}N\to\pi \Lambda}^{1},& 
\end{alignat}
and similarly for $g(s,\theta)$. Partial-wave expansion of isospin amplitudes is given by 
\begin{alignat}{2}
f^I_{kj}(s,\theta) =& \sum_{\ell=0}^{\infty} \left[ (\ell+1) R^{I,kj}_{\ell + }(s)
+ \ell R^{I,kj}_{\ell-}(s) \right] P_\ell \left( \theta\right), \\
g^I_{kj}(s,\theta)=& \sum_{\ell=1}^{\infty} \left[ R^{I,kj}_{\ell + }(s)
- R^{I,kj}_{\ell-}(s) \right] P^1_\ell \left( \theta \right),
\end{alignat}
where $P_\ell \left( \theta \right)$ is a  Legendre polynomial 
and  $P^1_\ell \left( \theta \right)= \sin \theta  d P_\ell \left( \theta \right) /d  \cos \theta $. 
The partial waves $R^{I,kj}_{\ell\tau }(s)$  ($\tau = \pm$)   
 are to be considered as $kj$ elements of the channel-space 
matrix $R_\ell (s)$ as  defined below in Eq.~(\ref{eq:smatrix}). In a given meson-baryon channel  
$\ell$ labels the relative orbital angular momentum and the total angular momentum is given by 
 $J=\ell + \tau /2 $. 
The orbital angular momentum $\ell$ coincides with the orbital angular momentum
of the initial $\bar{K}N$ state in $R^{I,kj}_{\ell \tau}(s)$ 
but it is not necessarily the orbital angular momentum of the other possible states.
For example, for the $I=1, \ell=0$ partial wave it is possible to have $\bar{K} \Delta(1232)$ 
in a $D$ wave state ($L=2$). 
A complete list of included channels is given in Section \ref{sec:fits}.
In terms of partial waves,  the total cross section is given by
 \begin{equation}
\begin{split}
\sigma (s) = \frac{4\pi}{q^2} \sum_{\ell=0}^{\infty} \left[  (\ell+1) |R_{\ell +}(s)|^2 +  \ell  \: |R_{\ell -}(s)|^2\right],
\end{split}
\end{equation}
where $R_{\ell \tau}(s)=\alpha^0 R^{0,kj}_{\ell \tau}(s) + \alpha^1 R^{1,kj}_{\ell \tau}(s)$.

\subsection{Partial-Wave Scattering Matrix} 
For a given partial wave we write the scattering amplitude as a  matrix in the channel-space 
\begin{equation}
S_\ell=\mathbb{I}+2iR_\ell(s)=\mathbb{I}+2i \left[C_\ell (s) \right]^{1/2} T_\ell(s) \left[C_\ell (s) \right]^{1/2},  \label{eq:smatrix}
\end{equation}
where $\mathbb{I}$ is the identity matrix, $C_\ell (s)$ is a 
diagonal matrix that accounts for the phase space and $T_\ell(s)$ 
is the analytical partial-wave amplitude matrix. 
We write $T_\ell(s)$ in terms of a $K$ matrix  \cite{kmatrix} to ensure 
unitarity 
\begin{equation}
T_\ell (s)= \left[ K(s)^{-1} -i \rho(s,\ell) \:  \right]^{-1}.\label{eq:kmatrix}
\end{equation}
For real $s$, $K(s)$ is a real symmetric matrix  and $\rho(s,\ell)$ 
is a diagonal matrix. To ensure that $\rho(s,\ell )$ is free from kinematical cuts 
and has only the square-root branch point demanded by unitarity, we write it as a dispersive integral over the 
 phase space matrix $C_\ell (s)$, \textit{a.k.a.} the Chew--Mandelstam representation,  
\begin{equation}
i \rho (s,\ell)  =   \frac{s-s_k}{\pi}
\int_{s_k}^\infty\frac{ C_\ell (s')  }{s'-s} \frac{ds'}{s'-s_k}. \label{eq:rho}
\end{equation}
Here $s_k$ is the threshold center-of-mass energy squared of the corresponding channel $k$
and we define 
\begin{equation}
 C_\ell (s) = \frac{q_k (s)}{q_0}\left[ \frac{r^2q^2_k(s)}{1+ r^2q^2_k(s) }\right]^{\ell}.  \label{C}
\end{equation}
The first factor on the \textit{r.h.s.} of Eq.~(\ref{C}) is related to the breakup momentum near threshold. For 
 a meson-baryon pair with masses $m_1$ and $m_2$ respectively, $s_k = (m_1 + m_2)^2$ , and 
\begin{equation}
\begin{split}
q_k(s)  &= \frac{\sqrt{(s - (m_1+m_2)^2)(s - (m_1 - m_2)^2)}}{2\sqrt{s}}  \\
&\simeq \frac{\sqrt{m_1 m_2}}{(m_1+m_2)} \sqrt{ s-s_k }. 
  \end{split}\label{eq:q2analytic}
\end{equation} 
The term in the square bracket ensures the  threshold behavior and introduces the effective interaction range parameter,   
$r=1 \: \text{fm}$. Finally,  $q_0=2 \: \text{GeV}$ is a  normalization factor 
for the momentum in the resonance region. 
Evaluation of the integral in Eq.~(\ref{eq:rho}) yields, 
\begin{equation}
\begin{split}
&i \rho (s,\ell)=  \frac{1}{q_0 r } 
\left[ - \frac{a^{\ell+1/2}_k (s-s_k)^\ell \sqrt{s_k-s}}{\left[ 1+a_k \: (s-s_k) \right]^\ell}
+ \frac{\Gamma(\ell+\frac{1}{2})}{\sqrt{\pi}  \Gamma(\ell+1) } \right. \\
&\times \left(   \left[ 1+a_k (s-s_k)\right] { _2F_1}\left[1,\ell+\frac{1}{2},-\frac{1}{2},\frac{1}{a_k(s_k-s)} \right] \right. \\
&\left. \left. - \left[ 3 + 2 \ell + a_k(s-s_k)  \right] {_2F_1}\left[1,\ell+\frac{1}{2},\frac{1}{2},\frac{1}{ a_k(s_k-s)} \right] \right)\right],
\end{split} \label{eq:rhoan}
\end{equation}
where 
$a_k=m_1 m_2r^2/(m_1+m_2)^2$. 
Notice that Eq.~(\ref{eq:rhoan}) does not require $\ell$ to be integer;
hence our amplitudes can be analytically continued both in the
$s$ and $\ell$ complex planes \cite{Gribov}. 
The physical limit of the amplitudes corresponds to  $s+i0$, hence 
 resonances close to the physical region ($s_p$ poles in the $T$ matrix)  appear 
at  negative values of $\text{Im} s $ when $s$ is continued below the unitary cut of $\rho (s,\ell )$.  
In Secs. \ref{sec:1res} and \ref{sec:1bkg}  we introduce the building blocks of the $K$ matrix 
 and in Section \ref{sec:2res} we show how these matrices are combined to build the $T$ matrix.

\subsection{The Single Pole in the $K$ Matrix } \label{sec:1res}
The formalism of Manley \textit{et al.} \cite{Manley} serves as a starting point for our model.
Given a partial wave  that appears in $n_C$ 
channels, it is straightforward to write the elements of the $K$ matrix
that may lead to a  pole in the amplitude,  

\begin{equation}
\left[ K_P (s)\right]_{kj} = x^P_k \: \frac{M_P}{M_P^2-s} \:  x^P_j \:. \label{eq:kr}
\end{equation}
Here $P$ labels the pole part of $K$. The pole is at a real value of  $s=M_P^2$ 
 and the residue is given in terms of couplings $x^P_k$ that may be related to partial decay widths. 
 To this end we write, 
\begin{equation}
x^P_k = y^P_k \slash \left[ |C_\ell(M^2_P)|\right]_{kk} \:, \label{eq:xs}
\end{equation}
where $\left( y^P_k\right) ^2 \equiv \Gamma^P_k$ 
is to be related to the Breit--Wigner   partial-decay width. To see this, we define 
\begin{equation}
\Sigma_P (s,\ell)= \sum_{k=1}^{n_C} \Sigma^P_k (s,\ell) =  
\sum_{k=1}^{n_C}   \left[ \rho(s,\ell) \right]_{kk}\left( x^P_k\right)^2.
\end{equation}
and  
\begin{equation}
\Gamma_P(s)  =  \sum_{k=1}^{n_C} \Gamma^P_k (s)= 
\sum_{k=1}^{n_C}  \theta\left(M_P^2-s_k\right)  \: \text{Re} \: \Sigma^P_k (s,\ell),  \label{gs} 
\end{equation}
which at $s = M_P^2$ reduces to 
\begin{equation}
\Gamma_P =   \sum_{k=1}^{n_C} \Gamma^P_k =  \sum_{k=1}^{n_C} 
\left( y^P_k\right)^2  \theta\left(M_P^2-s_k\right). \label{eq:gammabw}
\end{equation}
From the relation between $K$ and $T$ matrices  in Eq.~(\ref{eq:kmatrix})   it follows that the $T$ matrix can be written
\begin{equation}
\left[ T_\ell(s) \right]_{kj} = x^P_k\: \mathcal{T}_P(s)\:  x^P_j \:, \label{eq:tmatrix}
\end{equation}
where 
\begin{equation}
\mathcal{T}_P(s) = \frac{M_P }{M_P^2-s - i M_P \Sigma_P (s,\ell) } \label{eq:Ta}\:.
\end{equation}
Thus  $\Gamma^P_k (s)$ in Eq.~(\ref{gs}) is the energy-dependent  Breit--Wigner partial width 
 for decay to the $k$-th channel.  The $K$-matrix pole  mass $M_P$ and the couplings $x^P_k$ 
 are real parameters that will be fitted by comparing the resulting $T$-matrix elements to the data. 
 The resonance pole of the $T$ matrix is given by the solution of equation 
 $M_P^2-s - i M_P \Sigma_P (s,\ell)=0$ with $\text{Im} s <0 $ on the Riemann sheet analytically 
 connected to the physical region $\text{Im} s \to 0^+$ 
  We note that  $i M_P \Sigma_P (s,\ell)$ contributes  
to both the real and the imaginary parts of the resonance pole.  

\subsection{Background Contribution to the $K$ Matrix} \label{sec:1bkg}
In addition to resonance poles, which are constrained by the direct channel unitarity, 
partial-wave amplitudes have dynamical cuts, \textit{a.k.a.} left hand cuts, which arise when 
  unitarity cuts in the cross-channels are projected onto the direct channel partial waves. 
In the direct channel physical region, in absence of anomalous thresholds, these non-resonant  
contributions add up to a smoothly varying background.  A  simple parameterization of these 
singularities in the direct channel is to use an expression analogous to that in Eq.~(\ref{eq:kr}), \textit{i.e.} use 
\begin{equation}
\left[ K_B (s)\right]_{kj} = x^B_k \: \frac{M_B}{M_B^2+s} \:  x^B_j \:. \label{eq:kb}
\end{equation}
The label $B$ distinguishes it from the pole contribution to the $K$-matrix.  
 The coefficients $x^B_k$ are defined by
\begin{equation}
x^B_k = y^B_k \slash \left[ C_\ell(4 \text{GeV}^2)\right]_{kk}\:, \label{eq:xb}
\end{equation}
where $y^B_k$ is a real number that will be fitted.
The parameters $y^B_k$ are normalized 
by the phase-space factor as was done for the 
pole $K_P$ matrix parameters, evaluated at an arbitrarily  chosen scale 
of  $4$ GeV$^2$. 
If Eq.~(\ref{eq:kb}) is used in place of $K_P$, then the $T$ matrix  becomes
\begin{equation}
\mathcal{T}_B(s) = \frac{M_B }{M_B^2+s - i M_B \Sigma_B (s,\ell) } \label{eq:Tb}\: ,
\end{equation}
where
\begin{equation}
\Sigma_B(s,\ell)= \sum_{k=1}^{n_C} \Sigma^B_k (s,\ell) =  \sum_{k=1}^{n_C} 
\left[ \rho(s,\ell) \right]_{kk}\left( x^B_k\right)^2.
\end{equation}
It follows that Eq.~(\ref{eq:Tb}) has a pole on the real axis at a negative 
value of $s$. As  discussed above  this becomes an effective parameterization  of  
non-resonant singularities that originate from exchange processes. 
Unlike $M_P$, the parameter $M_B$ in the background parameterization of $K_B$ can have any sign 
(which roughly corresponds to the attractive or repulsive effect of the exchange forces).  
  
\subsection{The General Case: Addition of Several $K$ Matrices} \label{sec:2res}
In general more than one pole and/or background terms are needed in a  given
partial wave.  Let's first spell out the result of addition of two $K$ matrices 
(we drop the $\ell$ index in what follows)
\begin{equation}
\left[ K(s) \right]_{kj} =  x^1_k\:K_1(s)\: x^1_j  +  x^2_k\:K_2(s)\: x^2_j . 
\end{equation}
The  corresponding $T$ matrix is given by \cite{Manley}
\begin{equation}
\begin{split}
\left[ T(s) \right]_{kj} =   \frac{1}{\mathcal{D}_2(s)} & 
\left[ \:  x^1_k \: c_{11} (s)\: x^1_j  +  x^1_k \:  c_{12} (s)\:  x^2_j \right. \\
+&\: \left. x^2_k \:c_{21} (s) \:  x^1_j+ x^2_k \: c_{22}  (s)\:  x^2_j \:  \right] \:,
\end{split}
\end{equation}
where 
\begin{alignat}{2}
c_{11}  (s)=& \: \mathcal{T}_1(s)  \: , \\
c_{22} (s) =& \: \mathcal{T}_2(s)\: ,  \\
c_{12} (s)=&\: c_{21}(s) = i \epsilon_{12} (s)\mathcal{T}_1(s) \mathcal{T}_2(s) \:, \\
\mathcal{D}_2(s) =& \: 1 + \left( \epsilon_{12} (s) \right)^2  \mathcal{T}_1 (s)\mathcal{T}_2(s) \: ,
\end{alignat}
and  $ \mathcal{T}_1(s)$ and $ \mathcal{T}_2(s)$ are given by either 
Eq.~(\ref{eq:Ta}) or by Eq.~(\ref{eq:Tb}) depending on whether $K_{1,2}$
corresponds to the pole or the background parameterization, respectively, and 
\begin{alignat}{2}
\epsilon_{11}(s) =& \sum^{n_C}_{k=1}\left[ \rho (s,\ell)\right]_{kk} \left( x^1_k \right)^2 = \Sigma_1(s,\ell)  \: ,\\
\epsilon_{22}(s) =& \sum^{n_C}_{k=1}\left[ \rho (s,\ell)\right]_{kk} \left( x^2_k \right)^2 = \Sigma_2(s,\ell)  \: , \\
\epsilon_{12}(s) =& \epsilon_{21} (s)=\sum^{n_C}_{k=1}\left[ \rho (s,\ell)\right]_{kk}\: x^1_k  \:  x^2_k \: .
\end{alignat}
The generalization to several pole/background components
\begin{equation}
\left[ K(s) \right]_{kj} =  \sum_a x^a_k\:K_a(s)\: x^a_j  \:,
\end{equation}
yields 
\begin{equation}
\left[ T(s) \right]_{kj} =  \frac{1}{\mathcal{D}(s)} \sum_{a,b} x^a_k \: c_{ab} (s)\: x^b_j  \:, \label{eq:T}
\end{equation}
where $\mathcal{D}(s)$ and  $c_{ab}(s)$ are given by the solution of the system of equations
\begin{widetext}
\begin{alignat}{3}
c_{aa}(s) =& \mathcal{T}_a(s) \left[ \mathcal{D}(s)+i \: 
\sum_{b}(1-\delta_{ab})\:  c_{ab}(s)\:  \epsilon_{ab} (s) \right], \label{eq:cc1}&\\
c_{ab}(s)= &  i\:  \mathcal{T}_a(s) \left[ c_{aa}(s) \:  \epsilon_{ab}(s)
+ \sum_{d} (1-\delta_{ad})\:(1-\delta_{bd}) \:  c_{ad}(s)\:  \epsilon_{bd}(s) \right] \: ; & \quad a \neq b, \label{eq:cc2}
\end{alignat}
\end{widetext}
where $ c_{ab} (s)= c_{ba}(s)$ and
\begin{equation}
\epsilon_{ab}(s) = \epsilon_{ba}(s)  =\sum^{n_C}_{k=1}\left[ \rho (s,\ell)\right]_{kk}\: x^a_k  \:  x^b_k. \label{eq:varepsilon}
\end{equation}
In fits we use up to six  pole and background components, $K_a$, of the $K$ matrix 
and up to 13 channels, \textit{cf.} Table \ref{tab:fits}. 
We note that resonance poles in the $T$ matrix  are determined by solutions of  $\mathcal{D}(s)=0$ 
in the unphysical Riemann sheets. It encapsulates the difference between  
$K$-matrix poles and $T$-matrix poles. 
 Explicit  solutions of  Eqs.~(\ref{eq:cc1}) and (\ref{eq:cc2}) 
are  given  in  Appendix \ref{sec:appendix}.  

\subsection{Analytic Structure of the $T$ Matrix} \label{sec:analyticstructure}
The $T$ matrix of the model  has the following singularities. It has right-hand cuts due to unitarity 
 whose branch points are placed at corresponding channel thresholds, $s_k$.  
 Unitarity gives the discontinuity  of the $T$-matrix elements across the right-hand cuts 
 and  determines continuation to complex values of $s$ below the real axis where resonance poles are located. 
 There should be no complex  poles on the first Riemann sheet, so the equation $\mathcal{D}(s)=0$ should 
 have no complex solutions in the physical, first sheet. 
The left-hand cuts are represented by poles, on the real axis on the first-sheet below direct channel thresholds. 
For a single-pole $K$ matrix, as shown in Sec.~\ref{sec:1res} the resonance pole of the  $T$ matrix 
is simply related to that of  $K$. The background model of $K_B$ results in a pole at a real negative value of $s$, 
approximating the left hand cut. 
In the general case, $\mathcal{D}(s)$ has a rather complicated structure and the best we can do is to 
check numerically that the singularities of $T$ are consistent with those described above. 
In the fits we enforce that any first-sheet pole  is far away from the physical region, \textit{i.e.} 
we require that it lies at $s<1 \:  \text{GeV}^2$.  
When several pole and background terms are combined, matching between a certain pole in $K_P$ 
and a resonance pole  in  $T$  is, in general,  lost. Not even the number 
of resonance poles of $T$  has to be the same as the number of input poles in $K_P$ 
We have taken advantage of this freedom by allowing for various combinations of pole vs. background terms 
and  to assess sensitivity of the data to the presence of certain resonances. 

\section{Results} \label{sec:results}

\subsection{Fits to the Single-Energy Partial Waves} \label{sec:fits}
\begin{table}
\caption{Summary of the fitted single-energy partial waves. Notation: 
$n_P$: number of pole $K$ matrices; 
$n_B$: number of background  $K$ matrices;
$n_C$: number of channels;
$N$: number of fitted single-energy points;
$n_p$: number of parameters;
$\text{dof}=N-n_p$: degrees of freedom.}  \label{tab:fits}
\begin{ruledtabular}
\begin{tabular}{c|cccccccc}
$\ell_{I\:2J}$ &\:  $n_P$\:& \:$n_B$\:&\: $n_C$ \:&\: $N$ \:& \: $n_p$\: &\: $\text{dof}$\: &\: \: $\chi^2/N$ &\: \: $\chi^2/\text{dof}$ \\
\hline 
$S_{01}$ & 4 & 2 &  7 & 360 &43 &317 &7.64  &  8.62\\
$P_{01}$ & 4  & 2 & 6  & 358 &42 &316 & 3.11 & 3.53 \\
$P_{03}$ &2& 2 & 8  &508 &36 &472 &1.52 &1.64 \\
$D_{03}$ & 3& 1 & 6  & 372& 28&344 &  2.25 & 2.43\\
$D_{05}$ & 2& 1 & 5  &302 &18 &284 & 0.67 & 0.71 \\
$F_{05}$ & 2& 1 & 8  &460 &27 &433 & 1.32 & 1.41\\
$F_{07}$ & 1& 1 & 4  & 208& 10& 198&  0.11& 0.11\\
$G_{07}$ & 1& 1 & 6  & 350& 14&336 & 1.24 & 1.29 \\
$S_{11}$ & 4& 2 & 10  &546 &66 &480 & 8.53& 9.70\\
$P_{11}$ & 2& 3 & 9  &546 & 50&496 & 1.68 & 1.84 \\
$P_{13}$ & 2& 4 & 11  & 722&72 & 650& 0.75 & 0.83 \\
$D_{13}$ & 1& 2 & 13  & 814&42 & 772& 0.88 & 0.93\\
$D_{15}$ & 2& 1 & 11  &714  &36 & 678 & 1.09 & 1.15 \\
$F_{15}$ & 2& 1 &12   & 782 &39 & 743& 0.29 & 0.30\\
$F_{17}$ & 1& 1 & 11  & 704& 24&680&  0.49 & 0.51  \\
$G_{17}$ & 1& 0 & 10  & 580 & 11& 569 &  0.10 & 0.10  \\
\end{tabular}
\end{ruledtabular}
\end{table} 
The experimental database in the resonance region with $2.19 < s < 4.70~ \mbox{GeV}^2$, 
which corresponds to kaon lab momentum of $0.288 < p_\text{lab} <   1.820 ~\mbox{GeV}/c$, 
 \cite{Daum1968,Andersson1970,Albrow1971,Conforto1971,Adams1975,Abe1975,Mast1976,Alston1978,
Armenteros1968,Armenteros1970,Jones1975,Griselin1975,
Conforto1976,Prakhov2009,Berthon1970a,Berthon1970b,Baxter1973,London1975,Manweiler2008,Baldini1988} 
contains approximately 8000 data points for the 
$\bar{K}N\to\bar{K}N$ channel ($K^- p \to K^- p$ and $K^- p \to \bar{K}^0 n$),
4500 for the $\bar{K}N \to \pi \Lambda$ channel ($K^- p \to \pi^0 \Lambda$), 
and 5000 for the $\bar{K}N \to \pi \Sigma$ channel 
($K^- p \to \pi^0 \Sigma^0$, $K^- p \to \pi^- \Sigma^+$, and $K^- p \to \pi^+ \Sigma^-$). 
This data set was analyzed in \cite{Manley13a}
and  single-energy partial waves were obtained  for ($\ell_{I\: 2J}$) up to $J=7/2$,
namely $S_{01}$, $P_{01}$, $P_{03}$, $D_{03}$, $D_{05}$, $F_{05}$, $F_{07}$, 
$G_{07}$, $S_{11}$, $P_{11}$, $P_{13}$, $D_{13}$, $D_{15}$, $F_{15}$, $F_{17}$, and $G_{17}$.  
In  \cite{Manley13b} these partial waves were described in terms of  a $K$-matrix model, which in what follows, 
 we refer to as the KSU model. From the model the  $\Lambda^*$ and $\Sigma^*$ spectrum was determined in 
 terms  of $T$-matrix poles. Our model is similar to the  KSU approach 
 as far as parameterization of the pole $K$ matrix, 
but differs in construction of the background.  Furthermore, in the KSU model 
unitarity constrains amplitudes only on the real axis, while in the present analysis unitarity is implemented in an analytical 
 way enabling unique continuation of the amplitudes beyond the physical sheet. 
We compare our results (resonances) to the KSU model in 
Section \ref{sec:poles}. 

\subsubsection{Channels} \label{sec:channels}
We fit the $T$-matrix elements to single-energy partial waves. When evaluating fit uncertainties  
 one should  keep in mind that extraction of partial waves from experimental data also 
  carries some model dependence \cite{Manley13a,Manley13b}. 
Consequently, in each partial wave we consider the same set of  channels as employed in \cite{Manley13a,Manley13b}. 
The possible initial (final) states correspond to the $k$ ($j$) labels in the $R^{I,kj}_{\ell\tau}(s)$ matrix.
All the channels are treated as two-body (meson-baryon) states and are labeled as follows:
\begin{itemize}
\item[(i)] if the state has the same orbital angular momentum ($\ell$) as the partial wave,
the channel is identified by the names of the meson and the baryon, e.g. $\bar{K}N$ or $\pi \Sigma$;
\item[(ii)] if the baryon has spin $3/2$, as it is the case of $\Sigma (1385)$, $\Delta (1232)$ and $\Lambda (1520)$ 
(in what follows $\Sigma^*$, $\Delta$, and $\Lambda^*$ respectively), 
the orbital angular momentum of the
initial state does not correspond to $\ell$ and a 
subindex  $L$ is added denoting  the angular momentum of the initial (final) state.
For example, in ${\bar K} N$ system the $S_{01}$ denotes the isoscalar, $l=0$ partial wave with total spin $J=1/2$. 
It may couple to $\pi \Sigma^*$ with orbital angular momentum $L=2$ ($D$ wave) which we label 
 as $\left[ \pi \Sigma^*\right]_D$;
\item[(iii)] if the state contains a spin one $\bar{K}^*$ and a nucleon, they can couple to spin $1/2$, which
we name $\bar{K}_1^*N$ or to spin $3/2$, which we name $\bar{K}_3^*N$.  The $\bar{K}_1^*N$ state
has the same orbital angular momentum as the  $\bar K N$ and the partial wave 
but the $\bar{K}_3^*N$ does not, hence we add a $L$ subindex to the last.
For example, the $S_{01}$ partial wave has as possible states $\bar{K}_1^*N$
and $\left[ \bar{K}_3^*N\right]_D$.
\end{itemize}
For every partial wave we include an additional meson-hyperon channel that collectively accounts 
for any missing inelasticity arising from channels not included explicitly. 
The kinematical variables for such a dummy channel are chosen arbitrarily as if it were a two-pion $\Lambda$  or $\Sigma$ state 
labeled as 
$\pi \pi \Lambda$  for $I=0$ and $\pi \pi \Sigma$ for $I=1$ partial waves.
All the channels incorporated in the model
have single-energy partial-wave data to fit except for the dummy channels 
$\pi\pi\Lambda$ and $\pi\pi\Sigma$ and the $\eta \Lambda$ and $\eta \Sigma$ channels in the $S$ waves. 
The full list of  initial (final) states for each partial wave is:
\begin{itemize}
\item[$S_{01}$:] $\bar{K}N$, $\pi \Sigma$, $\eta \Lambda$, $\bar{K}^*_{1}N$, 
$\left[ \bar{K}^*_{3}N\right]_D$, $\left[ \pi \Sigma^*\right]_D$, $\pi \pi \Lambda$;
\item[$P_{01}$:] $\bar{K}N$, $\pi \Sigma$, $\bar{K}^*_{1}N$, 
$\left[ \bar{K}^*_{3}N\right]_P$, $\left[ \pi \Sigma^*\right]_P$, $\pi \pi \Lambda$;
\item[$P_{03}$:] $\bar{K}N$, $\pi \Sigma$,
$\bar{K}^*_{1}N$, $\left[ \bar{K}^*_{3}N\right]_P$, 
$\left[ \bar{K}^*_{3}N\right]_F$, $\left[ \pi \Sigma^*\right]_P$, 
$\left[ \pi \Sigma^*\right]_F$, $\pi \pi \Lambda$;
\item[$D_{03}$:] $\bar{K}N$, $\pi \Sigma$, 
$\bar{K}^*_{1}N$, $\left[ \pi \Sigma^*\right]_S$, 
$\left[ \pi \Sigma^*\right]_D$, $\pi \pi \Lambda$;
\item[$D_{05}$:] $\bar{K}N$, $\pi \Sigma$, $\left[ \pi \Sigma^*\right]_D$, 
$\left[ \pi \Sigma^*\right]_G$, $\pi \pi \Lambda$;
\item[$F_{05}$:] $\bar{K}N$, $\pi \Sigma$, $\bar{K}^*_{1}N$, 
$\left[ \bar{K}^*_{3}N\right]_P$, $\left[ \bar{K}^*_{3}N\right]_F$, 
$\left[ \pi \Sigma^*\right]_P$, $\left[ \pi \Sigma^*\right]_F$, $\pi \pi \Lambda$;
\item[$F_{07}$:] $\bar{K}N$, $\pi \Sigma$, $\bar{K}^*_{1}N$, $\pi \pi \Lambda$;
\item[$G_{07}$:] $\bar{K}N$, $\pi \Sigma$, $\bar{K}^*_{1}N$, 
$\left[ \bar{K}^*_{3}N\right]_D$, $\left[ \bar{K}^*_{3}N\right]_G$, $\pi \pi \Lambda$;
\item[$S_{11}$:] $\bar{K}N$, $\pi \Sigma$, $\pi \Lambda$,
$\eta \Sigma$, $\bar{K}^*_{1}N$, $\left[ \bar{K}^*_{3}N\right]_D$, 
$\left[ \pi \Sigma^*\right]_D$, $\left[ \pi \Lambda^*\right]_P$, 
$\left[ \bar{K} \Delta \right]_D$, $\pi \pi \Sigma$;
\item[$P_{11}$:] $\bar{K}N$, $\pi \Sigma$, $\pi \Lambda$,
$\bar{K}^*_{1}N$, $\left[ \bar{K}^*_{3}N\right]_P$, $\left[ \pi \Sigma^*\right]_P$, 
$\left[ \pi \Lambda^*\right]_D$, $\left[ \bar{K} \Delta \right]_P$, $\pi \pi \Sigma$;
\item[$P_{13}$:] $\bar{K}N$, $\pi \Sigma$, $\pi \Lambda$,
$\bar{K}^*_{1}N$, $\left[ \bar{K}^*_{3}N\right]_P$, $\left[ \pi \Sigma^*\right]_P$,  
$\left[ \pi \Lambda^*\right]_D$, 
$\left[ \pi \Sigma^*\right]_F$, $\left[ \pi \Lambda^*\right]_S$,
$\left[ \bar{K} \Delta \right]_P$, $\pi \pi \Sigma$;
\item[$D_{13}$:] $\bar{K}N$, $\pi \Sigma$, $\pi \Lambda$,
$\bar{K}^*_{1}N$, $\left[ \bar{K}^*_{3}N\right]_S$, $\left[ \bar{K}^*_{3}N\right]_D$, 
$\left[ \pi \Sigma^*\right]_S$, $\left[ \pi \Sigma^*\right]_D$,
$\left[ \pi \Lambda^*\right]_P$, $\left[ \pi \Lambda^*\right]_F$, 
$\left[ \bar{K} \Delta \right]_S$, $\left[ \bar{K} \Delta \right]_D$, $\pi \pi \Sigma$;
\item[$D_{15}$:] $\bar{K}N$, $\pi \Sigma$, $\pi \Lambda$,
$\bar{K}^*_{1}N$, $\left[ \bar{K}^*_{3}N\right]_D$, $\left[ \pi \Sigma^*\right]_D$, 
$\left[ \pi \Sigma^*\right]_G$,
$\left[ \pi \Lambda^*\right]_P$, $\left[ \pi \Lambda^*\right]_F$, 
$\left[ \bar{K} \Delta \right]_D$, $\pi \pi \Sigma$;
 \item[$F_{15}$:] $\bar{K}N$, $\pi \Sigma$, $\pi \Lambda$,
 $\bar{K}^*_{1}N$, $\left[ \bar{K}^*_{3}N\right]_P$, $\left[ \bar{K}^*_{3}N\right]_F$, 
 $\left[ \pi \Sigma^*\right]_P$, $\left[ \pi \Sigma^*\right]_F$, 
 $\left[ \pi \Lambda^*\right]_D$, $\left[ \pi \Lambda^*\right]_G$, $\left[ \bar{K} \Delta \right]_P$, $\pi \pi \Sigma$;
\item[$F_{17}$:] $\bar{K}N$, $\pi \Sigma$, $\pi \Lambda$, 
$\bar{K}^*_{1}N$, $\left[ \bar{K}^*_{3}N\right]_F$, 
$\left[ \bar{K}^*_{3}N\right]_H$, $\left[ \pi \Sigma^*\right]_F$, 
$\left[ \pi \Lambda^*\right]_D$, $\left[ \pi \Lambda^*\right]_G$, 
$\left[ \bar{K} \Delta \right]_F$, $\pi \pi \Sigma$;
\item[$G_{17}$:] $\bar{K}N$, $\pi \Sigma$, $\pi \Lambda$,
$\bar{K}^*_{1}N$, $\left[ \bar{K}^*_{3}N\right]_G$,
$\left[ \pi \Lambda^*\right]_F$, $\left[ \pi \Lambda^*\right]_H$, 
$\left[ \bar{K} \Delta \right]_D$, $\left[ \bar{K} \Delta \right]_G$, $\pi \pi \Sigma$;
\end{itemize}

The $K \Xi$ channel is not considered in our model because it was not included in the single-energy 
partial-wave analysis of \cite{Manley13a}. 
This channel was not incorporated in \cite{Manley13a} because
the amount of experimental data for that reaction is not enough to perform a reliable partial-wave extraction. 

\subsubsection{Parameters and Fitting Strategy}
The parameters of the model that are fitted to the single-energy partial-wave data are 
 the $K$-matrix parameters,  -- \textit{i.e.} $M_P$'s for pole  and $M_B$'s  for background,
and the pole and background couplings $y_i^P$'s,  $y_i^B$'s, 
as given in \textit{cf.}  Eqs.~(\ref{eq:kr}) and (\ref{eq:kb}). 
The summary of the fit results is given in Table \ref{tab:fits}, where for each partial wave we provide 
the number  of background ($n_B$) and pole ($n_P$) $K$ terms, 
the number of channels ($n_C$), the  number of data points ($N$), the total number of
parameters ($n_p$), the number of degrees of freedom ($\text{dof}$) and the resulting $\chi^2$'s.
Due to the large number of parameters,  fits have been performed 
with different strategies and optimization methods until a sufficiently satisfactory solution was obtained.
Most of the partial waves,  
\textit{i.e.}  $P_{13}$, $D_{03}$, $D_{05}$, $D_{13}$, $D_{15}$, 
$F_{05}$, $F_{07}$, $F_{15}$, $F_{17}$, $G_{07}$ and $G_{17}$,
could be fitted using  MINUIT \cite{MINUIT} only, while the other,  
\textit{i.e.}  $S_{01}$, $S_{11}$, $P_{01}$, $P_{11}$, and $P_{03}$,
required more sophisticated methods based on a genetic algorithm \cite{genetic} combined 
with MINUIT to increase accuracy  as described in \cite{genetic}.
The masses and couplings of the pole $K$ matrices 
have been guided to yield optimal values for  $M_P$ and $\Gamma_P$ in Eq.~(\ref{eq:gammabw}) 
penalizing  fits that yielded unnatural parameters (such as disproportionate values for the couplings)
while the background parameters have been allowed to run freely.

\subsubsection{Error Estimation} \label{sec:fiterrors}
We have computed the
statistical errors of the partial waves parameters and $T$-matrix poles
employing the bootstrap technique \cite{NumericalRecipes}.
This calculation is rather straightforward but computationally demanding. 
It consists of generating, in our case,   
50 data sets by randomly sampling  the experimental 
points according to their uncertainties and independently fitting each data sample. 
The uncertainty for each fitted parameter is given by the standard deviation from the average in 50 fits. 
For each set of parameters we compute the partial waves, observables and $T$-matrix poles
and, again, we estimate the error as the standard deviation.

If the model has problems reproducing a specific partial wave
(reflected in large $\chi^2/\text{dof}$ values)
we perform an additional error estimation by  \textit{pruning} this partial wave. 
That is, we randomly remove $20\%$ of the data points and fit the remaining $80\%$ of the data. 
This procedure is repeated 20 times and the standard deviation 
 gives an estimate of the systematic error, these systematic errors have not been propagated 
to either observables or poles.

Finally, due to the fact that we are fitting single-energy partial waves, 
our error analysis misses correlations between
 partial waves as well as systematic uncertainties in the 
measured differential cross sections and polarization observables. 
The latter, in those experiments that report them, average 
to approximately $\pm10\%$. 

\subsubsection{Fits} \label{sec:fitsquality}
\begin{figure*}
\begin{center}
\begin{tabular}{cc}
\rotatebox{0}{\scalebox{0.4}[0.4]{\includegraphics{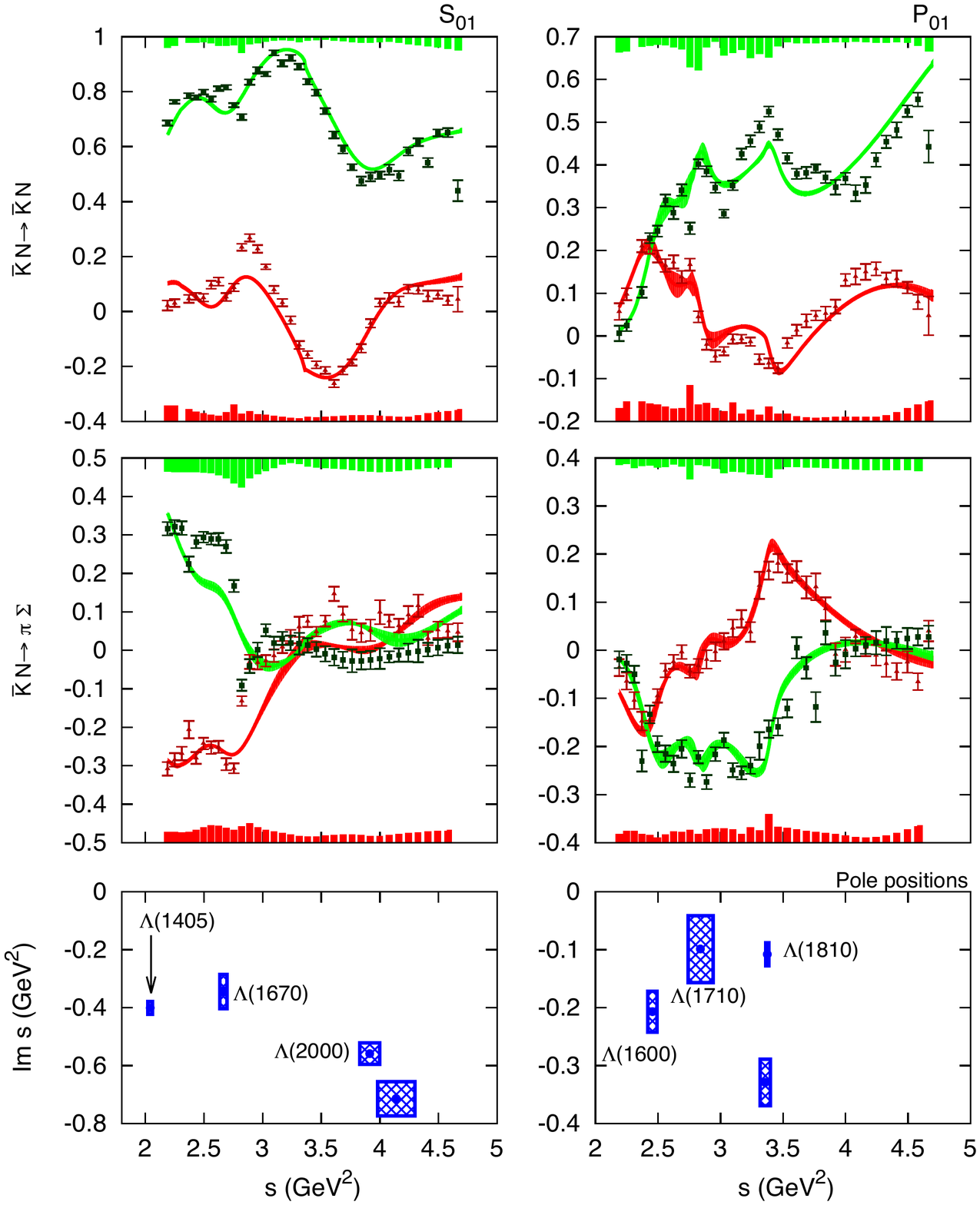}}} & 
\rotatebox{0}{\scalebox{0.4}[0.4]{\includegraphics{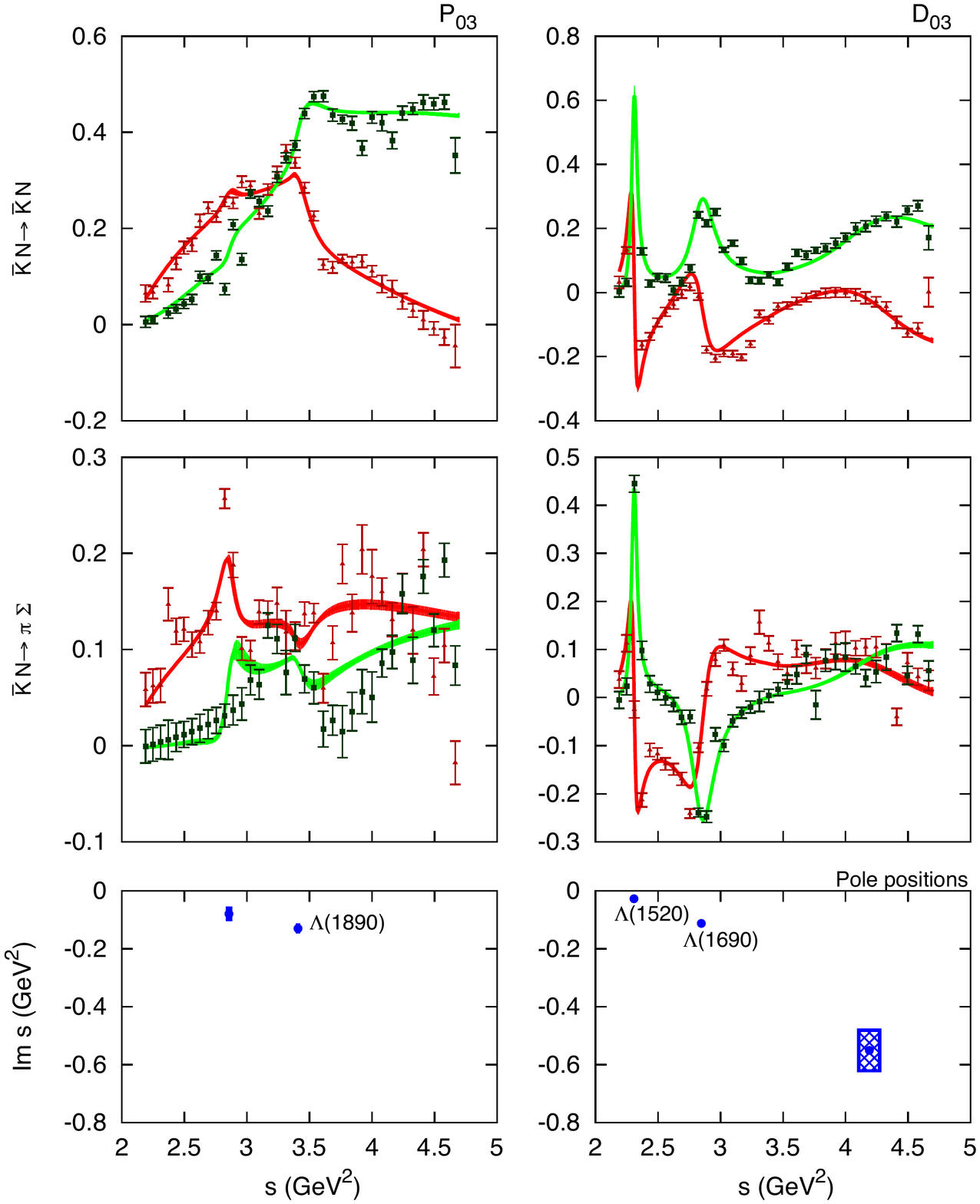}}}
\end{tabular}
\caption{(color online). Partial waves $S_{01}$ (left column), 
$P_{01}$ (center-left column), 
$P_{03}$ (center-right column), and $D_{03}$ (right column) 
together with the $T$-matrix pole positions (last row) compared 
to the single-energy partial waves from KSU analysis \cite{Manley13a} for 
channels $\bar{K}N\to\bar{K}N$ and $\bar{K}N\to\pi\Sigma$ 
(real part: red triangles; imaginary part: green squares). 
Red band stands for the real part of the partial wave 
and green band for the imaginary part of our model.
For the $S_{01}$ and $P_{01}$ waves we provide an estimation of the systematic error:
bottom-red histogram for the real part of the partial wave and top-green for the imaginary.
The resonances (poles of the $T$ matrix) computed 
are the closest to the physical axis in the corresponding Riemann sheet.
One additional pole in the $S_{01}$ partial wave at $2.45 -i \:  0.47$ GeV$^2$ is not shown 
and believed to be an artifact of the fits (see Sec.~\ref{sec:lambdaresonances}).
Another pole in the $D_{03}$ partial wave at $\left(4.24\pm 0.48\right) -i \left(2.38\pm 0.58 \right)$ GeV$^2$
is not shown in the bottom-right figure.
Error bars for $\Lambda(1520)$, $\Lambda(1690)$  and $\Lambda(1890)$ 
are smaller than the size of the dots.} \label{fig:PW_S01P01}
\end{center}
\end{figure*}

\begin{figure*}
\begin{center}
\begin{tabular}{cc}
\rotatebox{0}{\scalebox{0.4}[0.4]{\includegraphics{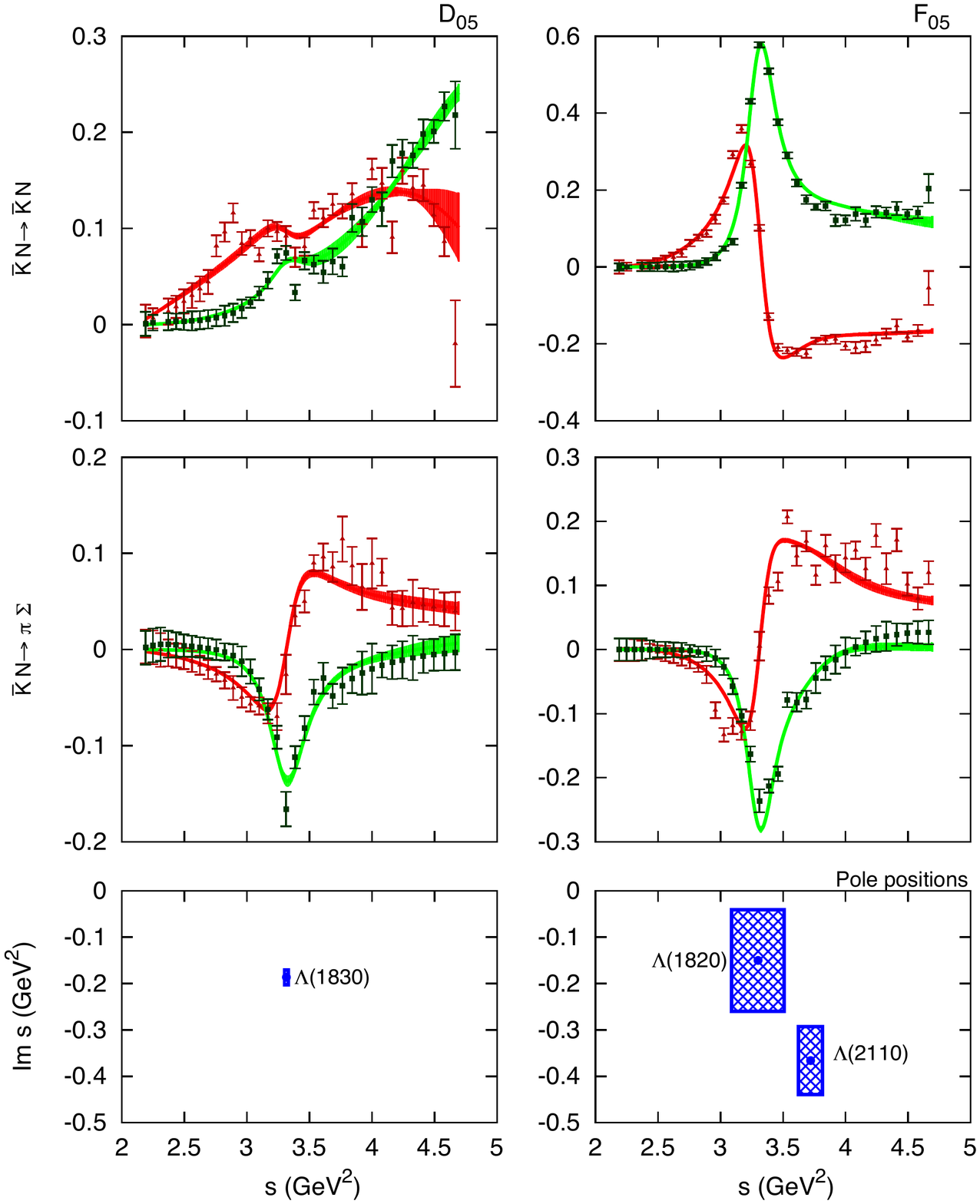}}} & 
\rotatebox{0}{\scalebox{0.4}[0.4]{\includegraphics{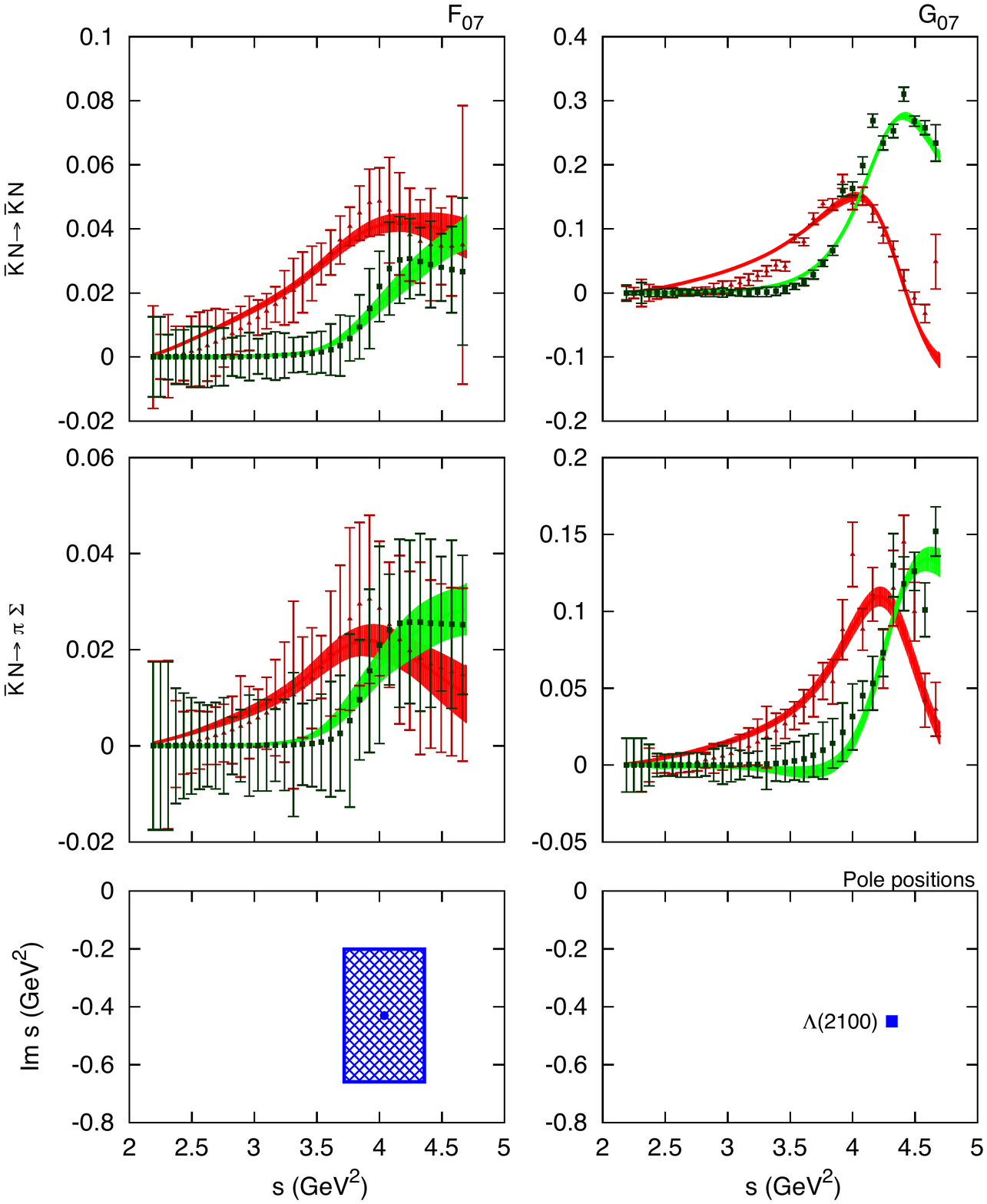}}}
\end{tabular}
\caption{(color online). 
Same as in Fig.~\ref{fig:PW_S01P01} for $D_{05}$ (left column), $F_{05}$ (center-left column), 
$F_{07}$ (center-right column), and $G_{07}$ (right column) partial waves.
An additional pole in the $D_{05}$ partial wave at 
$\left(4.75\pm 0.19\right) -i \left(1.24\pm 0.41 \right)$ GeV$^2$ 
is not shown in the bottom-left figure. The $\Lambda(1820)$ state is found to be 
highly correlated with a close resonance $\Lambda(2110)$ 
located at higher energy and deeper in the complex plane.} \label{fig:PW_D05F05}
\end{center}
\end{figure*}

\begin{figure*}
\begin{center}
\begin{tabular}{cc}
\rotatebox{0}{\scalebox{0.4}[0.4]{\includegraphics{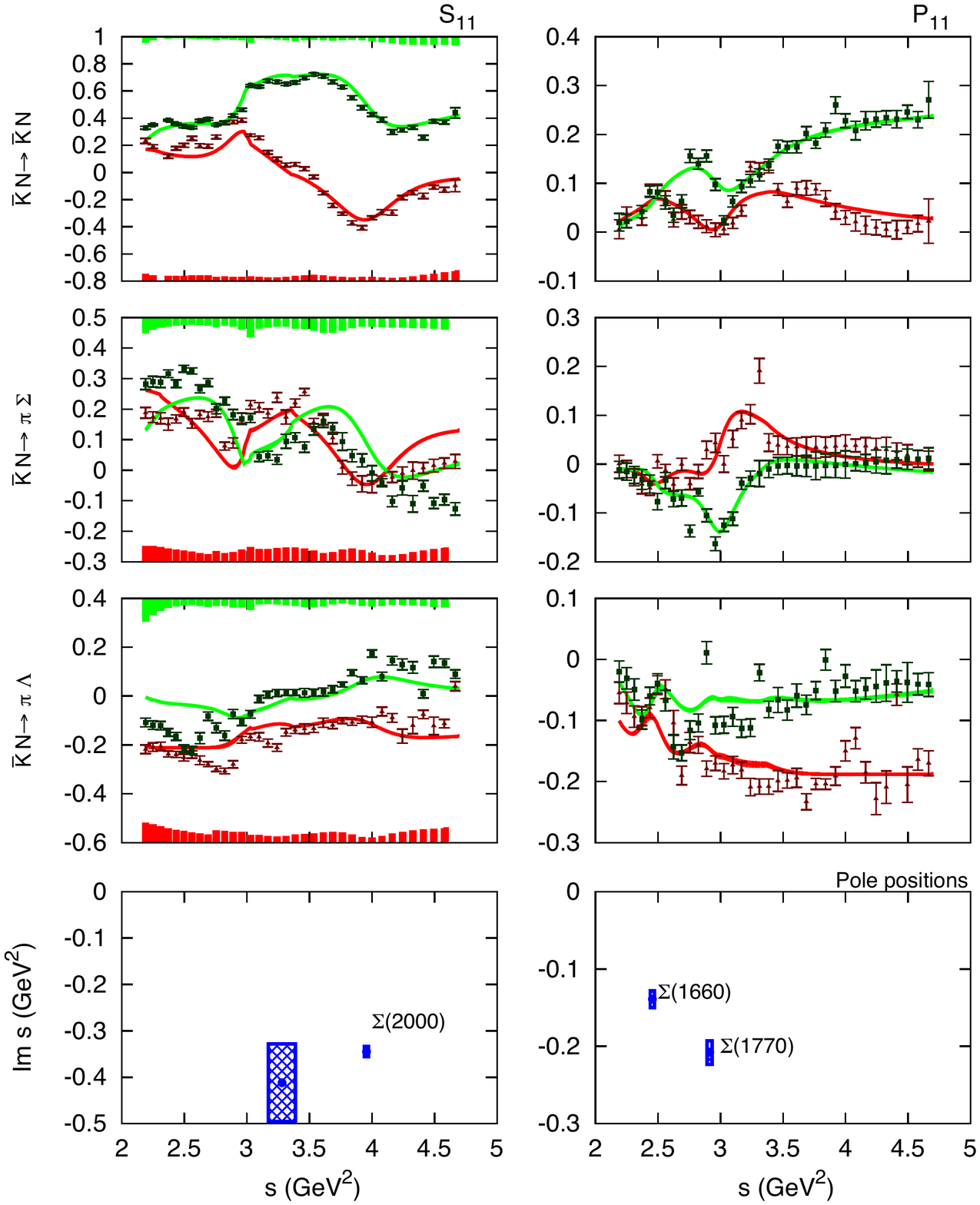}}} & 
\rotatebox{0}{\scalebox{0.4}[0.4]{\includegraphics{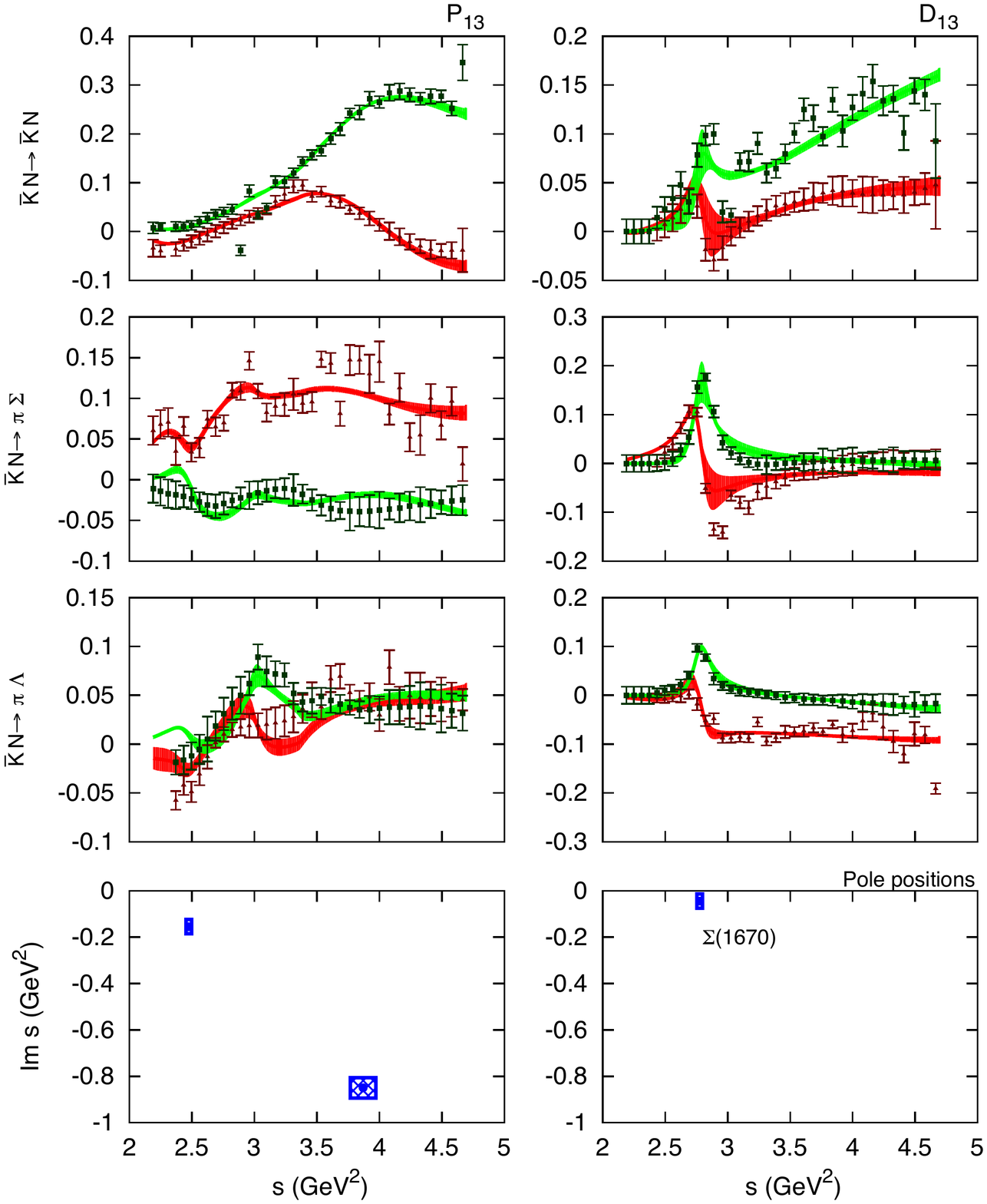}}}
\end{tabular}
\caption{(color online). Partial waves $S_{11}$ (left column), $P_{11}$ (center-left column), 
$P_{13}$ (center-right column), and $D_{13}$ (right column) 
together with the pole positions (last row) compared to single-energy partial waves from KSU analysis for 
channels $\bar{K}N\to\bar{K}N$, $\bar{K}N\to\pi\Sigma$, and $\bar{K}N\to\pi \Lambda$ 
(real part: red triangles; imaginary part: green squares). 
Red band stands for the real part of the partial wave and green band 
for the imaginary part for our model.
For the $S_{11}$ wave we provide an estimation of the systematic error:
bottom-red histogram for the real part of the partial wave and top-green for the imaginary.
}  \label{fig:PW_S11P11}
\end{center}
\end{figure*}

\begin{figure*}
\begin{center}
\begin{tabular}{cc}
\rotatebox{0}{\scalebox{0.4}[0.4]{\includegraphics{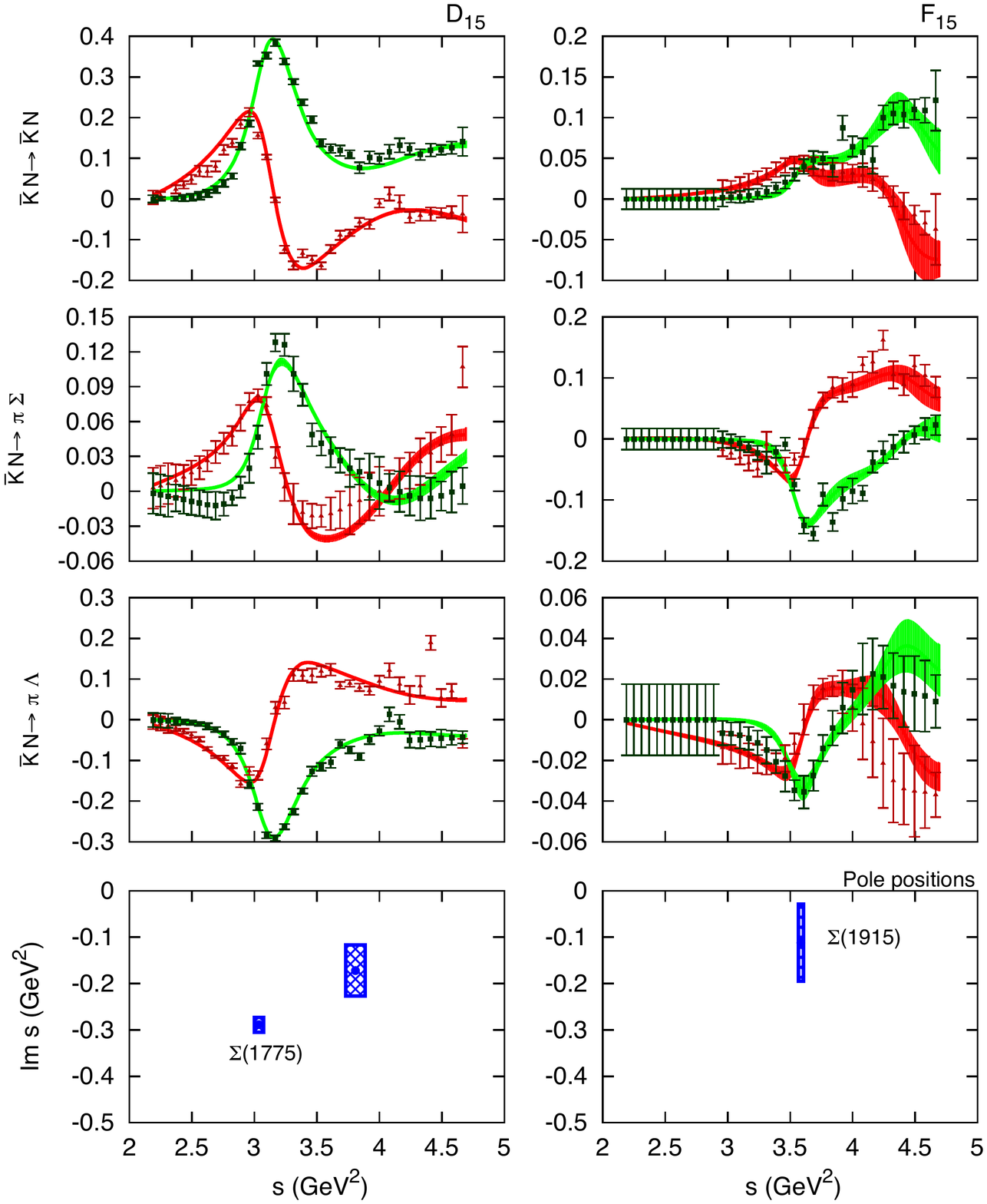}}} &
\rotatebox{0}{\scalebox{0.4}[0.4]{\includegraphics{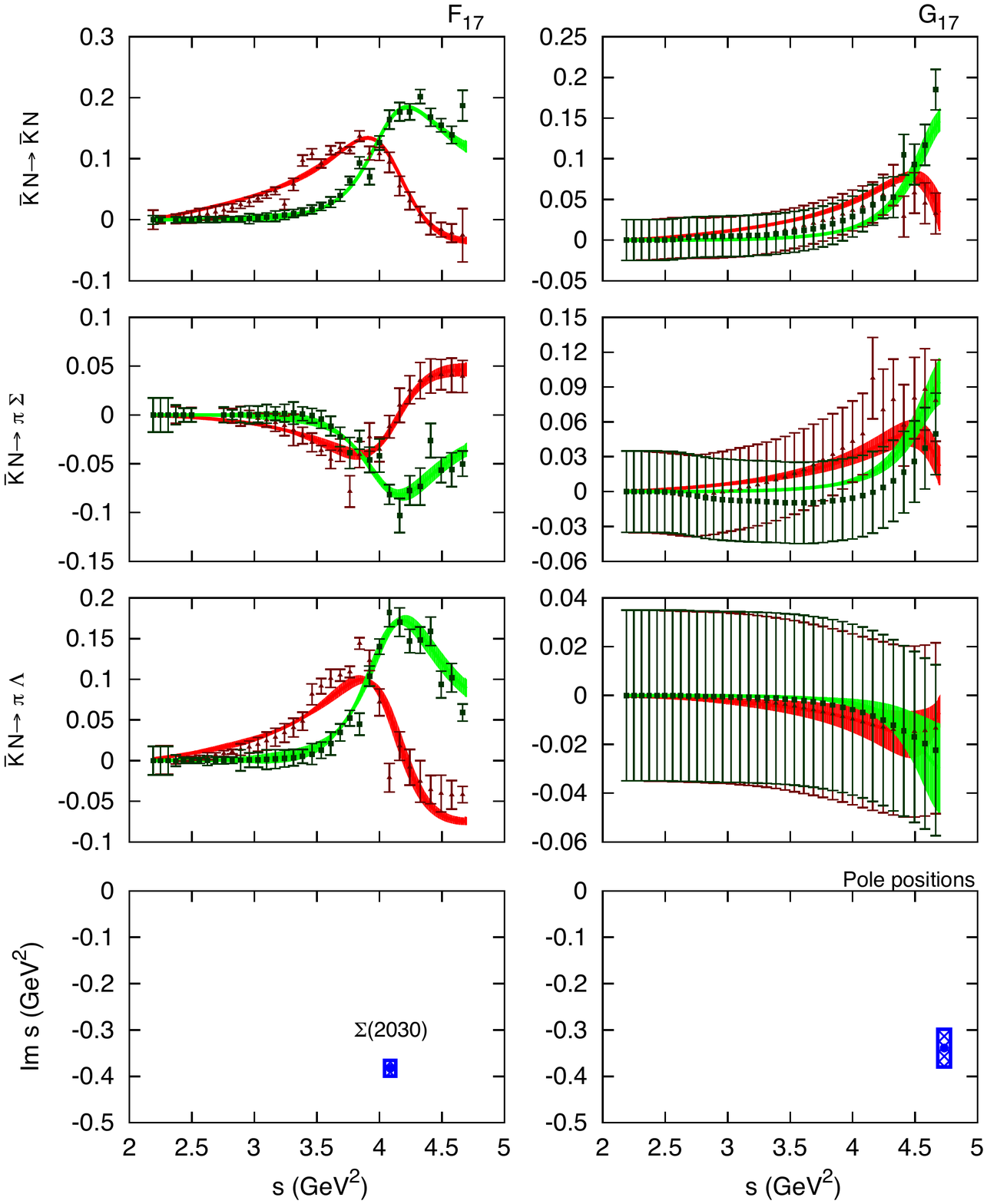}}}
\end{tabular}
\caption{(color online). 
Same as in Fig.~\ref{fig:PW_S11P11} for $D_{15}$ (left column), $F_{15}$ (center-left column),
$F_{17}$ (center-right column), and $G_{17}$ (right column)  partial waves.
An additional pole in the $F_{15}$ partial wave at 
$\left(4.346\pm 0.026\right) -i \left(0.993\pm 0.018 \right)$ GeV$^2$ 
is not shown in the bottom center-left figure. }\label{fig:PW_D15F15}
\end{center}
\end{figure*}

In Figs.~\ref{fig:PW_S01P01}--\ref{fig:PW_D15F15} 
we compare our fits to the KSU single-energy partial waves
for the channels where there are experimental data available, namely 
$\bar{K}N \to \pi \Lambda$  for $I=1$ and
$\bar{K}N\to\bar{K}N$ and $\bar{K}N \to \pi \Sigma$ for both isospins.
The bottom plots in each figure 
show the position of the $T$-matrix, resonance poles in the Riemann sheet closest to the threshold for the given channel.  
The values of the  pole parameters  are given in Tables \ref{tab:poles0} and \ref{tab:poles1} 
and discussed in Section \ref{sec:poles}.

The $\chi^2/\text{dof}$'s for most of the fits are quite reasonable (see Table \ref{tab:fits})
and provide a good description of the data as shown in 
Figs.~\ref{fig:PW_S01P01}--\ref{fig:PW_D15F15}. The exceptions are 
 the  $S_{01}$, $S_{11}$, $P_{01}$ and $D_{03}$ partial waves. 
 
The $S_{01}$ and $S_{11}$ fits were specially cumbersome. Even with the aid of a genetic algorithm
the parameters could get trapped in a local minima and fits needed to be repeated more 
 than 30 times 
to reach the $\chi^2$'s presented in Table \ref{tab:fits}. It is worth noting 
  that these two partial waves are the most affected by systematic errors and database inconsistencies.
For both partial waves we estimate the systematic uncertainty by data pruning and refitting 
 as was described in Section \ref{sec:fiterrors}.
These systematic errors are shown in Figs.~\ref{fig:PW_S01P01} and \ref{fig:PW_S11P11}
as vertical bars (see figure captions for more details).

 The $S_{01}$ for $\bar{K}N \to \bar{K}N$ has a complicated shape. 
 It is rather flat and between 2 and 3 GeV$^2$  the imaginary part suddenly drops,  
 which is followed by a  bump and another drop. 
 These variations are difficult to reproduce with an analytical  parameterization, and results 
  in the large $\chi^2/\text{dof}$. Nevertheless, the model seems to describe the general features of both 
  $\bar{K}N \to \bar{K}N$  and  $\bar{K}N \to \pi \Sigma$. 

One of the main features of the $S_{01}$ partial wave is the appearance of
the $\Lambda(1405)$ resonance below the $\bar{K}N$ threshold  \cite{PDG2014}.
This behavior of the $S_{01}$ wave in this mass region is often attributed to existence 
of two poles,  \cite{Lambda1405old,Mai1405} located at
$1429^{+8}_{-7}-i \: 12 ^{+2}_{-3}\:  \text{MeV}$  and
$1325^{+15}_{-15}-i\: 90 ^{+12}_{-18} \:  \text{MeV}$
\cite{Mai1405,Lambda1405}.
Our model is built to cover a wide energy range and  cannot account for the
fine details of the near-threshold effects. 
For example, the detailed analysis the poles in $\Lambda (1405)$ region required 
 constraints from  $\pi \Sigma K^+$ photoproduction off the proton \cite{Mai1405}.
If we do not restrict the fit to obtain a resonance in the region where $\Lambda(1405)$ should appear
we obtain $\chi^2/\text{dof}\sim6$ while no resonance poles appear in  the $\Lambda(1405)$ region. 
Hence, we enforce an effective $\Lambda (1405)$ resonance that accounts for both states
by penalizing fits that do not generate a pole in this region. 
The enforcement of this pole
results in a more rigid model and a larger  $\chi^2$.

The $S_{11}$ partial wave has the highest   $\chi^2/\text{dof}$.
Overall, out of the three channels shown in Fig.~\ref{fig:PW_S11P11}, only the data 
 $\bar{K}N \to \bar{K}N$ can be reasonably well described, except for the real part in the 
region  between 2.5 and 3 GeV$^2$. 
The $\bar{K}N \to \pi \Sigma$  channel of the $S_{11}$ partial wave is reproduced in shape but not in magnitude.
The same is true for $\bar{K}N\to \pi \Lambda$. Because of the disagreement with the $S_{11}$ data 
we cannot accurately reproduce the total cross section data for 
$K^-p\to \pi^0 \Lambda$ and  $s<3$ GeV$^2$  (\textit{cf.}  Section \ref{sec:experiment}). 

The difficulties encountered when using a highly constrained analytical model  
can have several origins. There could be missing  resonances or background features in the model,
other channels, or there could be inherent problems related to the single-energy extraction. For example,
the rapid variation of the partial waves in certain energy regions 
that the model tries to smear out, may have underestimated uncertainties. 
The single-energy partial waves were obtained from experimental data in 10 MeV bins, 
hence rapid variations from one bin to another should be taken with care because
a different binning of the data would impact the variation.
The uncertainties associated to binning can be assessed 
by pruning the data and refitting them as was described in Section \ref{sec:fiterrors}.
Finally we note that our channel set overlaps with that of  \cite{Manley13b} 
which are, for example, not the same as   used by Kamano \textit{et al.} \cite{Kamano14}. 

When uncertainties and visual inspection are considered,
$P_{01}$ and $D_{03}$ (Fig.~\ref{fig:PW_S01P01}) yield acceptable results. 
$P_{01}$ shows large uncertainties mostly derived from the difficulty of the imaginary parts 
to follow the several oscillations of single-energy partial waves in the
$2.5-4$ GeV$^2$ range
for the $\bar{K}N\to\bar{K}N$ and $\bar{K}N\to\pi \Sigma$ channels, 
that the model tries to average.
In the case of the $D_{03}$ partial wave most of the $\chi^2/\text{dof}$
is due to the difficulties of the model to follow
the rapid variation of the single-energy partial-wave data points 
in the region of the $\Lambda(1520)$.
The change in the partial wave is more rapid than produced by the model. 
We will return to this discrepancy in $D_{03}$ 
when comparing to $K^-p\to K^-p$ observables
in Section \ref{sec:experiment}. 

Some of the partial waves show clear signs of over-fitting (very low $\chi^2/\text{dof}$), 
e.g. $F_{07}$, $F_{15}$, $F_{17}$, $G_{17}$.
The $F_{07}$ and $G_{17}$ are straightforward to understand, as  the data have
large uncertainties. 
The $F_{15}$ and $F_{17}$ cases are different.
Because the number of channels for each partial wave is fixed by the single-energy data,
the only freedom is in the number of pole and background $K$ matrices.
We cannot change the number of parameters one by one until we get the optimal amount of them.
For example, the partial wave $F_{17}$ with 11 channels and two $K$ matrices has
24 parameters (two masses and 22 couplings). If we remove one $K$ matrix we
drop the number of parameters to 12 (one mass and 11 couplings). 
With this new model
the partial wave still yields a good $\chi^2/\text{dof} \sim 0.9$ but the most relevant channels
that are straightforwardly connected to experimental data,
\textit{i.e.} $\bar{K}N\to\bar{K}N$, $\bar{K}N\to\pi \Sigma$, and $\bar{K}N\to\pi \Lambda$
are poorly described ($\chi^2/\text{dof} \sim 2.5$). 
Hence, we prefer the model with two $K$ matrices.
A similar situation happens with the $F_{15}$ partial wave. 

In Section \ref{sec:poles}  additional details on the fitting procedure 
and results are discussed in connection with the $T$-matrix poles determination.

\subsection{$T$ Matrix Poles} \label{sec:poles}
\begin{figure}
\begin{center}
\subfigure[\  $\Lambda^*$ resonances.]{
\rotatebox{0}{\scalebox{0.3}[0.3]{\includegraphics{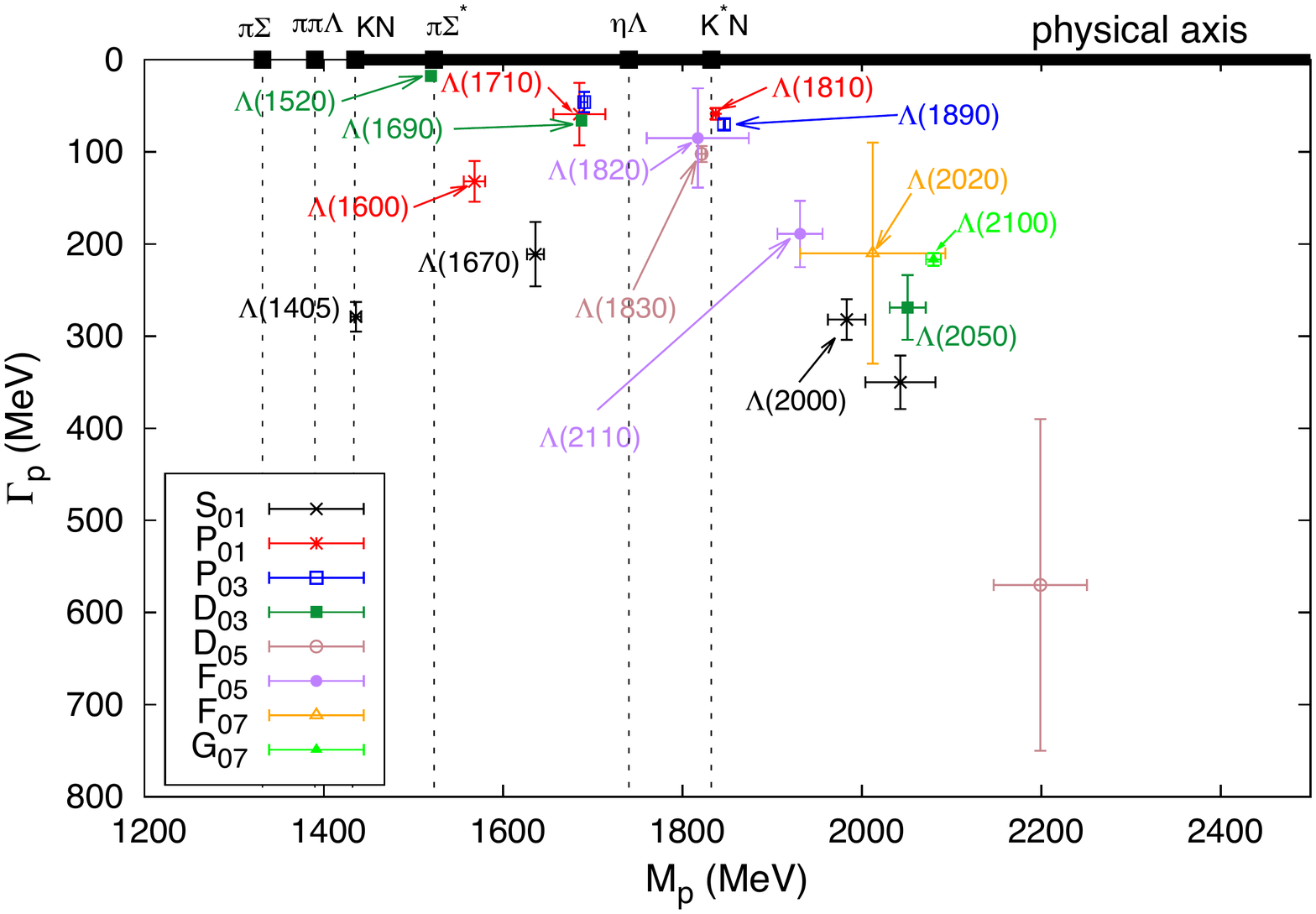}}} \label{fig:poles0}}
\subfigure[\ $\Sigma^*$ resonances. ]{
\rotatebox{0}{\scalebox{0.3}[0.3]{\includegraphics{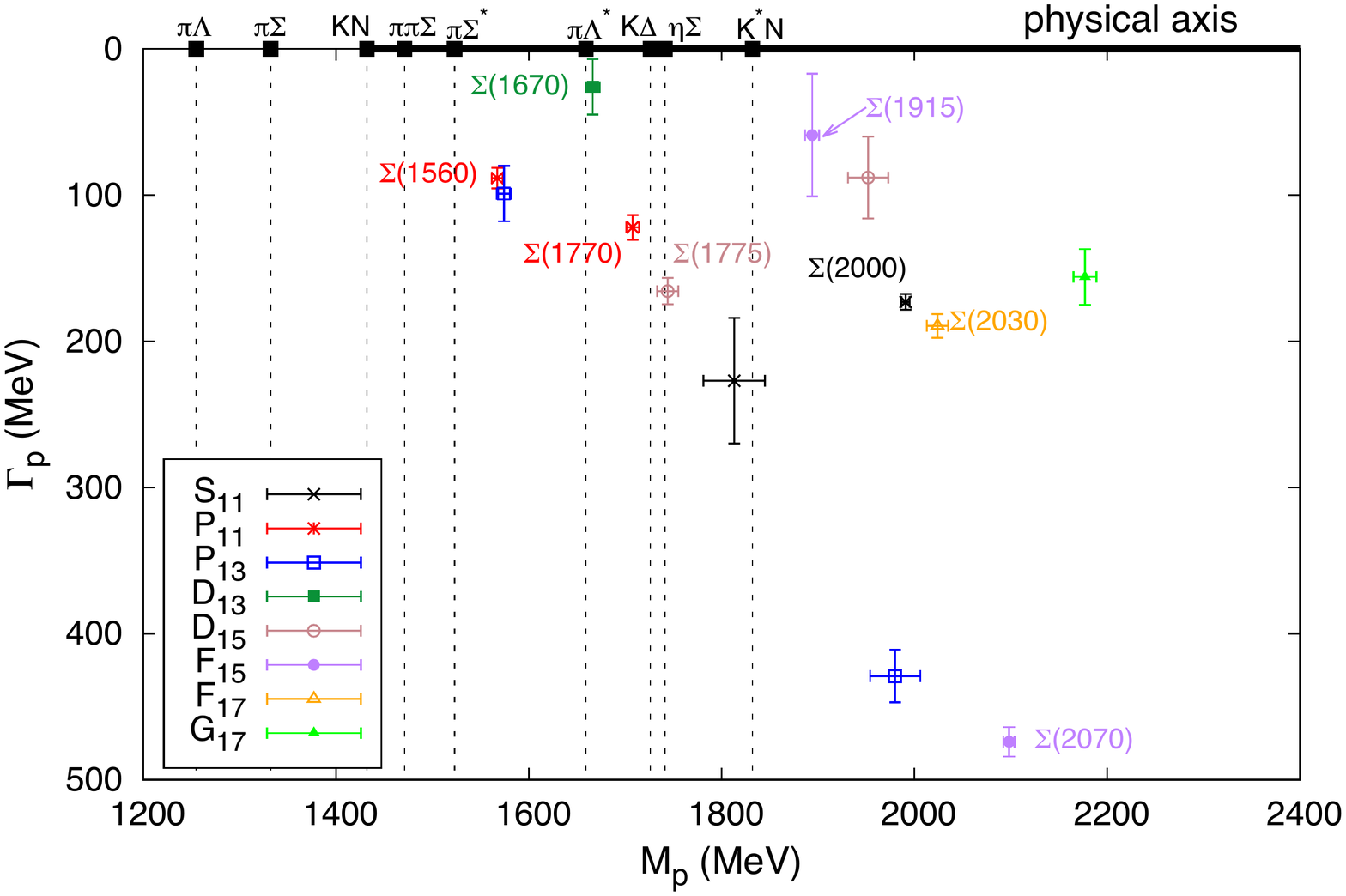}}}   \label{fig:poles1}}
\caption{(color online). Poles for $I=0$ (a) and $I=1$ (b) partial waves from 
Tables \ref{tab:poles0} and \ref{tab:poles1} except those 
with a very large imaginary part and those
marked with \textdaggerdbl ~(believed to be artifacts of the fits). 
Poles are computed in the unphysical Riemann sheet where all the available cuts have been crossed 
(nearest Riemann sheet to the physical amplitude) and their $\ell_{I\:2J}$ quantum numbers are provided.
The different thresholds are highlighted as vertical dashed lines 
and in the physical axis as filled black boxes where  $K^*$ stands for $K^*(892)$,
$\Delta$  for $\Delta (1232)$, $\Sigma^*$ for $\Sigma (1385)$ and $\Lambda^*$ for $\Lambda(1520)$.
The last is treated as a stable state and therefore an accessible decay channel 
in the $I=1$ channels although in the $I=0$ it is a resonance
 whose properties emerge from our analysis.} \label{fig:poles}
 \end{center}
\end{figure}

\begin{figure}
\centering
\subfigure[\  $\Lambda^*$ resonances.]{
\rotatebox{0}{\scalebox{0.3}[0.3]{\includegraphics{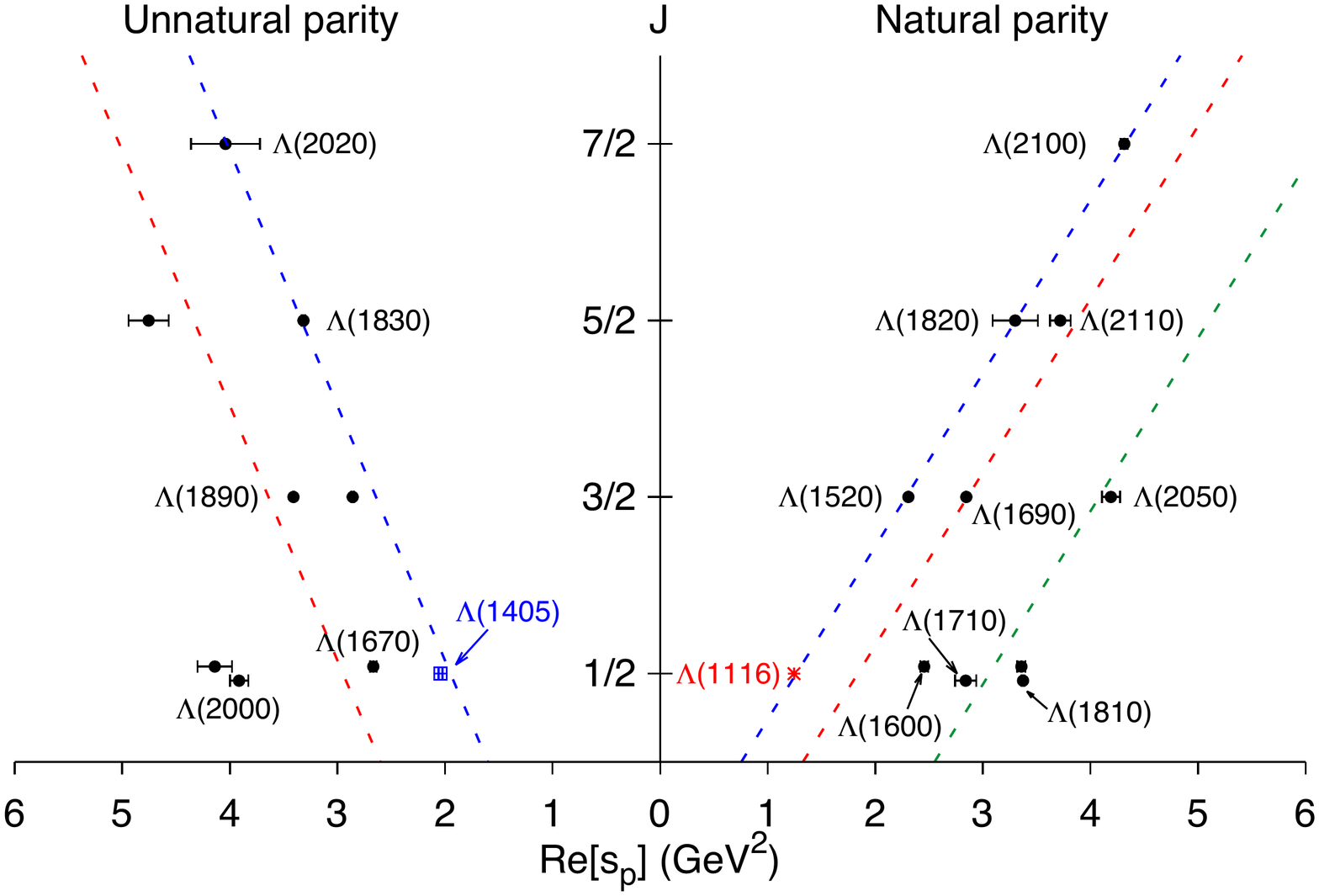}}} \label{fig:poles0regge}}
\subfigure[\ $\Sigma^*$ resonances. ]{
\rotatebox{0}{\scalebox{0.3}[0.3]{\includegraphics{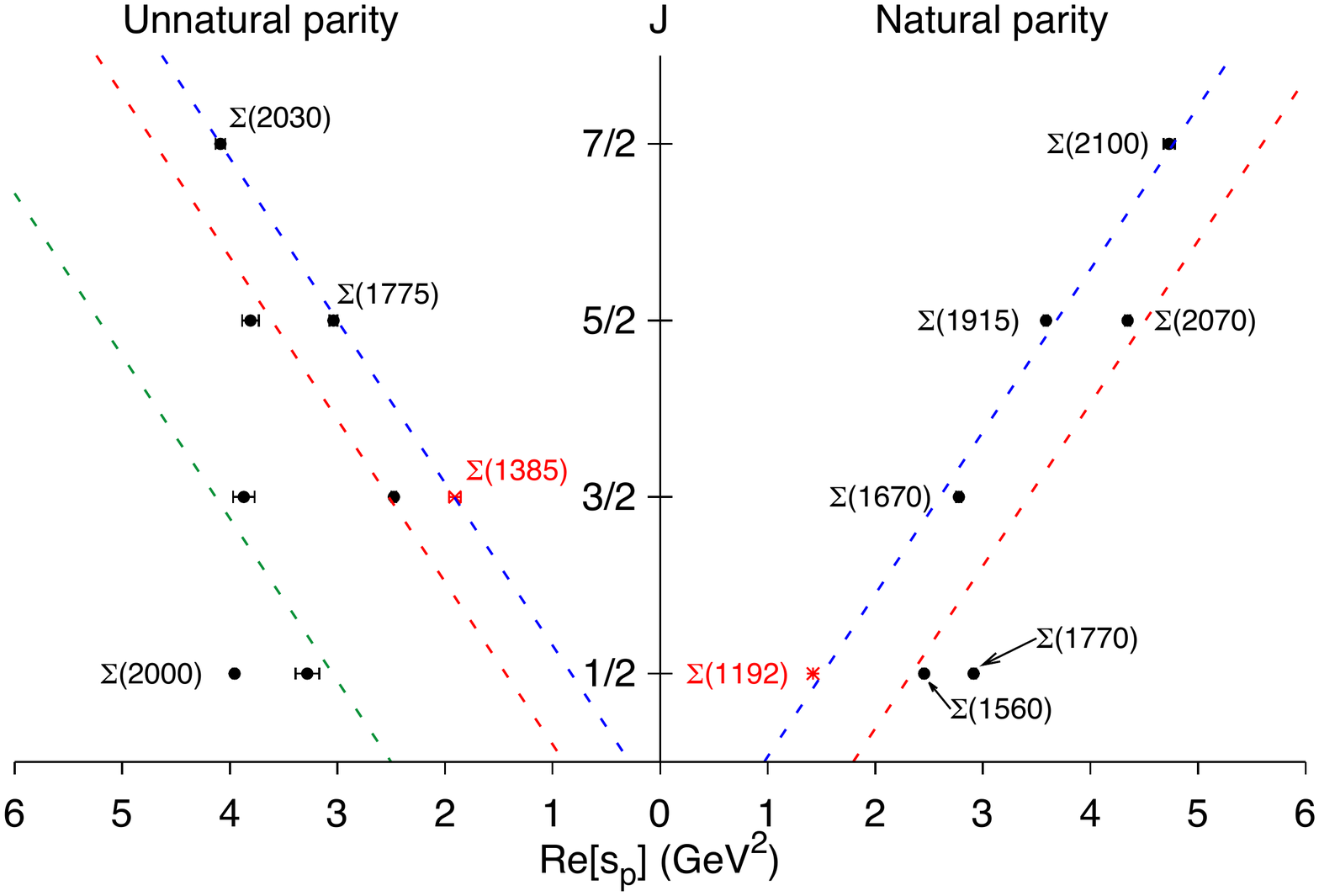}}}  \label{fig:poles1regge}}
\caption{(color online).  Chew--Frautschi plot for the  $\Lambda^*$ (a) and  $\Sigma^*$ (b) resonances.
Poles are displayed according to positive (natural) or negative (unnatural) naturality ($\eta$).
Poles colored in red,
\textit{i.e.} $\Lambda(1116)$, $\Sigma(1192)$ and $\Sigma(1385)$, 
are taken from RPP \cite{PDG2014}. 
The $\Lambda(1405)$ is displayed in blue to highlight that in our approach it is an effective state
that mimics two actual resonances (see text).
Dashed lines are displayed to guide the eye through the Regge trajectories. 
Blue lines guide the eye through the parent Regge trajectories while 
red and green guide through the daughter trajectories.} \label{fig:polesregge}
\end{figure}

The structure found in the partial waves is due to the appearance of poles (resonances)
in the $T$ matrix when extended to 
the unphysical Riemann sheets. These resonances are 
shown at the bottom in Figs.~\ref{fig:PW_S01P01}--\ref{fig:PW_D15F15}.
The poles are obtained by computing zeros of  $\mathcal{D}(s)=0$ 
in the nearest unphysical Riemann sheet defined by the crossing of all the available unitarity cuts.
$\mathcal{D}(s)$ is defined in Eq.~(\ref{eq:ds}).
In Tables \ref{tab:poles0} and \ref{tab:poles1} 
we summarized the obtained pole positions,
--in the usual notation of masses and widths--, and we compare
our results to the KSU model \cite{Manley13b}
and models $A$ and $B$ from Kamano \textit{et al.} \cite{Kamano15} 
(referred to as KA and KB models in what follows). We also give 
 a possible relation to the resonances listed by the  RPP   \cite{PDG2014}.
The poles in the analyses of  Kamano \textit{et al.}  are based 
on a dynamical coupled-channel model described in  \cite{Kamano14}.
Because we use the same single-energy partial waves as the KSU model one
would expect a fairly good agreement between the two analyses. 
There indeed is an agreement for some of the well-established resonances, but
several important discrepancies are found in  the remaining states, which we discuss 
in this  section.
In Fig.~\ref{fig:poles} we show the resonances from Tables
 \ref{tab:poles0} and \ref{tab:poles1} 
 (except those with very large imaginary part and those believed to be artifacts of the fits)  and
in Fig.~\ref{fig:polesregge} we show the real part of the pole positions on the  Chew--Frautschi plot. 

\begin{table*}
\caption{Summary of $\Lambda^*$ pole masses ($M_p=\text{Re}\: \sqrt{s_p}$) 
and widths ($\Gamma_p=-2 \: \text{Im}\: \sqrt{s_p}$) in MeV.
Our poles are depicted in Fig.~\ref{fig:poles}
unless they have a very large imaginary part or are considered artifacts. 
In \cite{Qiang10} the $\Lambda (1520)$ pole was obtained at ($M_p=1518.8$, $\Gamma_p=17.2$).
Ref.~\cite{Kamano15} implements two models labeled 
as KA and KB (see text).
$I$ stands for isospin, $\eta$ for naturality, $J$ for total angular momentum, 
$P$ for parity, and $\ell$ for orbital angular momentum. 
For baryons, $\eta=+$, natural parity, if $P=(-1)^{J-1/2}$
and $\eta=-$, unnatural parity, if $P=-(-1)^{J-1/2}$.
Resonances marked with \textdagger ~are unreliable due to systematics
and lack of good-quality $\chi^2/\text{dof}$.
Resonances marked with \textdaggerdbl ~are most likely artifacts of the fits.} \label{tab:poles0}
\begin{ruledtabular}
\begin{tabular}{ccccccccccc}
& \multicolumn{2}{c}{This work}
& \multicolumn{2}{c}{KSU from \cite{Manley13b}}  
& \multicolumn{2}{c}{KA  from \cite{Kamano15}}
& \multicolumn{2}{c}{KB from \cite{Kamano15}}
& \multicolumn{2}{c}{RPP \cite{PDG2014}} \\
$I^\eta \: J^P \: \: \ell$ & 
$M_p$ &$\Gamma_p$&
$M_p$ &$\Gamma_p$&
$M_p$ &$\Gamma_p$&
$M_p$ &$\Gamma_p$ & Name & Status\\
\hline
$0^-\: \frac{1}{2}^-\:S$ 
               & $1435.8 \pm 5.9^\dagger$ & $279 \pm 16$&1402&49&---& ---&---&---&$\Lambda (1405)$& ****\\
               & $ 1573^\ddagger$ & $300$&---&---&---& ---&1512&370&---& ---\\
               & $ 1636.0\pm 9.4^\dagger $ & $211 \pm 35$&1667& 26&1669&18&1667&24&$\Lambda (1670)$& ****\\
               & --- & ---& 1729& 198&---&---&---&---&  $\Lambda (1800)$&***\\
               & $1983 \pm 21^\dagger$ & $282 \pm 22$ &1984&233 & ---& ---&---&---& $\Lambda (2000)$&*\\
               & $ 2043 \pm 39^\dagger$ & $350 \pm 29$&---&---&---& ---&---&---&---& ---\\
$0^+\:\frac{1}{2}^+\:P$ & $1568 \pm 12$&$132 \pm 22 $   &1572& 138&1544&112&1548&164& $\Lambda (1600)$&***\\
                & $1685  \pm  29^\dagger$&$59 \pm 34$ &1688&166&---&---&---&--- &$\Lambda (1710)$&*\\
                & $1835 \pm 10^\ddagger$&$180 \pm 22$ &---&---&---&---&---&---&---& ---\\                
                & $1837.2 \pm 3.4^\dagger$&$58.7\pm 6.5$ &1780&64&---&---&1841&62 & $\Lambda (1810)$&***\\
                &---&---&2135&296&2097&166&---&---& ---&---\\               
$0^-\:\frac{3}{2}^+\:P$ & $1690.3 \pm 3.8 $&$ 46.4 \pm 11.0$ & ---& ---&---&---&1671&10 &---&---\\
                & $1846.36 \pm 0.81$&$70.0 \pm 6.0$ &1876 &145&1859&112&---&---& $\Lambda (1890)$ &****\\ 
                & ---&--- &2001 &994&---&---&---&---& --- &---\\ 
$0^+\:\frac{3}{2}^-\:D$ & $1519.33 \pm 0.34$&$17.8 \pm 1.1$ & 1518&16 &1517&16&1517&16& $\Lambda (1520)$&****\\
                & $1687.40 \pm 0.79$&$66.2 \pm 2.3$ &1689&53&1697&66&1697&74& $\Lambda (1690)$&****\\ 
                & $2051 \pm 20$&$269 \pm 35$ &1985&447&---&---&---&---&$\Lambda (2050)$&* \\ 
                & $2133 \pm 120^\ddagger$&$1110  \pm  280$&---&--- &---&---&---&---&$\Lambda (2325)$& *\\
$0^-\:\frac{5}{2}^-\:D$ & $1821.4 \pm 4.3 $&$102.3 \pm 8.6$ &1809&109&1766&212&---&---&$\Lambda (1830)$&****\\
                &---&--- &1970&350&1899&80&1924&90& ---&--- \\
                &$2199 \pm 52 $&$  570 \pm 180$&--- &---&---&---&---&---& ---& ---\\
$0^+\:\frac{5}{2}^+\:F$ & $1817 \pm 57$&$  85 \pm 54 $ &1814&85&1824&78&1821&64& $\Lambda (1820)$&****\\
                & $1931 \pm 25$&$189\pm 36$ &1970&350&---&---&---&---&$\Lambda (2110)$&***\\     
$0^-\:\frac{7}{2}^+\:F$ & ---&--- &---&---&1757&146&---&---&---&---\\
               & $2012 \pm 81 $&$210 \pm 120 $ &1999&146&---&---&2041&238&$\Lambda (2020)$&*\\
$0^+\:\frac{7}{2}^-\:G$& $2079.9 \pm 8.3 $&$216.7 \pm 6.8 $&2023&239&---&---&---&---& $\Lambda (2100)$&****\\
\end{tabular}
\end{ruledtabular}
\end{table*}

\begin{table*}
\caption{Summary of $\Sigma^*$ pole masses ($M_p=\text{Re}\: \sqrt{s_p}$) and widths 
($\Gamma_p=-2 \: \text{Im}\: \sqrt{s_p}$) in MeV.
Our poles are depicted in Fig.~\ref{fig:poles} 
unless they have a very large imaginary part. Notation is the same as in Table \ref{tab:poles0}.
Resonances marked with \textdagger ~are unreliable due to systematics
and lack of good-quality  $\chi^2/\text{dof}$.} \label{tab:poles1}
\begin{ruledtabular}
\begin{tabular}{ccccccccccc}
& \multicolumn{2}{c}{This work}
& \multicolumn{2}{c}{KSU from \cite{Manley13b}}  
& \multicolumn{2}{c}{KA  from \cite{Kamano15}}
& \multicolumn{2}{c}{KB from \cite{Kamano15}}
& \multicolumn{2}{c}{RPP \cite{PDG2014}} \\
$I^\eta \: J^P \: \:  \ell$ & 
$M_p$ &$\Gamma_p$&
$M_p$ &$\Gamma_p$&
$M_p$ &$\Gamma_p$&
$M_p$ &$\Gamma_p$ & Name & Status\\
\hline
$1^- \: \frac{1}{2}^-\:S$ & ---&--- &1501&171&---&---&1551&376&$\Sigma (1620)$&*\\      
               &--- &--- &1708&158&1704&86&---&---& $\Sigma (1750)$&***\\
               &$1813 \pm 32^\dagger $&$ 227 \pm   43$  & ---&---&---&---&---&---&  ---&---\\        
               &---&---&1887&187&---&---&---&---&  $\Sigma (1900)$&*\\ 
               &$1990.8 \pm 4.3^\dagger $&$173.1 \pm 5.4$ &---&---&---&---&1940&172 & $\Sigma (2000)$&*\\ 
               & ---&--- &2040&295&---&---&---&---&---&---\\ 
$1^+ \: \frac{1}{2}^+\:P$ 
               & $1567.3 \pm 5.7$&$88.4 \pm 7.0$&---&---&1547&184&1457&78& $\Sigma (1560)$&**\\
               &---&---&---&---&---&---&---&---&$\Sigma (1660)$&***\\
               &$1707.7 \pm 6.6$&$122.1 \pm 8.5$&1693&163&1706&102&---&---& $\Sigma (1770)$&*\\                              
               &---&---  &1776&270&---&---&---&---& $\Sigma (1880)$&**\\               
               &---&--- &---&---&---&---&2014&140& ---&---\\  
$1^- \: \frac{3}{2}^+\:P$ & $1574.1 \pm 7.2$&$99 \pm 19$&--- &---&---&---&---&---&---&--- \\
               &---&---&1683&243&---&---&---&---&--- &--- \\
               & ---&---&1874&349&---&---&---&---&---&--- \\                                 
               & $1980 \pm  26$&$429\pm 18$ &---&---&---&---&---&---&---&--- \\                  
$1^+ \: \frac{3}{2}^-\:D$ &---&---  &---&---&1607&252&1492&138& $\Sigma (1580)$ &*\\
               & $1666.3 \pm 7.0$&$26 \pm 19$  &1674&54&1669&64&1672&66&$\Sigma (1670)$&****\\
               & ---&--- &---&---&---&---&---&---&$\Sigma (1940)$&***\\
$1^- \: \frac{5}{2}^-\:D$ & $1744 \pm 11$&$165.7 \pm9.0 $  &1759&118&1767&128&1765&128& $\Sigma (1775)$&****\\
               & $1952 \pm 21$&$88 \pm 28$&2183&296&---&---&---&---& ---&--- \\
$1^+ \: \frac{5}{2}^+\:F$ &---  &---&---&--- &---&---&1695&194&---&--- \\
               & $1893.9 \pm 7.2$&$59 \pm 42$&1897&133 &1890&99&---&--- & $\Sigma (1915)$&****\\
               &$2098.2 \pm 5.8$&$474 \pm 10$ &2084&319&---&---&---&---&$\Sigma (2070)$&*\\   
$1^- \: \frac{7}{2}^+\:F$ & $2024  \pm  11$&$189.5 \pm   8.1$ &1993&176&2025&130&2014&206& $\Sigma (2030)$&****\\
$1^+ \: \frac{7}{2}^-\:G$ & $2177 \pm 12$&$156 \pm19$  &2252&290&---&---&---&---& $\Sigma (2100)$&*\\     
\end{tabular}
\end{ruledtabular}
\end{table*}

As explained in Section \ref{sec:analyticstructure}, in our model there are no poles 
 on the first Riemann sheet except for those on the real axis below thresholds 
 parameterizing the left-hand cut. 
These poles, in most of the cases were found to be  far away from the physical region. 
The poles closest to the physical region are found in 
$F_{05}$ at $-0.45$ GeV$^2$, 
$D_{13}$ at $-0.41$ GeV$^2$,
$P_{03}$ at $0.08$ GeV$^2$,
$S_{11}$ at $0.31$ GeV$^2$,
$S_{01}$ at $0.38$ GeV$^2$, and
$P_{13}$ at $0.88$ GeV$^2$,
and they all produce a smooth behavior in the  physical region. 

The resonance poles are mainly responsible for giving
structure to the partial waves on the real axis. 
Therefore, when the fit is not very good the model \textit{tries}  to smear the structures
that it is not able to reproduce.  If we take a set of parameters
(far from the best-fit parameters but not too far) 
in a certain partial wave and we do a pole search
it is likely that we find fewer resonances than for the best fit. 
As $\chi^2/\text{dof}$ improves, more resonances appear. 
If we overfit the data, we start to identify as structure some variations 
in the data that could potentially be identified as a statistical noise instead of genuine resonances. 
Hence, pole extraction from under-fitted and over-fitted waves has to be treated with care.

\subsubsection{$\Lambda^*$ Resonances} \label{sec:lambdaresonances}
All the $\Lambda^*$ resonances obtained are summarized in Table \ref{tab:poles0}
and almost all are displayed in 
Figs.~\ref{fig:poles0} and \ref{fig:poles0regge} (see respective captions for details).
Throughout this section pole masses and widths are reported in MeV unless stated otherwise.

$\mathbf{S_{01}}$ \textbf{poles.} 
Besides the $\Lambda (1405)$  (which was imposed as explained in Section \ref{sec:fits})
we find four resonances in our best fit of the  $S_{01}$  partial wave.
The first one at $1573 -i \: 300/2$ is close to 
the one obtained by model KB at $1512-i\: 370/2$ and is not obtained by any other model.
We believe it is an artifact of the fit
because when we perform the bootstrap to obtain the error bars, 
it does disappears from most of the fits.
Hence, we do not quote an error bar for it in Table  \ref{tab:poles0} and we do not show it
in Figs.~\ref{fig:PW_S01P01}, \ref{fig:poles0}, and \ref{fig:poles0regge}.
Two of the other poles can be associated with $\Lambda (1670)$ and $\Lambda (2000)$ states 
in the RPP. The $\Lambda (2000)$ pole agrees with KSU analysis and is not found either by KA or KB.
The $\Lambda (1670)$ has a  four-star status in the RPP.
The mass we obtain is within a reasonable  range when compared to the 
KSU, KA, and KB analyses, although 
our width is larger with a sizable uncertainty. 
This pole appears in the energy region where our model does not reproduce properly the abrupt
change in the $\bar{K}N\to\pi \Sigma$ channel around 2.8 GeV$^2$ 
(see left column in Fig.~\ref{fig:PW_S01P01}) so the width we obtain is not very reliable.
We also find a higher-energy resonance that no other analysis finds. 
Further confirmation of its existence is needed.
Neither us nor KA nor KB find a pole close to the $1729 -i\: 198/2$,  $\Lambda (1800)$, found in KSU analysis.
This pole has a three-star status and, considering the large systematic uncertainties 
of the $S_{01}$ wave its status might need to be reconsidered in the RPP.
Nevertheless, because of the systematic uncertainties and the high $\chi^2/\text{dof}$ 
we cannot make definitive statements on the $S_{01}$ pole locations.

$\mathbf{P_{01}}$ \textbf{poles.} 
In the $P_{01}$ partial wave we find four resonances. 
The lowest-lying is at $1568 - i132\slash 2$, which corresponds to the three-star $\Lambda (1600)$.
All the analysis, KSU, KA, KB and us, agree on the location of this resonance within their uncertainties 
making it a very well established state.
However, when we try to identify which Regge trajectory it belongs to
(see Fig.~\ref{fig:poles0regge}) it looks like it does not match the general pattern.
This signals that the $\Lambda (1600)$ resonance is of different nature than the other resonances
we are finding and that it is not an ordinary three-quark state.
The $\Lambda (1710)$ state was first introduced by KSU analysis and we find a pole at similar mass
but closer to the real axis. This state needs further confirmation through an independent analysis
given that we obtain a smaller width than KSU model and we both fit the same single-energy partial waves.
However, our result together with those of KSU and \cite{rhohyperon} reinforce the hypothesis that 
there are two poles in the  $P_{01}$ partial wave for $\sqrt{s}<1.9$ GeV.
In  \cite{rhohyperon} a non-three-quark nature is suggested for both states.
We also obtain a pole at $1837.2-i 58.7\slash 2$ that can be identified as the $\Lambda (1810)$ state
and is in very good agreement with the pole at $1841-i 62\slash 2$ obtained by the KB model.
The reliability of the $\Lambda (1810)$ pole position
can be questioned due to the appearance of a pole at
$1835-i\:180/2$ that looks like an artifact linked to the opening of the $\bar{K}^*N$ threshold.
The $\Lambda (1710)$ and $\Lambda (1810)$ extractions are not very reliable
given the high value of the $\chi^2/\text{dof}$,  the discrepancies with other analysis and 
how they do not fit within the Regge trajectories
in Fig.~\ref{fig:poles0regge} ($J=1/2$, natural parity) while the higher-lying resonances do.

$\mathbf{P_{03}}$ \textbf{poles.} 
The $P_{03}$ is a very interesting case regarding the interplay of resonances,  
fits and Regge trajectories.
First it has to be noted that this particular partial wave is dominated by inelasticities,
which make the extraction of the single-energy partial waves and the poles very
challenging. This is notorious if we see how scattered are the data in
the $\bar{K}N\to \pi \Sigma$ channel in Fig.~\ref{fig:PW_S01P01} (center-right column).
During the fitting process
we first obtained a solution with $\chi^2/\text{dof}=1.65$ with poles located at 
$3.580-i0.213$ GeV$^2$ ($1893 - i113\slash 2$) and $3.3169-i1.450$ GeV$^2$
($1862 - i779\slash 2$). 
In the KSU analysis, two poles were  also obtained, located at  
$1876- i145\slash 2$ and $2001- i994\slash 2$.
This first solution was smoother than the one we report 
and it did not show the apparent peak at 3 GeV$^2$ 
in the $\bar{K}N\to \pi \Sigma$ channel.
The first pole is a good candidate for the $\Lambda (1890)$ state and
is compatible with KSU analysis.
The second looks like an artifact because its mass is smaller than 
for the first pole and its width is larger. Also, it is very different from what
was obtained by KSU model,
suggesting that this second pole may be an artifact in both analyses.
Hence, we exchanged one of the pole $K$ matrices with a background $K$ matrix
to check what was the effect in the $\chi^2$ and the appearance of poles in the $T$ matrix.
The results were systematically worse and the $T$ matrix still presented two poles.
So, we conclude that data require the existence of two poles.
The location of the second pole was not satisfactory so we performed a new fit influenced by the
expected Regge behavior in Fig.~\ref{fig:poles0regge} guiding the fit to provide a pole with $M_p^2$ 
within 2 and 3 GeV$^2$ that would fill in the $3\slash 2^+$ gap in the unnatural parity parent trajectory.
We note that we did not impose any restriction in the imaginary part of the pole.
In this way we obtained the solution presented in Fig.~\ref{fig:PW_S01P01} with a
marginally better $\chi^2/\text{dof}=1.64$
and the resonances shown in Fig.~\ref{fig:poles} with more reasonable widths.
The real part of the parent Regge trajectory was slightly improved,
although, as shown in Fig.~\ref{fig:poles0regge},  
there is some tension between what we expect from linear Regge behavior.
For all these reasons, we consider this second fit to be more reliable and it is the one we report.
If we compare our $P_{03}$ poles to those in KSU and KA, we find a reasonable agreement with
the masses for the four-star $\Lambda (1890)$ state 
although our width is significantly smaller than in the other analyses. 
Guided by our Regge analysis we report a $P_{03}$ state at $1690$ MeV. 
In \cite{lambda1680} a similar $P_{03}$ state with mass 1680 MeV was found, although with a larger width.
The KB model reports
a state close in mass, at $1671$ MeV with a very small width, however, 
this result should be taken with care because for the same model no $\Lambda (1890)$ is obtained.
We found no evidence of the large-width state at $2001$ MeV reported by KSU analysis.
As a conclusion, we are convinced that the two poles have to be present in this partial wave
and lie in the regions where we obtained them, although the error bars might be underestimated
given that the $\chi^2/\text{dof}$ is larger than one.

$\mathbf{D_{03}}$ \textbf{poles.} 
This partial wave is modeled with three pole $K$ matrices and one background $K$ matrix.
Four $T$-matrix poles are obtained.
Two of them correspond to well-established states: $\Lambda (1520)$ and $\Lambda (1690)$.
These extractions agree very well with those in KSU, KA, KB and \cite{Qiang10} 
(which only computes $\Lambda (1520)$), as it should
for such well-established states. 
Any difference can be associated with model details. 
The third pole obtained can be matched to the $\Lambda (2050)$ state, which was
first obtained in the KSU analysis although it is not found in either KA or KB. 
However, we obtain a very different pole position, 
which can be understood if we realize that the
deeper in the complex plane we need to go to find a resonance,
the more important analyticity and model dependence become. 
Finally, we obtain a higher-energy and deep in the complex plane pole  
($M_p=2133$, $\Gamma_p=1110$).
It is likely that this state is an artifact of the fits although its
quantum numbers and mass would befit the one-star $\Lambda (2325)$ in RPP
(but not its width, which is reported to be $\sim 150$ MeV).

$\mathbf{D_{05}}$ \textbf{poles.} 
The four-star $\Lambda (1830)$ is obtained in the $D_{05}$ 
partial wave and our result agrees with the one obtained by
KSU model. Model KA also obtains this pole, although at smaller mass (1766) 
and larger width ($\Gamma_p \slash 2=106^{+47}_{-31}$).
However, the associated uncertainites are not small enough to consider the disagreement worrisome.
We obtain a second pole as KA and KSU model do, but
the three analyses find this pole at very different locations.
Hence, we can conclude that this second pole in the partial wave 
does exist but its exact position is debatable.

$\mathbf{F_{05}}$ \textbf{poles.} 
According to the RPP, the $F_{05}$ partial wave contains one four-star resonance, $\Lambda (1820)$,
and one three-star resonance, $\Lambda (2110)$. This is not obvious from Fig.~\ref{fig:PW_D05F05} because
the partial wave looks like a one well-isolated resonance instead of the combination of two states.
All the analyses find the $\Lambda (1820)$ at the same location within uncertainites. 
The $\Lambda (2110)$ is  a good example of how a resonance can 
show up in a partial wave without a bump when it is deep in the complex plane. 
The fact that both our analysis and KSU require the $\Lambda (2110)$ 
ratifies its three-star status, although the exact location is debatable.

$\mathbf{F_{07}}$ \textbf{and} $\mathbf{G_{07}}$ \textbf{poles.} 
Both the KSU model and us fit the same single-energy partial waves from \cite{Manley13a},
hence we are both biased by such extraction and we should be obtaining similar results 
for the simplest cases.
$F_{07}$ and $G_{07}$ partial waves present 
a clear resonant structure (see Fig.~\ref{fig:PW_D05F05})
that can be well reproduced with just one pole $K$ matrix and one background $K$ matrix. 
Both analyses yield similar resonance positions compatible within uncertainties.
The $\Lambda (2020)$ ($F_{07}$) state obtained in KSU, awarded a one-star status by the RPP,
gains further confirmation on existence and pole position
by both our analysis and KB. 

\subsubsection{$\Sigma^*$ Resonances}
All the $\Sigma^*$ resonances obtained are summarized in Table \ref{tab:poles1}
and displayed in Figs.~\ref{fig:poles1} and \ref{fig:poles1regge}  (see respective captions for details).
Throughout this section pole masses and widths are reported in MeV unless stated otherwise.

$\mathbf{S_{11}}$ \textbf{poles.} 
Our fit to the $S_{11}$ partial wave has large uncertainties.
Hence, resonances existence, their location and errors should be taken with care.
For example, the resonance that we get at $1813-i227\slash 2$ has large error bars both for the real and the imaginary part
and no other analysis finds a similar state. It is a state that should not be taken as well founded.
Contrary to KSU and KB (KA) analyses we find no evidence of the $\Sigma (1620)$ ($\Sigma (1750)$) state. 
Also in \cite{sigma1620} no evidence of $\Sigma (1620)$ was found.
We find a resonance compatible with KB analysis whose most likely RPP assignment is
$\Sigma (2000)$ and we do not find any evidence of the $\Sigma (1900)$ state.
Our model does not incorporate $\rho$-hyperon channels, which 
were found in the model of \cite{rhohyperon}
to couple strongly to 
$\Sigma(1620)$, $\Sigma(1750)$, and $\Sigma(1900)$ states.  
This fact could be the reason why we do not find such states in our analysis.

$\mathbf{P_{11}}$ \textbf{poles.} 
In the $P_{11}$ partial wave we find two resonances that we match to the 
$\Sigma (1560)$ and the $\Sigma (1770)$ states in RPP.
The KSU analysis does not find a resonance that can be matched 
to $\Sigma (1560)$ while our analysis, KA, and KB do,
although with very different values of the mass and the width.
The $\Sigma (1770)$ state is also found by KSU and  KA models, the latter agreeing 
with our analysis for both the mass and the width. 
The KSU analysis provides a larger width and a smaller mass, but not far from ours.
None of the analyses finds evidence of the three-star state $\Sigma (1660)$.
However, as suggested by \cite{p01molecule},
additional information from three-body decay channels
(\textit{e.g.} $\pi \bar{K} N$ and $\pi \pi \Sigma$),
might be important to establish the existence/properties of $\Sigma (1660)$.
In  \cite{sigma1620} a $\Sigma (1635)$ is found to be necessary while we find two states 
in the same energy region at 1567 and 1708 MeV.
Neither our calculation nor KA nor KB  find the higher energy state that KSU assigns to $\Sigma(1880)$.

$\mathbf{P_{13}}$ \textbf{poles.} 
States that contribute to the $P_{13}$ are controversial. 
We find two resonances in this partial wave, KSU also finds two resonances at different locations
and KA and KB find no resonances. The strongest argument in favor of the existence of these states
comes from the unnatural parity  daughter  $\Sigma^*$ Regge trajectories in Fig.~\ref{fig:poles1regge}, 
which requires two states at the approximate masses we report.

$\mathbf{D_{13}}$ \textbf{poles.} 
To describe the $D_{13}$ partial wave we employed one pole 
and two background $K$ matrices.  We find only one resonance at
$1666.3-i26\slash 2$ that corresponds to the four-star $\Sigma (1670)$ resonance. 
The same state is also found in the KSU, KA, and KB analyses with a larger width 
on average, although all compatible within errors.
In \cite{Kamano15} a low-lying state in both the KA and KB models with very large width was found, 
that can be matched to one-star resonance $\Sigma (1580)$.
Neither we nor the other analyses, KSU, KA, or KB obtain 
the three-star $\Sigma (1940)$ state, which sheds doubts on its existence. 
However, Fig.~\ref{fig:poles1regge} presents a gap in the natural parity daughter trajectory
suggesting that $\Sigma (1940)$ should be there.
In \cite{rhohyperon}, $\Sigma (1940)$ was found to couple to the $K^* \Xi$ channel,
which was not included in neither in the KSU, KA, KB, or present analyses. This 
 could  explain why
none of the global $\bar{K}N$ coupled-channel analyses finds it.
This state requires further experimental information and analysis before any definitive statement can be made.

$\mathbf{D_{15}}$ \textbf{poles.}
We find two resonances in the $D_{15}$ partial wave.
One corresponds to the four-star $\Sigma (1775)$ state,
which was also found by KSU, KA, and KB. KSU also finds a second resonance 
in this partial wave, but it appears
at a very different location. Hence, the existence of this second state is dubious.

$\mathbf{F_{15}}$  \textbf{poles.} 
KSU, KA, and we agree within errors on the $\Sigma (1915)$ state for the $F_{15}$ partial wave and 
we get a similar result for $\Sigma (2070)$  to the one of KSU.

$\mathbf{F_{17}}$ \textbf{poles.} 
The $F_{17}$ partial wave provides a very clean resonant signal 
and all the analysis obtain reasonably compatible results as expected 
for $\Sigma (2030)$, a four-star state.

$\mathbf{G_{17}}$ \textbf{poles.} The $G_{17}$ partial wave has too large uncertainties 
to be able to make a conclusive determination.
However, the mass we obtain fits very nicely within the natural parity Regge trajectory in Fig.~\ref{fig:poles1regge}. 

\subsubsection{Regge Trajectories}
From Fig.~\ref{fig:polesregge} it is apparent that there is an alignment of the resonances in Regge
trajectories. We are employing the real part of the extracted poles
and not Breit--Wigner masses as has been customary  \cite{reggetrajectories}.
It should be noted that each line displayed in Fig.~\ref{fig:polesregge}
actually contains two degenerate Regge trajectories, \textit{e.g.} in Fig.~\ref{fig:poles0regge} the 
parent trajectory, $I^\eta=0^+$ 
  $\Lambda (1116)$ and $\Lambda (1820)$ correspond to one trajectory and
$\Lambda (1520)$ and $\Lambda (2100)$ correspond to another.
The conclusions we can derive from Fig.~\ref{fig:polesregge}  are as follows: 
\begin{itemize}
\item[(i)] We have a fairly accurate and comprehensive picture
of the $Y^*$ spectrum for the parent Regge trajectories up to $J=7/2$. 
\item[(ii)] Our knowledge of the first daughter Regge trajectories up to $J=5/2$ is also good
except for the lowest ($J=1/2$) natural parity $\Lambda^*$ state associated to the $P_{01}$ partial wave,
for the gap at $J=3/2$ (connected to the $\Sigma (1940)$ resonance) in the $\Sigma^*$ 
natural trajectory associated to the $D_{13}$ partial wave, 
and the possible existence of the  a $S_{11}$ state that 
would constitute its lowest-energy state. 
\item[(iii)] The $\Lambda (1600)$ pole position is very well established and it does
not fit within the daughter $0^+$ linear Regge trajectory. 
Its nature seems to be different from that of the other resonances that do follow
the linear Regge trajectories, signaling a non-three quark nature.
\end{itemize}

\subsection{Comparison to Experimental Data} \label{sec:experiment}
\begin{figure}
\rotatebox{0}{\scalebox{0.4}[0.4]{\includegraphics{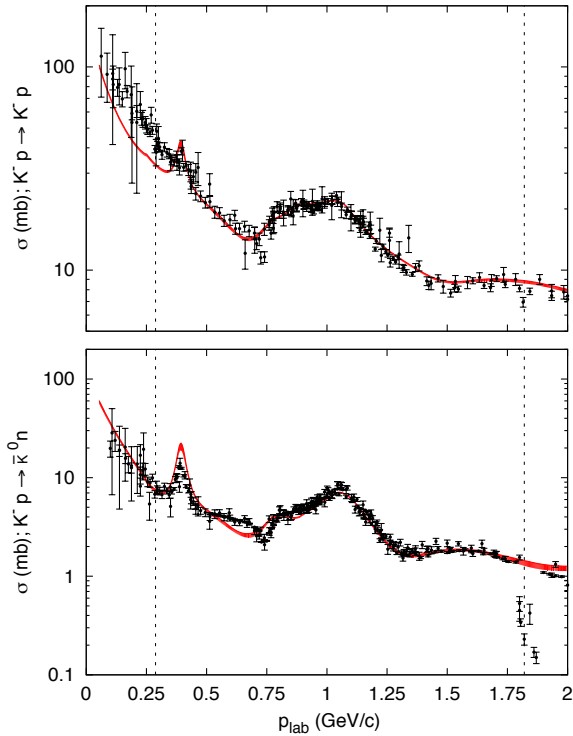}}}
\caption{(color online). Cross sections  for $K^-p\to K^-p, \bar{K}^0n$ processes. 
The vertical dashed lines mark the energy range 
where the single-energy partial waves have been fitted.} \label{fig:xseclog1}
\end{figure}

\begin{figure}
\rotatebox{0}{\scalebox{0.4}[0.4]{\includegraphics{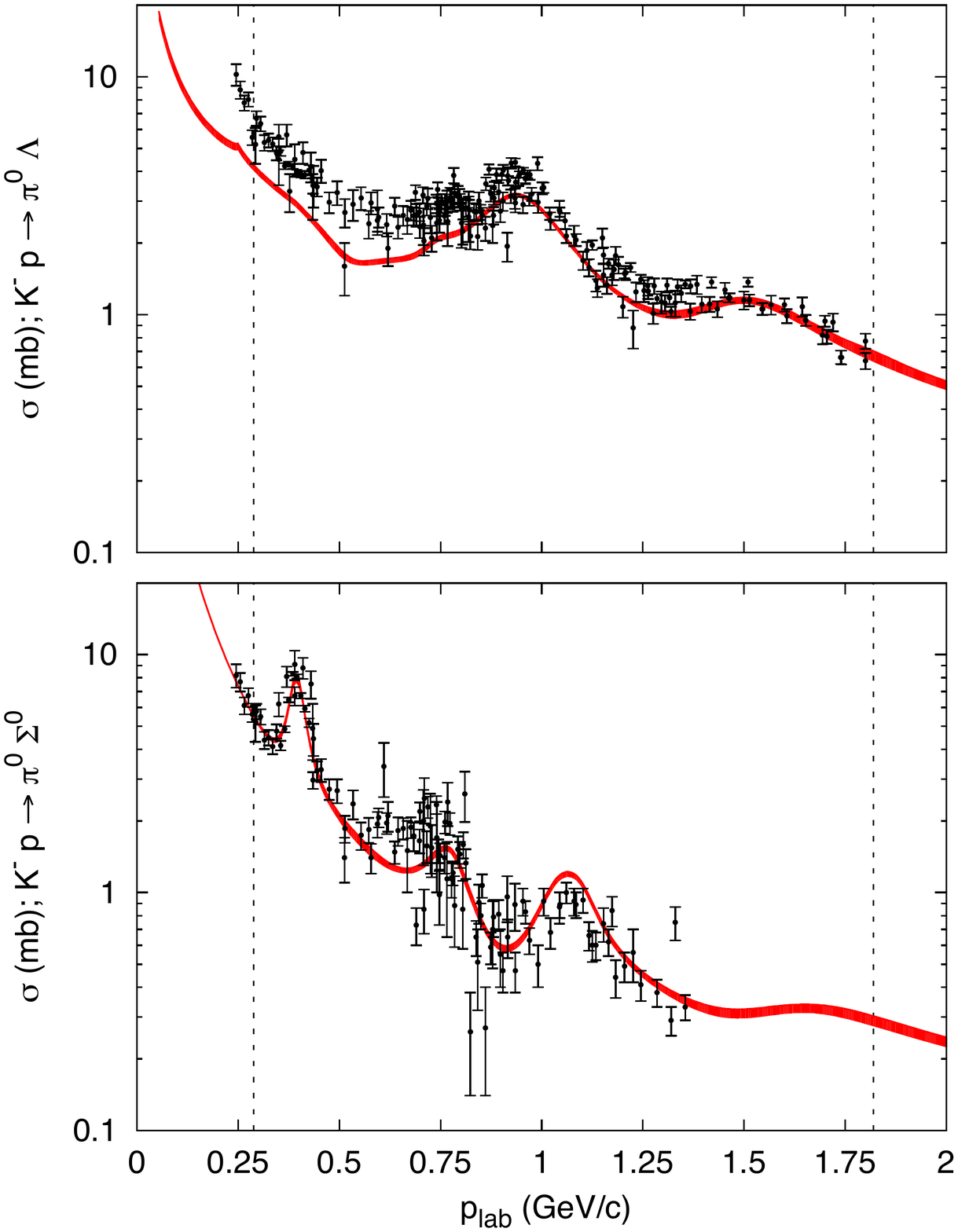}}}
\caption{(color online). Cross sections for $K^-p\to \pi^0 \Lambda, \pi^0 \Sigma^0$ processes.
The vertical dashed lines mark the energy range 
where the single-energy partial waves have been fitted.} \label{fig:xseclog3}
\end{figure}

\begin{figure}
\rotatebox{0}{\scalebox{0.4}[0.4]{\includegraphics{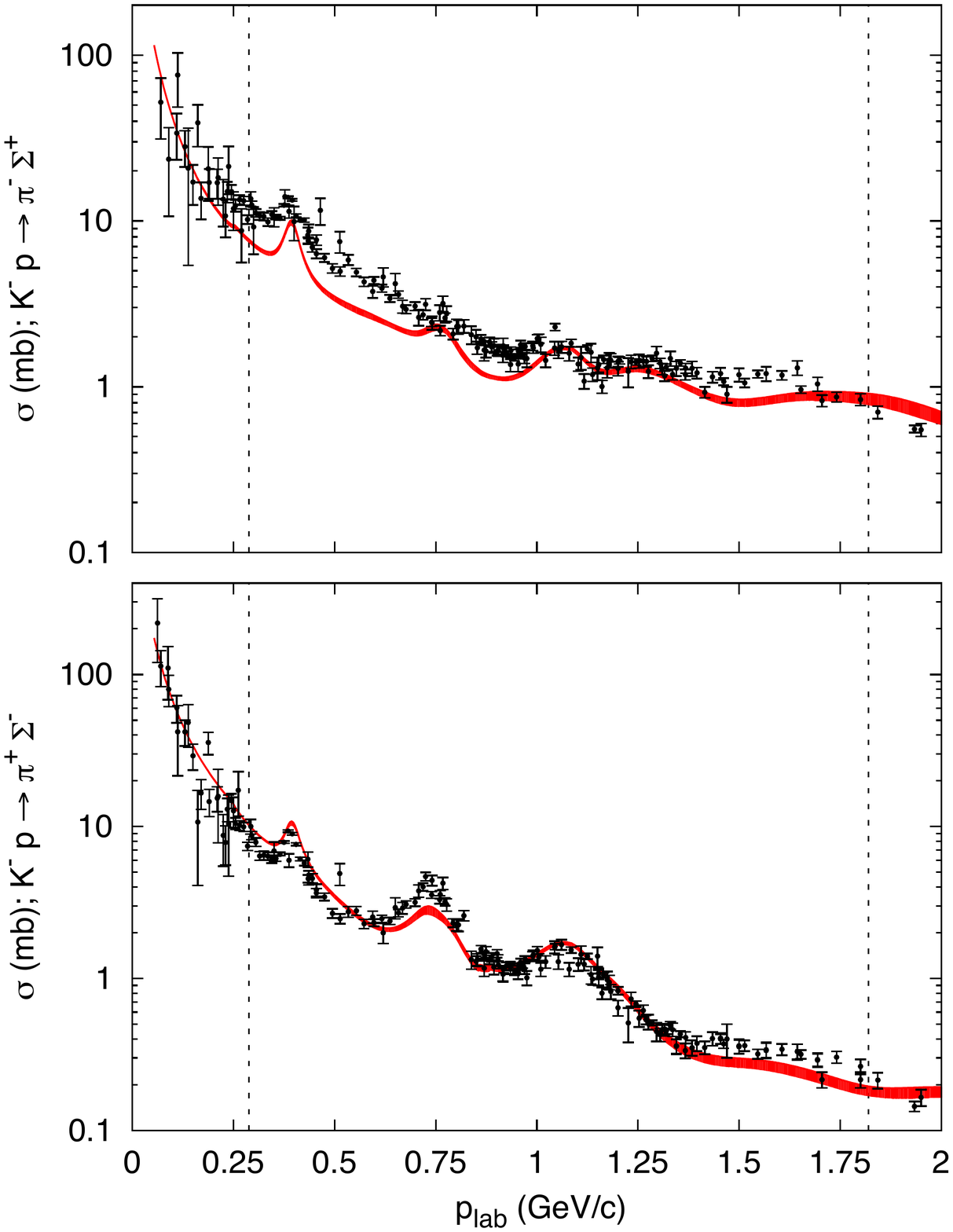}}}
\caption{(color online). Cross sections for $K^-p\to \pi^- \Sigma^+,  \pi^+ \Sigma^-$ processes.
The vertical dashed lines mark the energy range 
where the single-energy partial waves have been fitted.} \label{fig:xseclog2}
\end{figure}

\begin{figure*}
\begin{center}
\begin{tabular}{cc}
\rotatebox{0}{\scalebox{0.42}[0.42]{\includegraphics{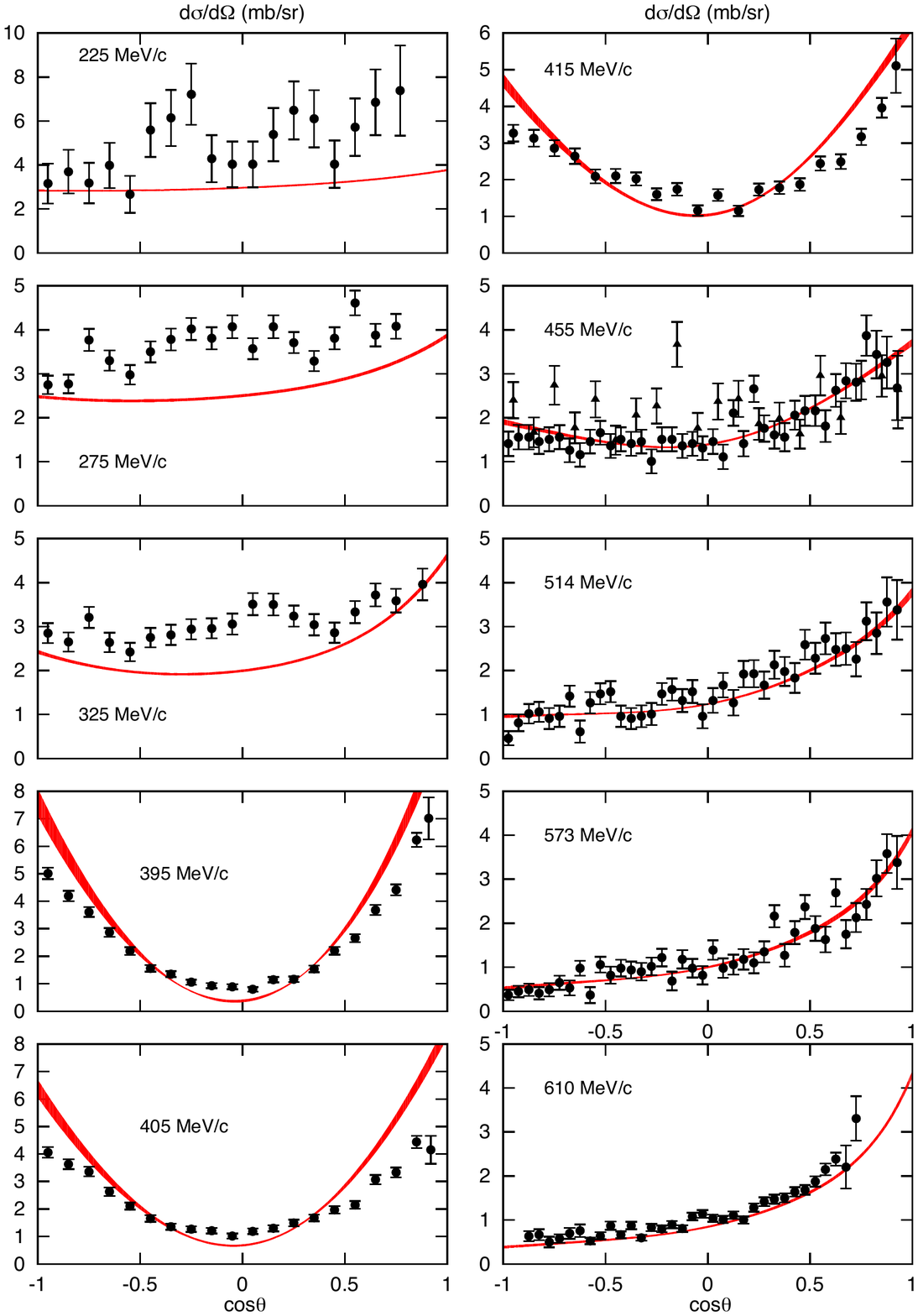}}} & 
\rotatebox{0}{\scalebox{0.42}[0.42]{\includegraphics{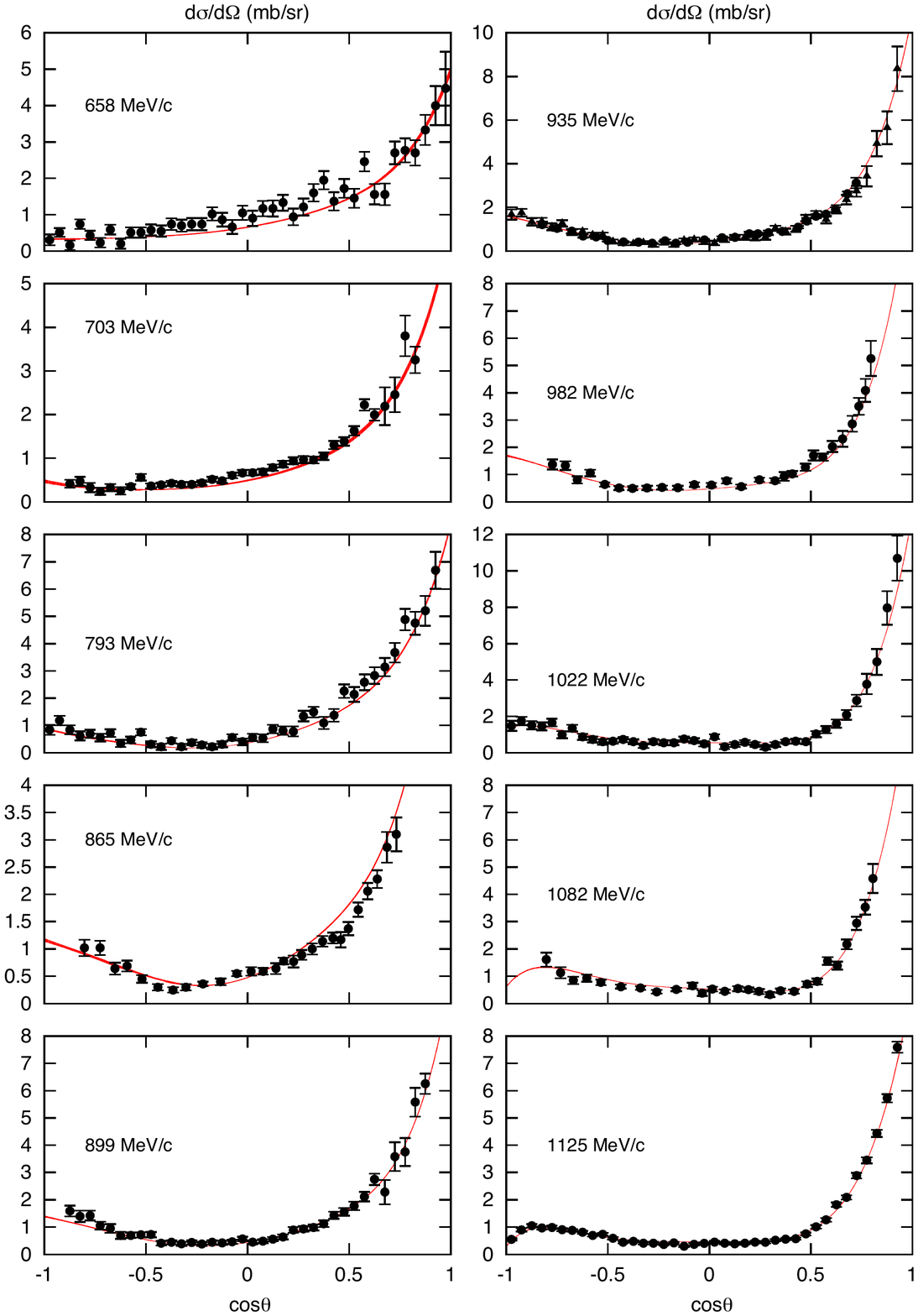}}}
\end{tabular}
\caption{(color online). Differential cross-section ($d\sigma /d\Omega$), for the $K^-p\to K^-p$ process 
in terms of the cosine of the center-of-mass scattering angle $\theta$. 
The width of the theory bands correspond to the errors propagated from the partial waves 
to the observables as explained in the text. 
The momentum of the incoming $K^-$ in the laboratory frame is shown for each plot.
Data from 
\cite{Andersson1970,Albrow1971,Conforto1971,Adams1975,Mast1976,Armenteros1970,Conforto1976}.} \label{fig:k-p1}
\end{center}
\end{figure*}

\begin{figure*}
\begin{center}
\begin{tabular}{cc}
\rotatebox{0}{\scalebox{0.42}[0.42]{\includegraphics{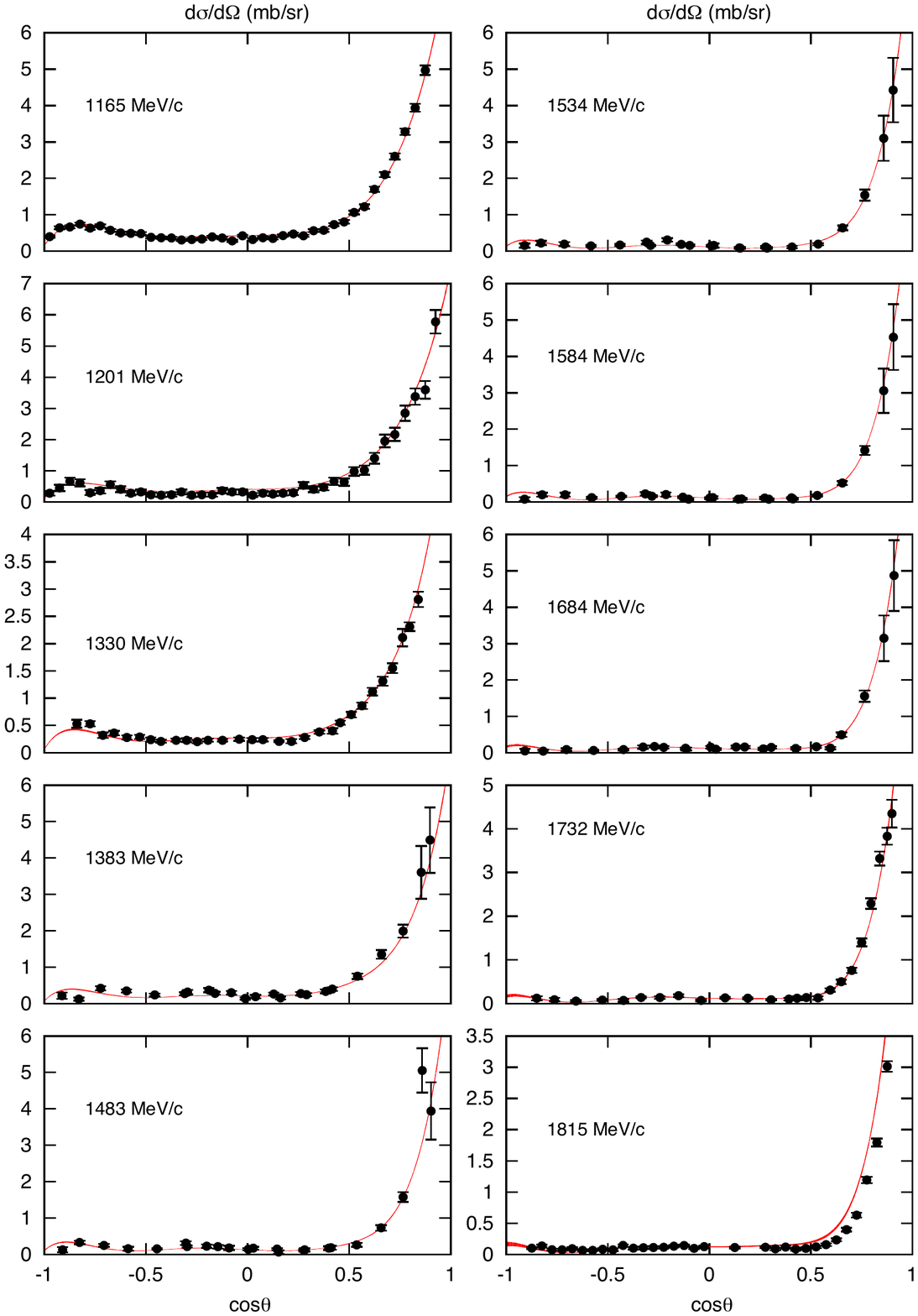}}} &
\rotatebox{0}{\scalebox{0.42}[0.42]{\includegraphics{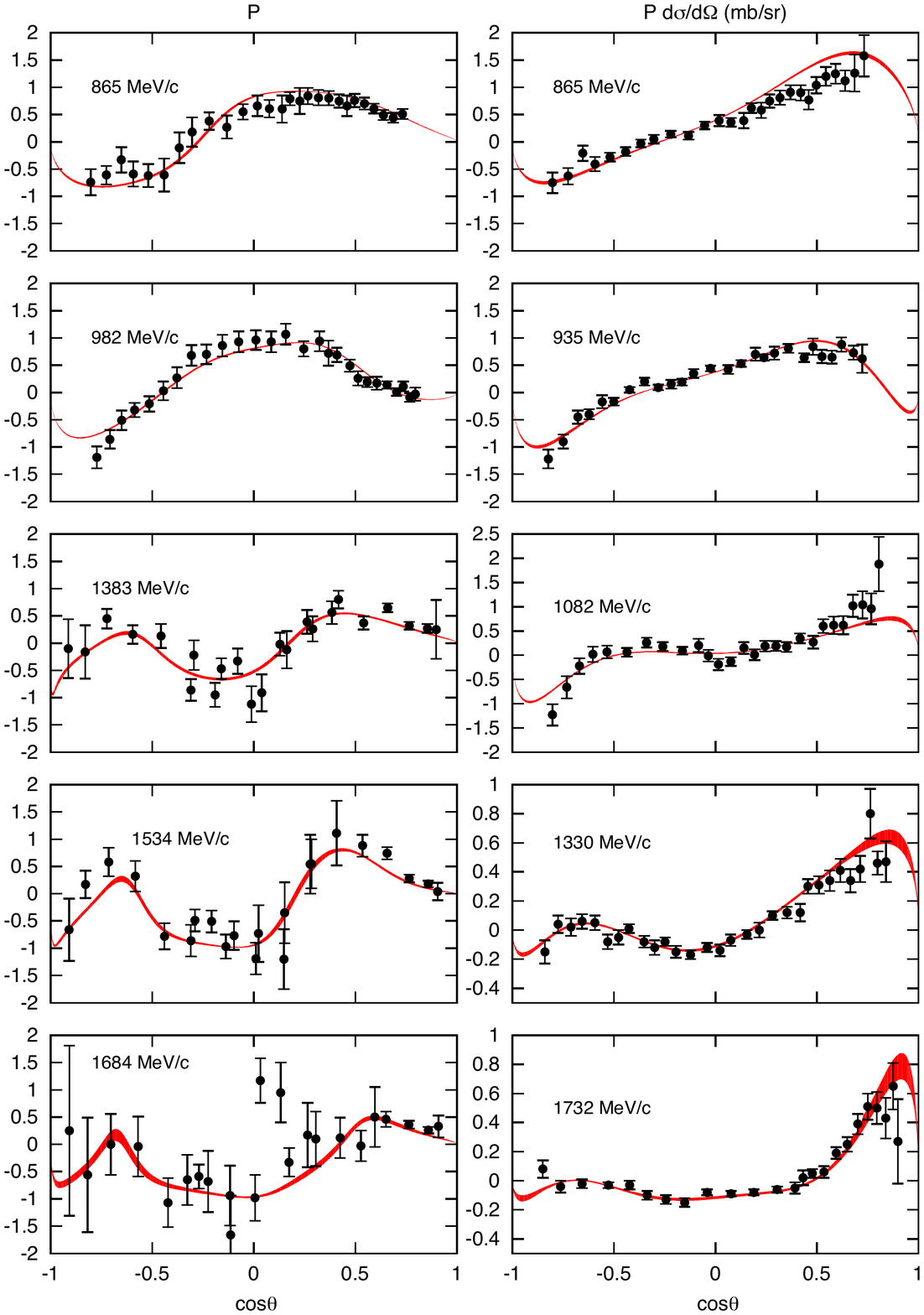}}}
\end{tabular}
\caption{(color online). Differential cross-section ($\frac{d\sigma}{d\Omega}$, left and left-center columns)
polarization asymmetry ($P$, right-center column), and $P\: \frac{d\sigma}{d\Omega}$ (right column)
for the $K^-p\to K^-p$ process in terms of the cosine of the center-of-mass scattering angle $\theta$. 
The width of the theory bands correspond to the errors propagated from the partial waves to the observables 
as explained in the text. The momentum of the incoming $K^-$ 
in the laboratory frame is shown for each plot.
Differential cross section data from 
\cite{Daum1968,Albrow1971,Abe1975,Andersson1970,Griselin1975,Conforto1976},
polarization data from \cite{Daum1968,Albrow1971,Andersson1970},
and $P\: \frac{d\sigma}{d\Omega}$ data from \cite{Albrow1971,Andersson1970}.} \label{fig:k-p2}
\end{center}
\end{figure*}

\begin{figure*}
\begin{center}
\begin{tabular}{cc}
\rotatebox{0}{\scalebox{0.42}[0.42]{\includegraphics{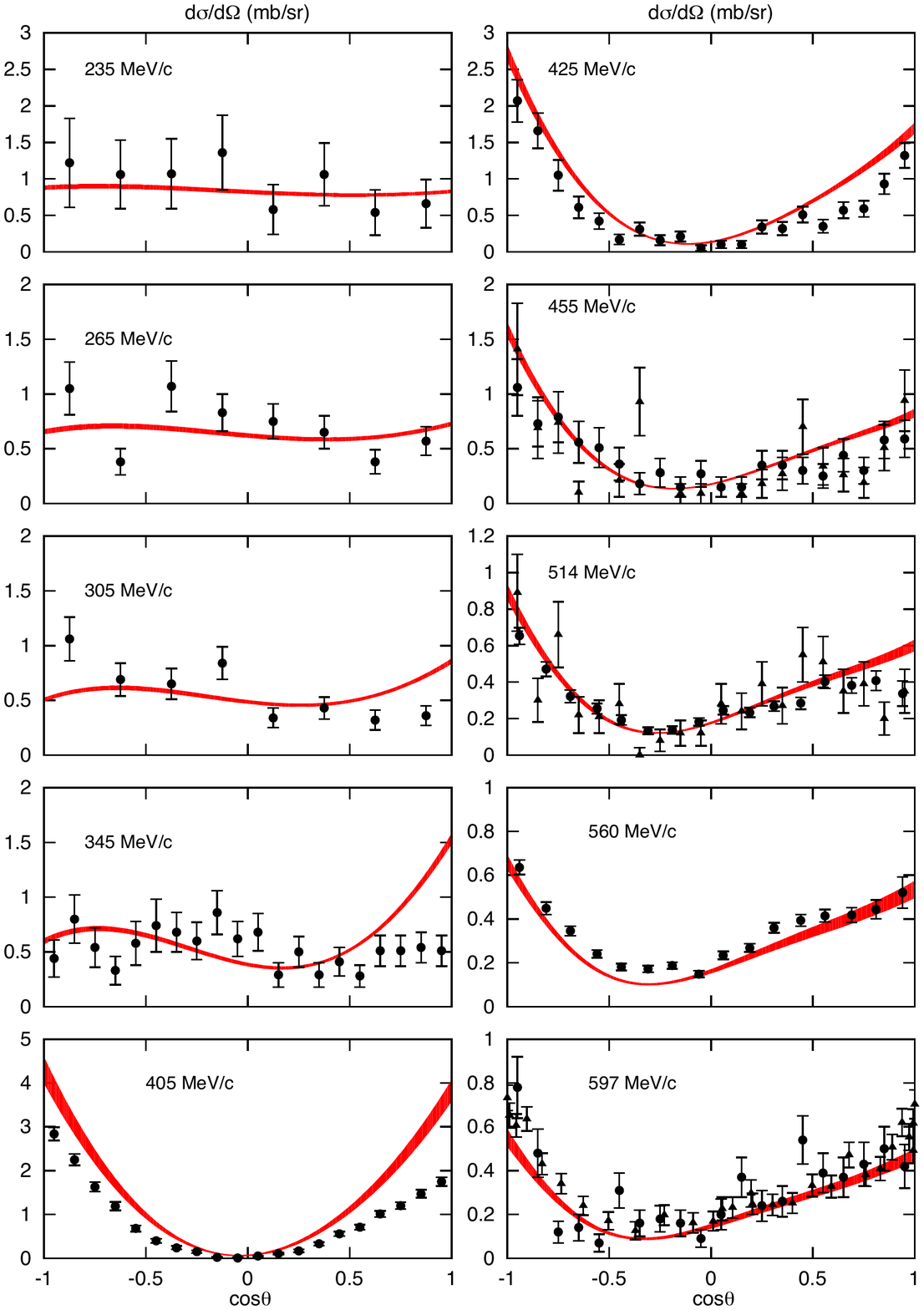}}} & 
\rotatebox{0}{\scalebox{0.42}[0.42]{\includegraphics{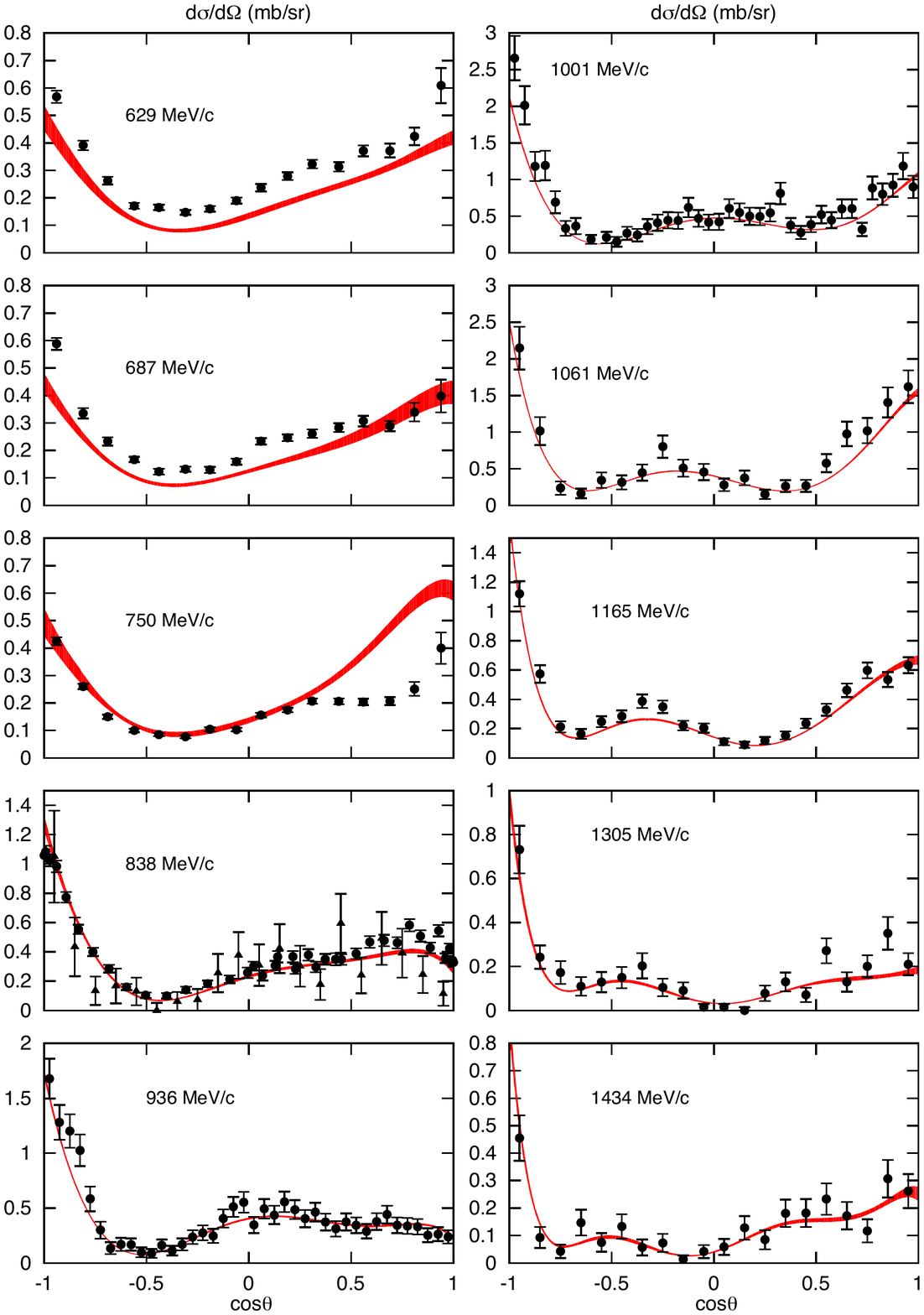}}}
\end{tabular}
\caption{(color online). Same as Fig.~\ref{fig:k-p1}  for the $K^-p\to \bar{K}^0 n$ process. Data from 
\cite{Daum1968,Mast1976,Armenteros1970,Alston1978,Jones1975,Griselin1975,Conforto1976,Prakhov2009}.} 
\label{fig:k0n1}
\end{center}
\end{figure*}

\begin{figure*}
\begin{center}
\begin{tabular}{cc}
\rotatebox{0}{\scalebox{0.42}[0.42]{\includegraphics{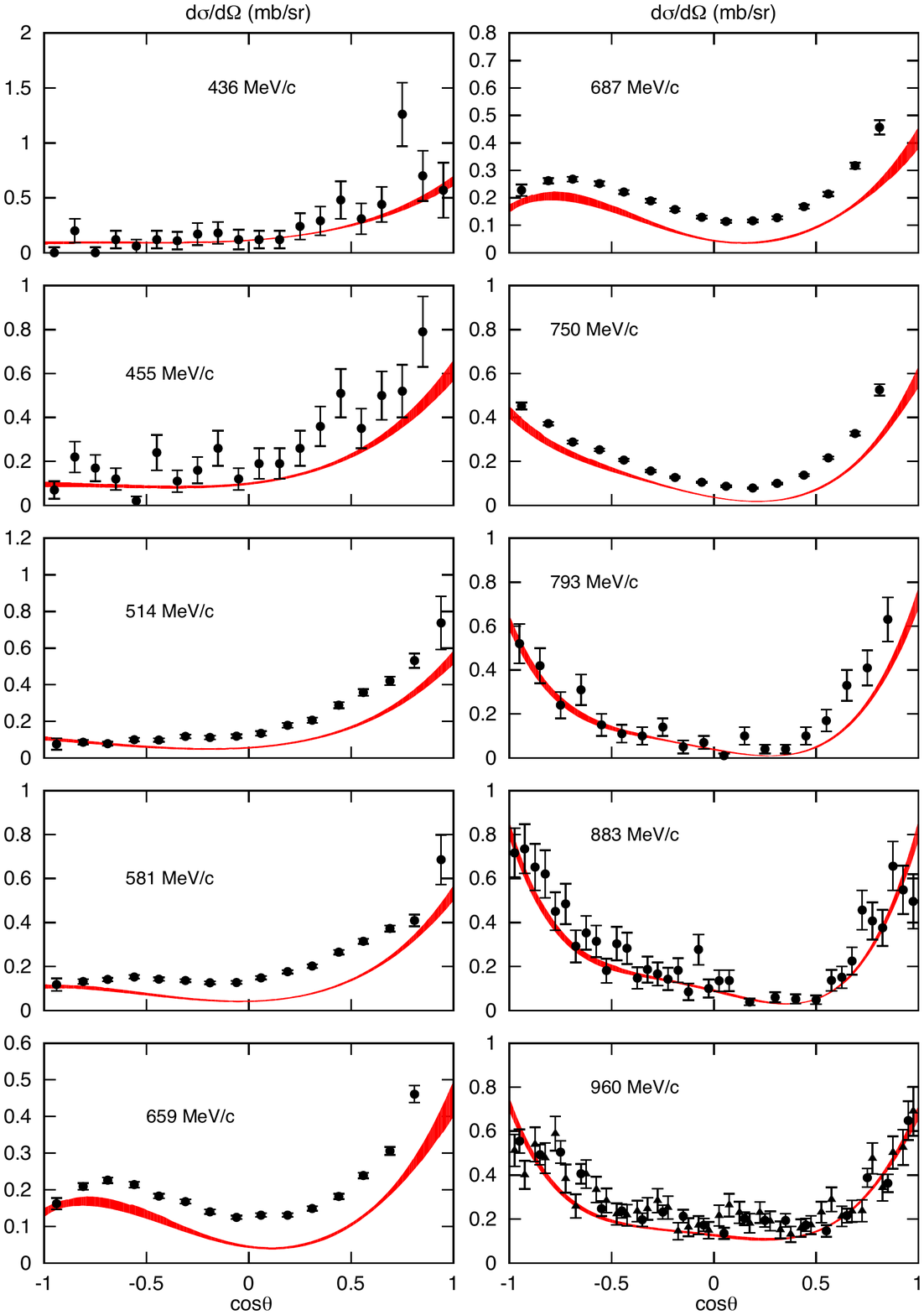}}} & 
\rotatebox{0}{\scalebox{0.42}[0.42]{\includegraphics{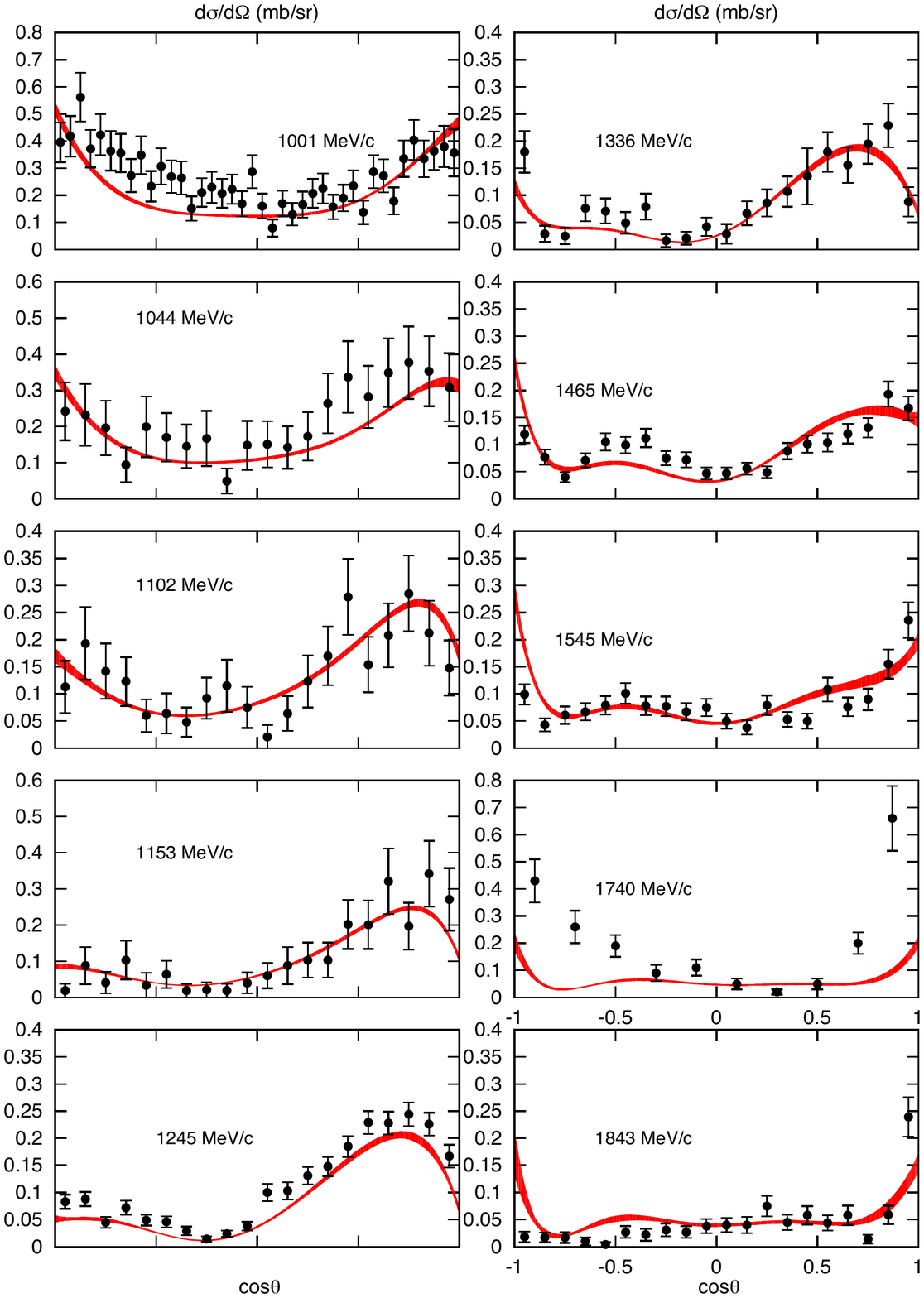}}}
\end{tabular}
\caption{(color online). Same as Fig.~\ref{fig:k-p1}  for the $K^-p\to \pi^0 \Lambda$ process.
Data from 
\cite{Armenteros1968,Armenteros1970,Jones1975,Griselin1975,Conforto1976,Berthon1970a,London1975,Prakhov2009}.} 
\label{fig:p0l01}
\end{center}
\end{figure*}

\begin{figure*}
\begin{center}
\begin{tabular}{cc}
\rotatebox{0}{\scalebox{0.42}[0.42]{\includegraphics{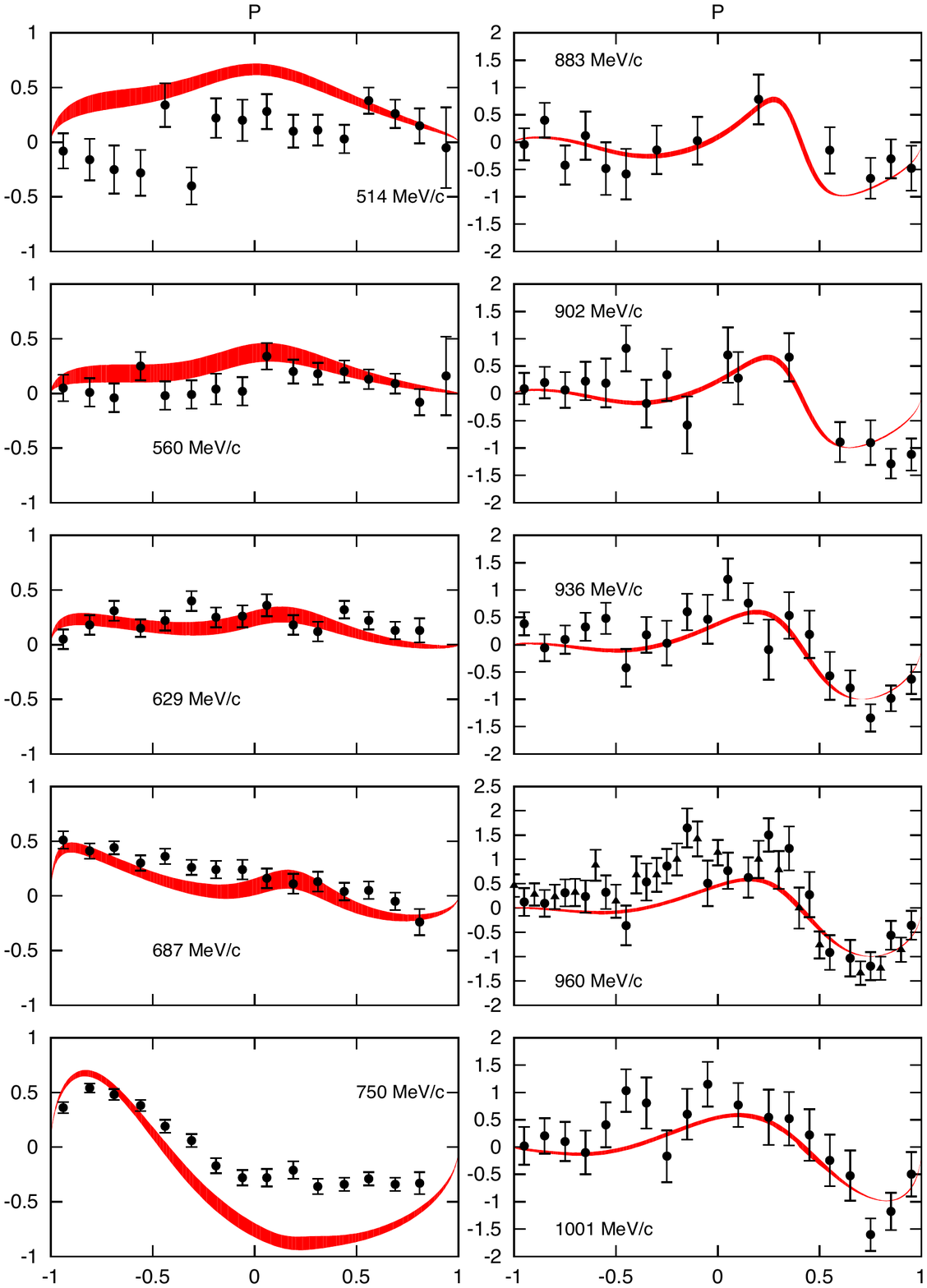}}} & 
\rotatebox{0}{\scalebox{0.42}[0.42]{\includegraphics{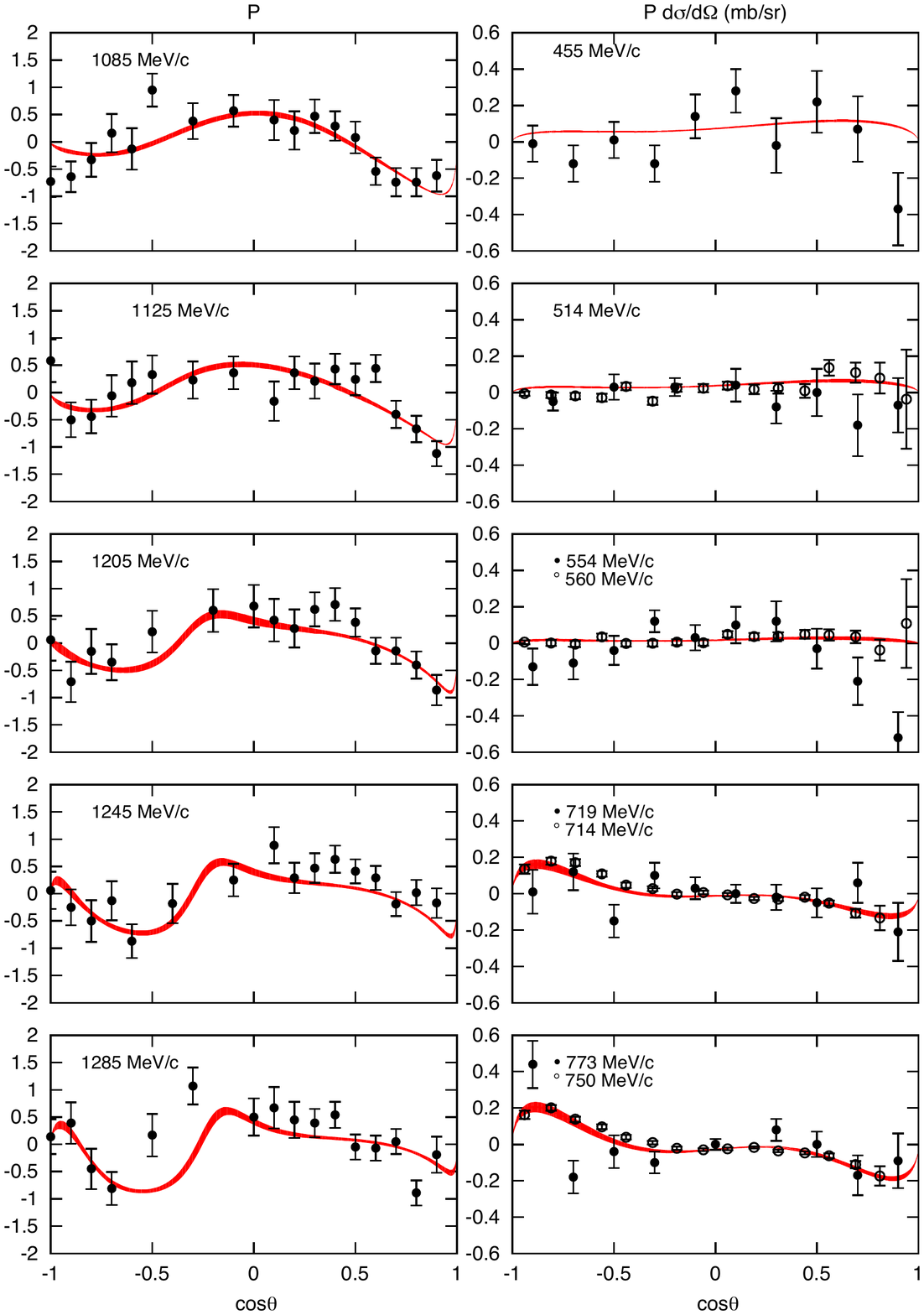}}}
\end{tabular}
\caption{(color online).  Polarization asymmetry ($P$, left, left-center and right-center columns),
and $P\: \frac{d\sigma}{d\Omega}$ (right column)
for the $K^-p\to \pi^0 \Lambda$ process in terms of the cosine of the center-of-mass scattering angle $\theta$.
Polarization data 
from \cite{Prakhov2009} for left column,
from \cite{Conforto1976} for left-center column,
and from \cite{Griselin1975} right-center  column.
$P\: \frac{d\sigma}{d\Omega}$ data from \cite{Armenteros1970} (solid circles)
and  \cite{Prakhov2009} (empty circles). For the last we have computed the error bars 
using standard error propagation from the $P$ and $d\sigma/d\Omega$ data.
For plots where data are shown at two different energies 
the theory band has been computed at both energies.} \label{fig:p0l02}
\end{center}
\end{figure*}

\begin{figure*}
\begin{center}
\begin{tabular}{cc}
\rotatebox{0}{\scalebox{0.42}[0.42]{\includegraphics{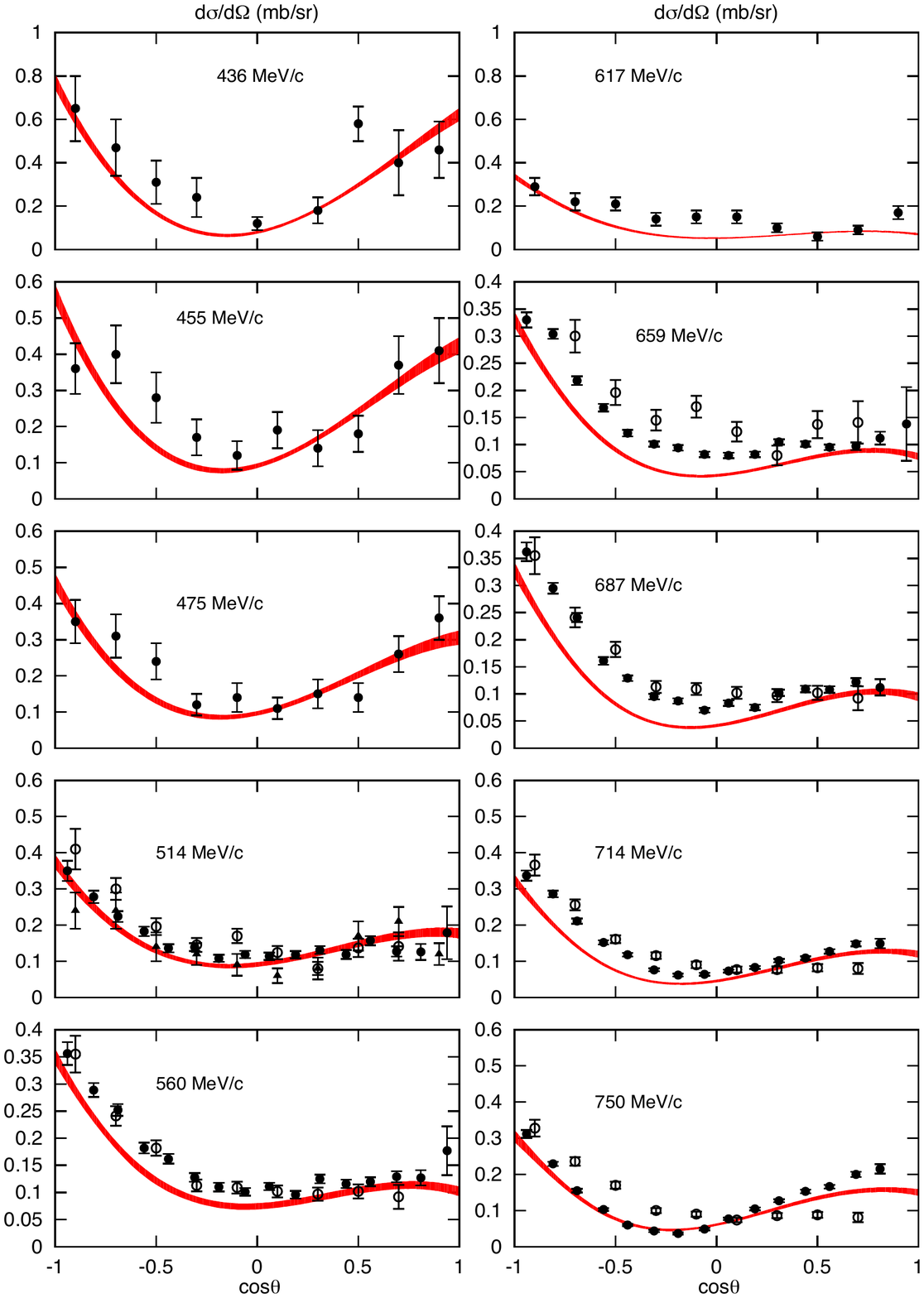}}} & 
\rotatebox{0}{\scalebox{0.42}[0.42]{\includegraphics{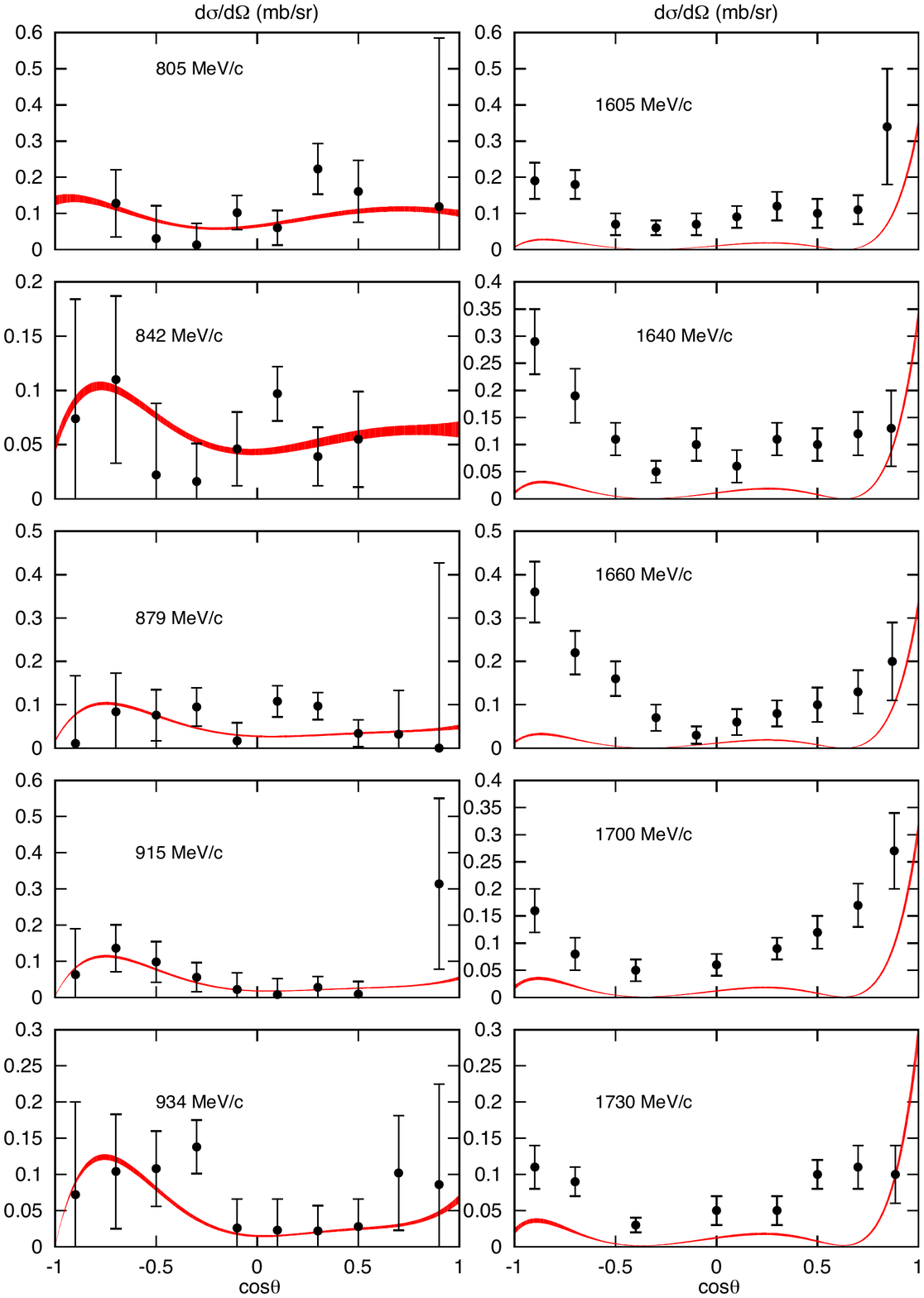}}}
\end{tabular}
\caption{(color online). Same as Fig.~\ref{fig:k-p1}  for the $K^-p\to \pi^0 \Sigma^0$ process.
Data at $p_\text{lab}=514, 560, 659, 687, 714$ and $750$ MeV$/c$ are from
\cite{Prakhov2009}  (solid circles), from \cite{Manweiler2008} (empty circles)
and from \cite{Armenteros1970} (solid triangles only at 514 MeV$/c$).
Remainder of the data from  \cite{Armenteros1970,Baxter1973,London1975}.} \label{fig:p0s01}
\end{center}
\end{figure*}

\begin{figure*}
\begin{center}
\begin{tabular}{cc}
\rotatebox{0}{\scalebox{0.42}[0.42]{\includegraphics{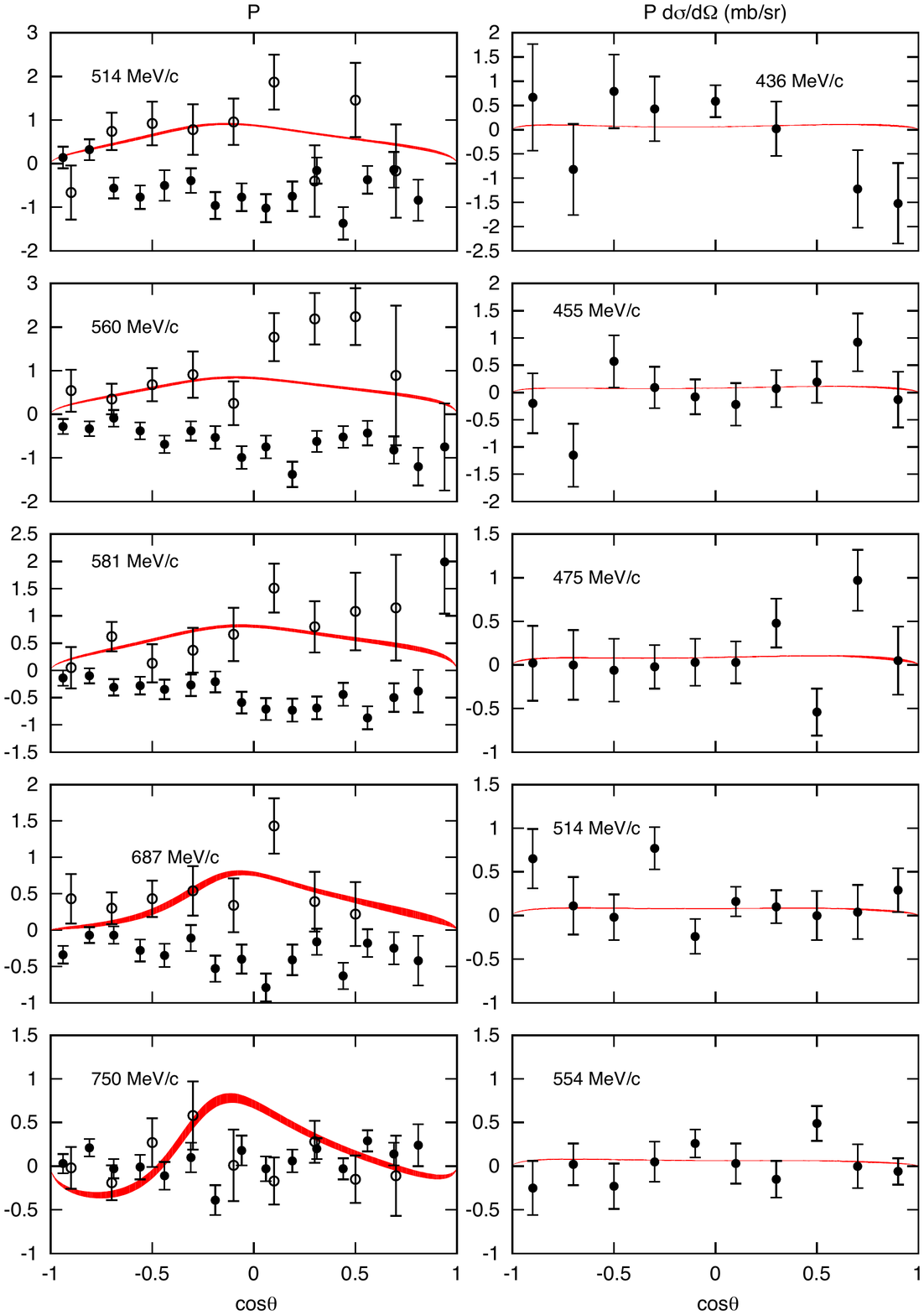}}} & 
\rotatebox{0}{\scalebox{0.42}[0.42]{\includegraphics{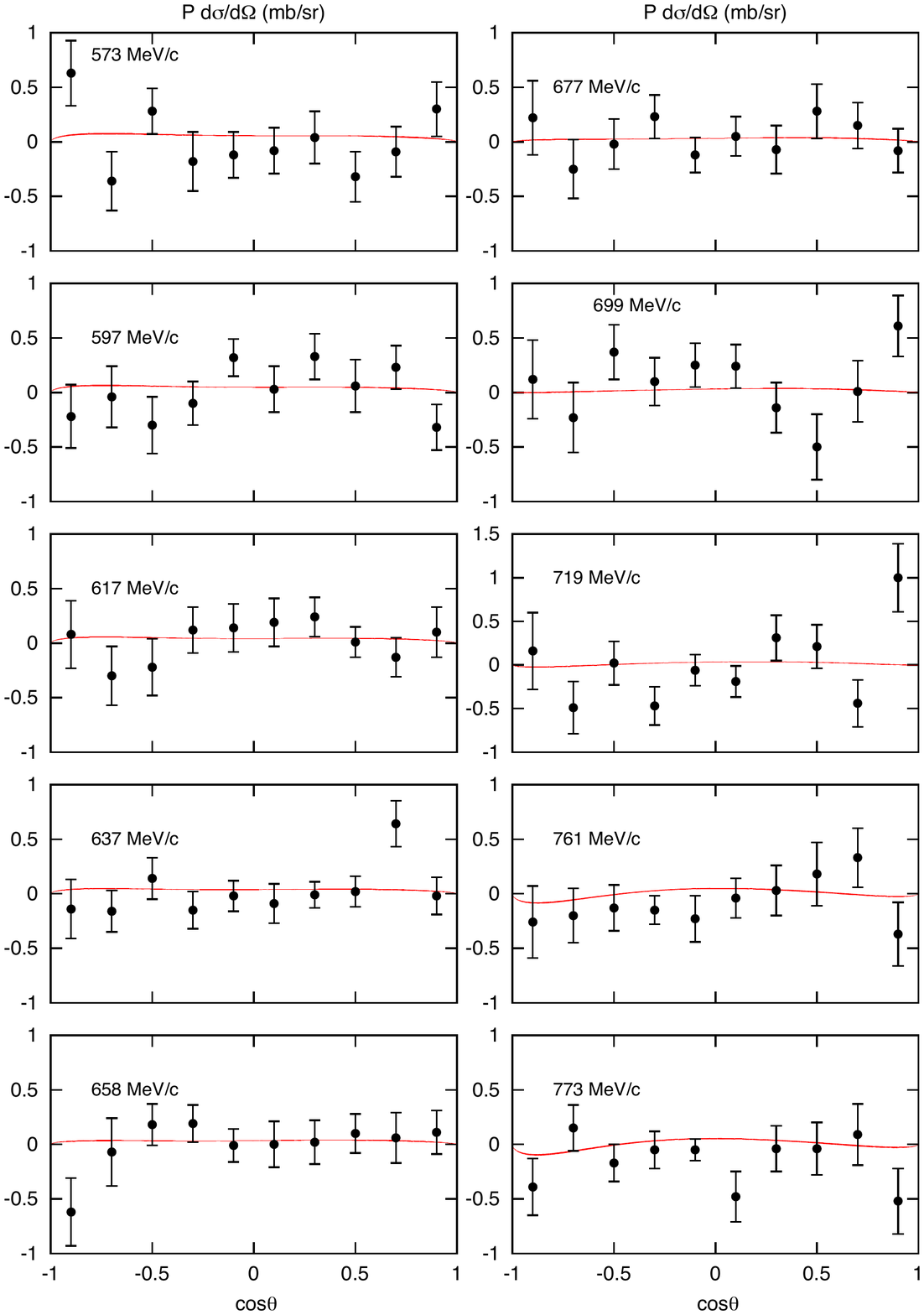}}}
\end{tabular}
\caption{(color online).  Polarization asymmetry ($P$, left column)
and $P\: \frac{d\sigma}{d\Omega}$ (left-center, right-center and right columns)
for the $K^-p\to \pi^0 \Sigma^0$ process in terms of the cosine of the center-of-mass scattering angle $\theta$.
Polarization data from \cite{Prakhov2009}, solid circles, and \cite{Manweiler2008}, empty circles.
$P\: \frac{d\sigma}{d\Omega}$ data from \cite{Armenteros1970}.} \label{fig:p0s02}
\end{center}
\end{figure*}

\begin{figure*}
\begin{center}
\begin{tabular}{cc}
\rotatebox{0}{\scalebox{0.42}[0.42]{\includegraphics{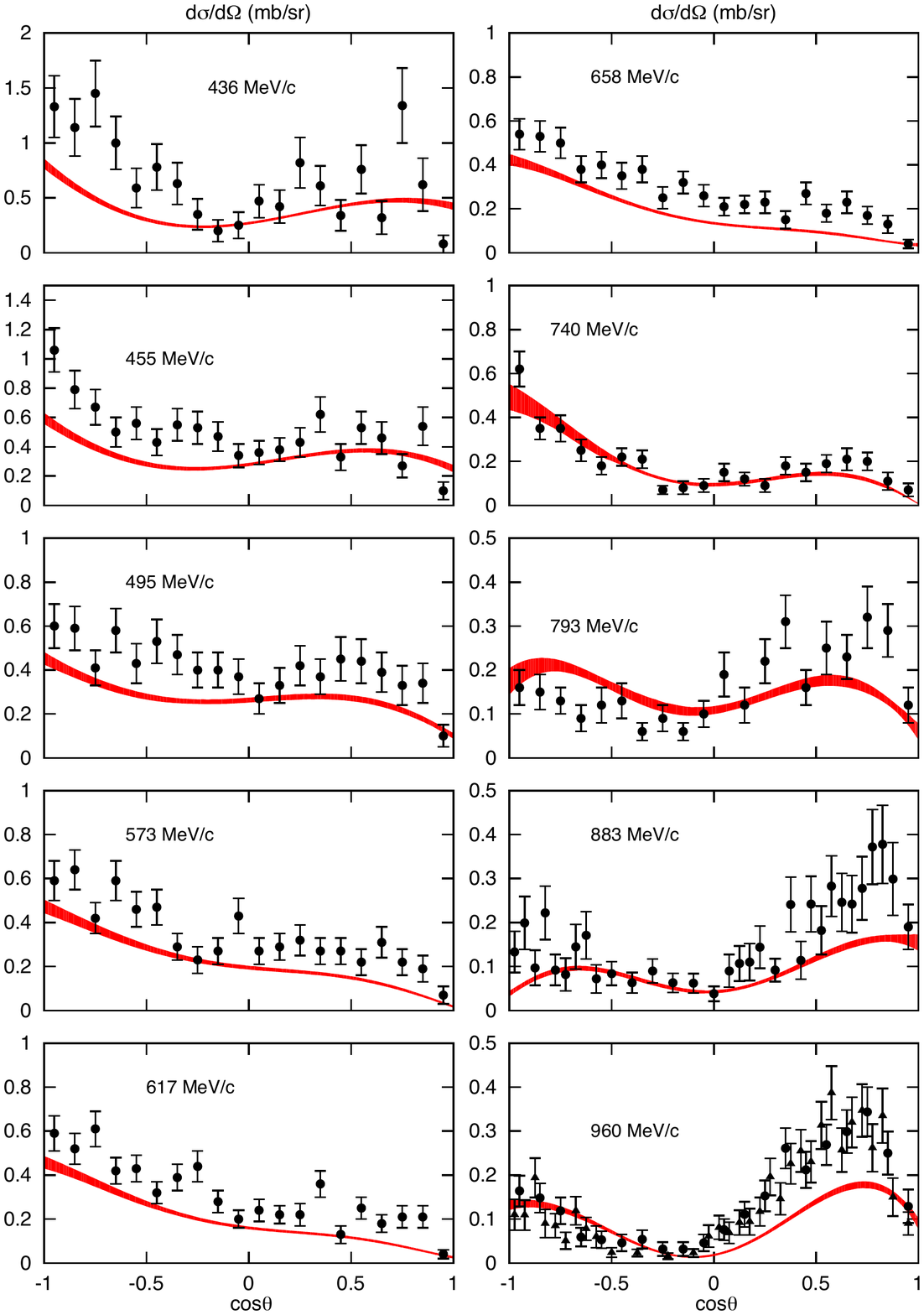}}} & 
\rotatebox{0}{\scalebox{0.42}[0.42]{\includegraphics{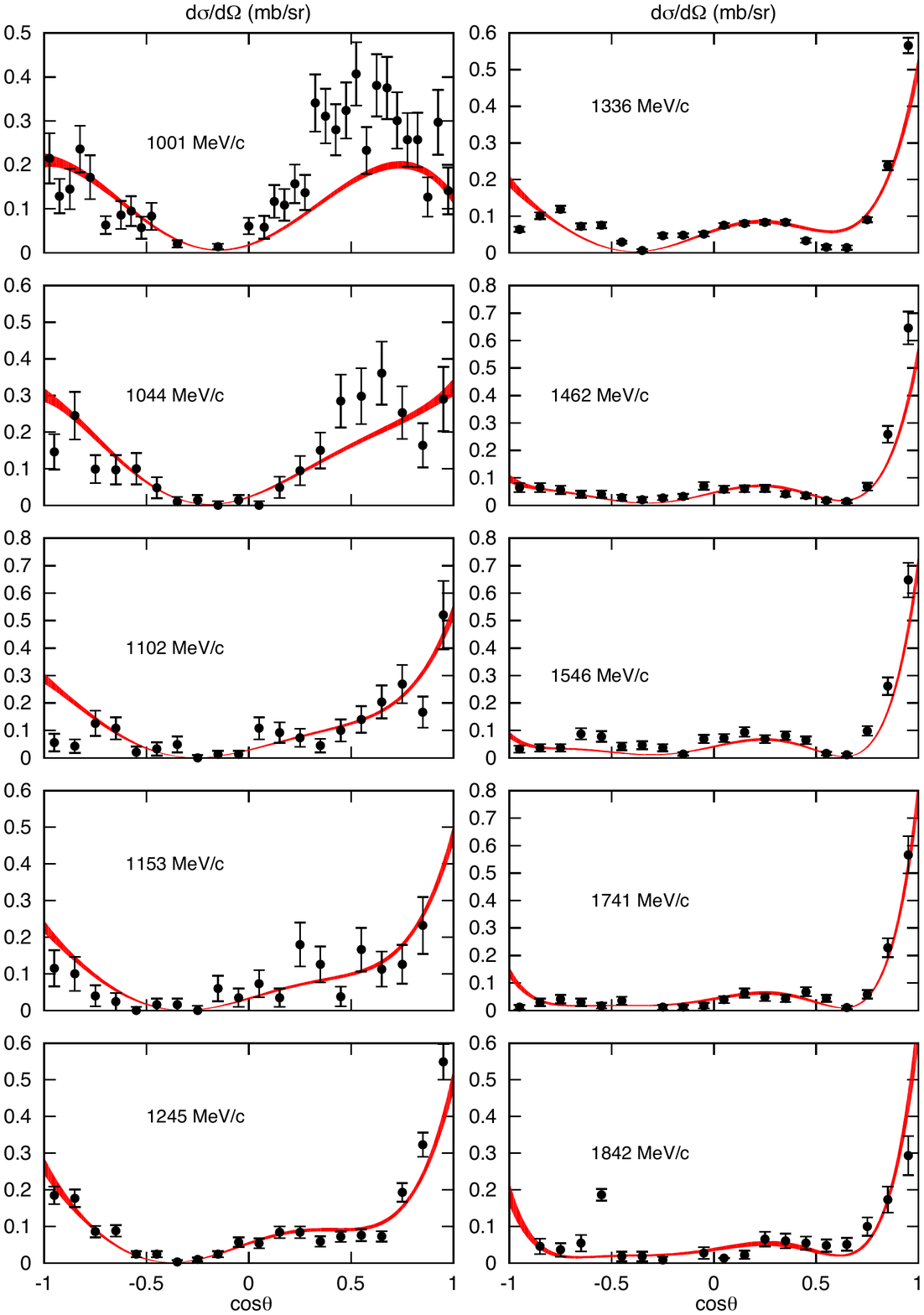}}}
\end{tabular}
\caption{(color online). Same as Fig.~\ref{fig:k-p1}  for the $K^-p\to \pi^- \Sigma^+$ process.
Data from \cite{Armenteros1968,Armenteros1970,Jones1975,Griselin1975,Conforto1976,Berthon1970b}.} \label{fig:p-s+1}
\end{center}
\end{figure*}

\begin{figure*}
\begin{center}
\begin{tabular}{cc}
\rotatebox{0}{\scalebox{0.42}[0.42]{\includegraphics{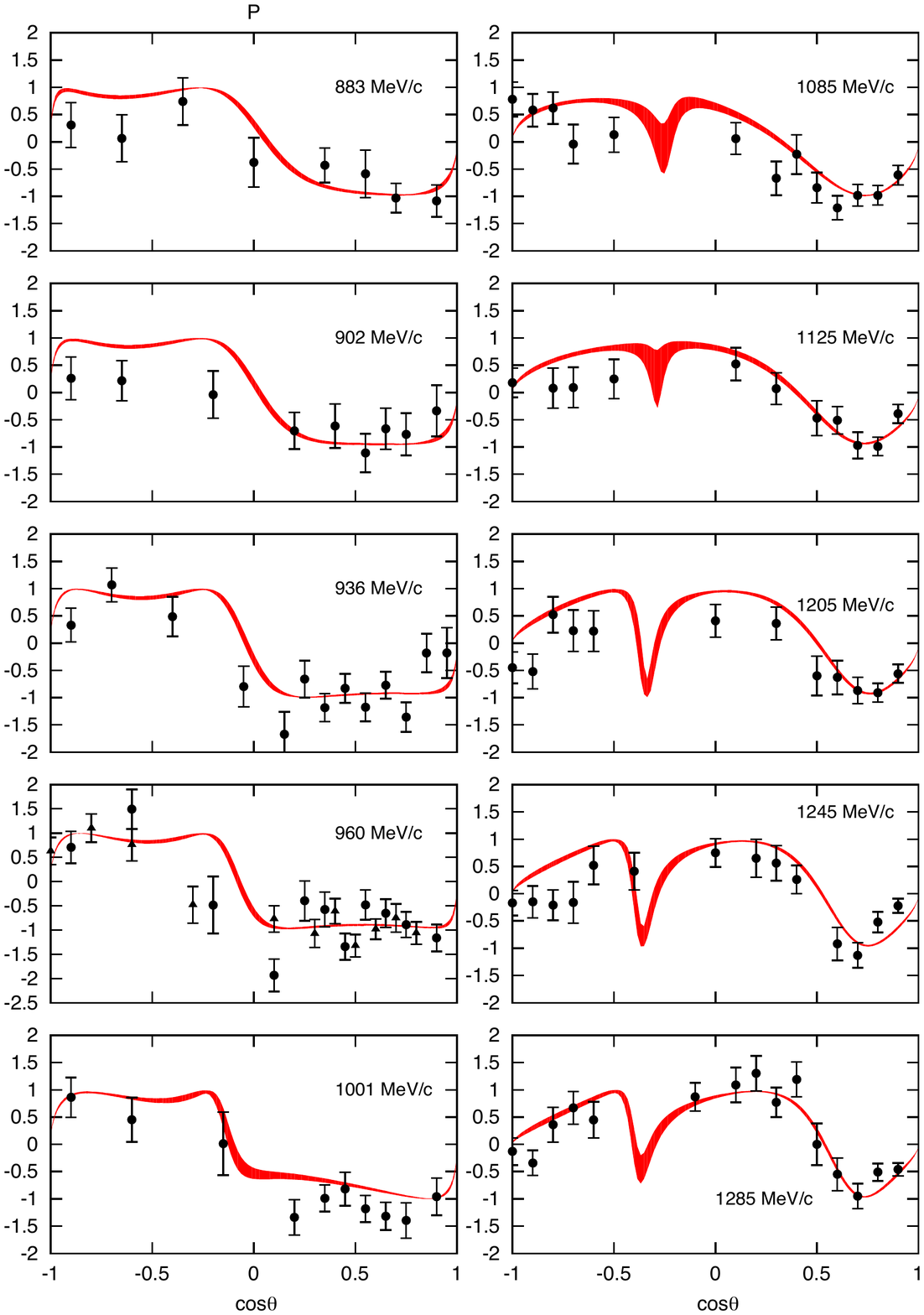}}} & 
\rotatebox{0}{\scalebox{0.42}[0.42]{\includegraphics{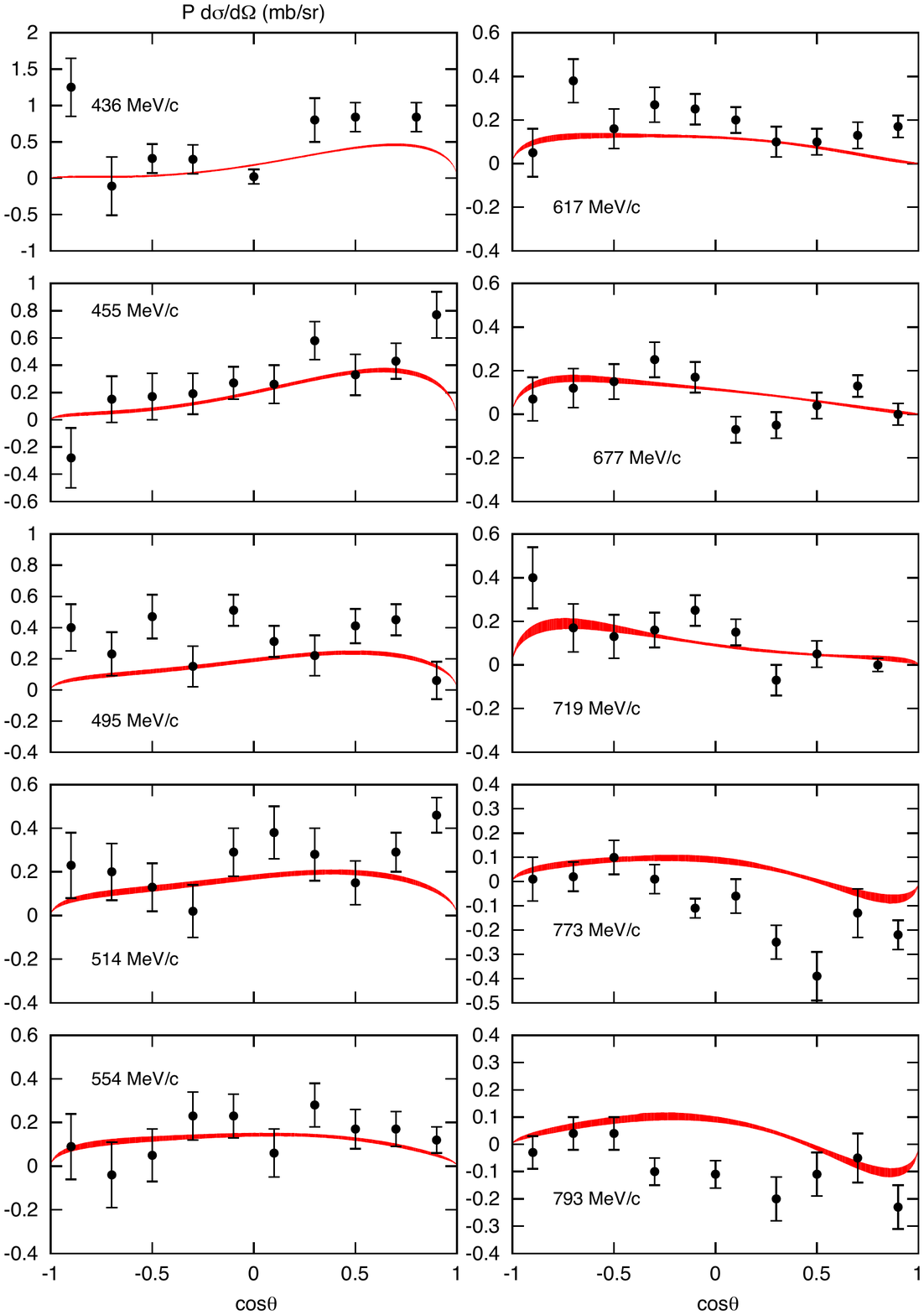}}}
\end{tabular}
\caption{(color online).  Polarization asymmetry ($P$, left and left-center columns)
and $P\: \frac{d\sigma}{d\Omega}$ (right-center and right columns)
for the $K^-p\to \pi^- \Sigma^+$ process in terms of the cosine of the center-of-mass scattering angle $\theta$.
Polarization data from \cite{Jones1975,Conforto1976}
and $P\: \frac{d\sigma}{d\Omega}$ data from \cite{Armenteros1970}.} \label{fig:p-s+2}
\end{center}
\end{figure*}

\begin{figure*}
\begin{center}
\begin{tabular}{cc}
\rotatebox{0}{\scalebox{0.42}[0.42]{\includegraphics{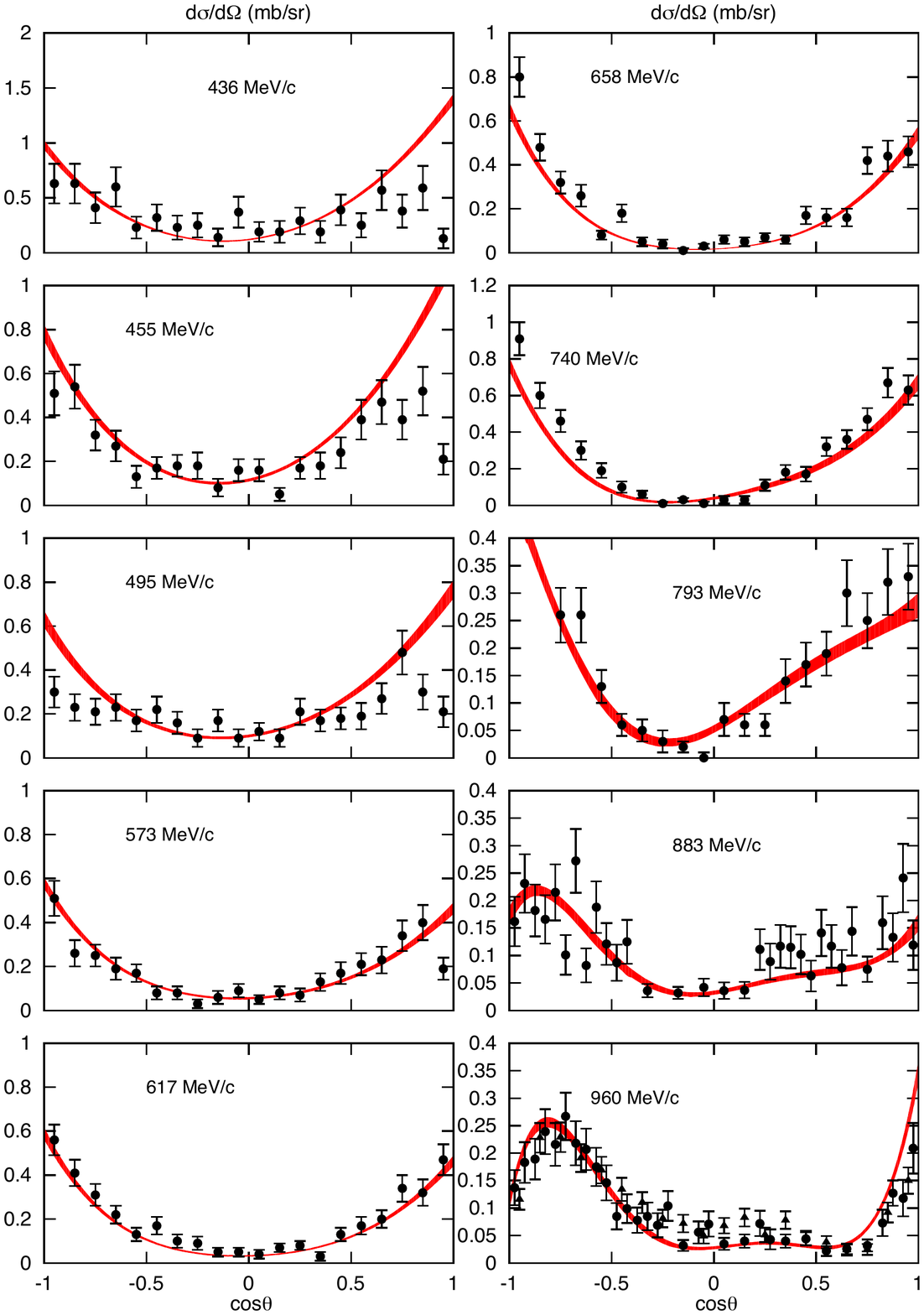}}} & 
\rotatebox{0}{\scalebox{0.42}[0.42]{\includegraphics{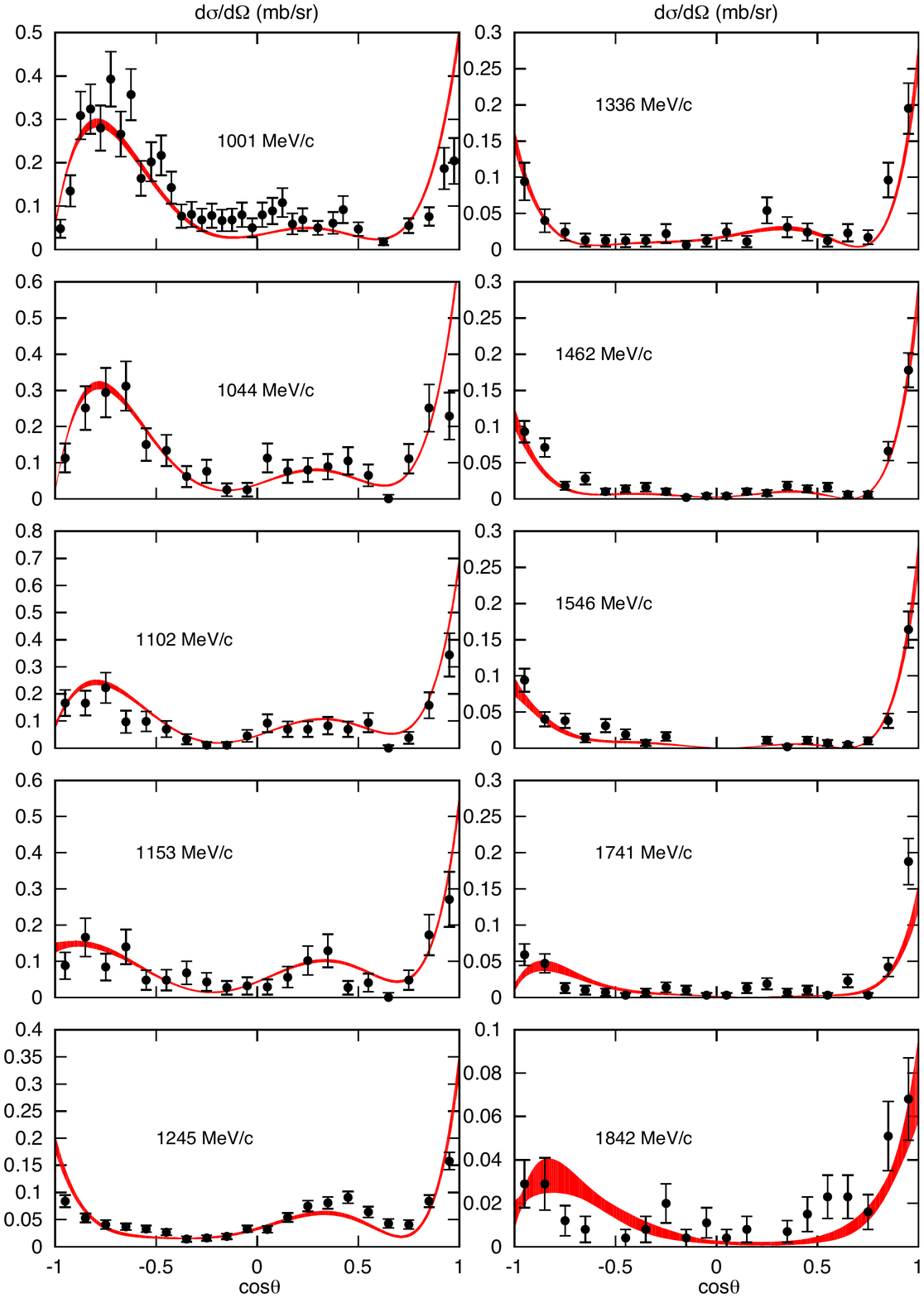}}}
\end{tabular}
\caption{(color online). Same as Fig.~\ref{fig:k-p1}  for the $K^-p\to \pi^+ \Sigma^-$ process.
Data from \cite{Armenteros1968,Armenteros1970,Jones1975,Griselin1975,Conforto1976,Berthon1970b}.} \label{fig:p+s-1}
\end{center}
\end{figure*}

In this section we compare our model to the data on total cross sections 
(Section \ref{sec:totalxsec}, Figs.~\ref{fig:xseclog1}, \ref{fig:xseclog3}, and \ref{fig:xseclog2}), 
differential cross sections, and polarization observables 
(Section \ref{sec:xsecandp}, Figs.~\ref{fig:k-p1}--\ref{fig:p+s-1})
for processes $\bar{K}N\to \bar{K}N$ 
\cite{Daum1968,Andersson1970,Albrow1971,Conforto1971,Adams1975,Abe1975,Mast1976,Alston1978,
Armenteros1968,Armenteros1970,Jones1975,Griselin1975,Conforto1976,Prakhov2009},
$\bar{K}N\to \pi \Lambda$ 
\cite{Armenteros1968,Armenteros1970,Jones1975,Griselin1975,
Conforto1976,Prakhov2009,Berthon1970a,Baxter1973,London1975,Baldini1988} 
and $\bar{K}N\to \pi \Sigma$
\cite{Armenteros1968,Armenteros1970,Jones1975,Griselin1975,
Conforto1976,Prakhov2009,Berthon1970b,Baxter1973,London1975,Baldini1988,Manweiler2008}.

Because we have fitted the single-energy partial waves, 
their correlations are not incorporated in our analysis, which translates into missing an important
piece of the error estimation in the observables. 
In order to account partly for that,
we performed the following simulation.
(i) For each partial wave we have picked randomly one of the 50 sets of parameters 
available from the bootstrap fits (\textit{cf.}  Section \ref{sec:fits}). 
(ii) We have computed each partial wave. 
(iii) We  have computed the observable. 
We have repeated this algorithm 1000 times to generate 
an average and standard deviation.
Systematics are not considered in either the theoretical or the experimental error bars displayed 
and might be of importance in what regards to the
cross sections, where a $\pm 10\%$ normalization effect is within experimental uncertainty.

\subsubsection{Total Cross Sections} \label{sec:totalxsec}

Figure  \ref{fig:xseclog1} shows the total cross sections for $K^-p\to K^-p$ and $K^-p\to \bar{K}^0n$, 
which are the ones that matter the most 
for the future analysis of heavy meson decays and
quasi-real diffractive photoproduction of $K\bar{K}$  on the proton
at GluEx and CLAS12 \cite{JLAB}.
Both processes are well reproduced in the whole energy range. 
The $K^-p\to K^-p$ is underestimated below $p_\text{lab} = 300$ MeV$/c$, although the general trend of the data
is well described. We will revisit this discrepancy in Section \ref{sec:xsecandp} 
where we compare to differential cross sections and polarizations.
 
Our results for $K^- p \to \pi^0 \Lambda$ and $K^- p \to \pi^0 \Sigma^0$ total cross sections are shown in 
Fig.~\ref{fig:xseclog3}.
The uncertainties in the $K^- p \to \pi^0 \Sigma^0$ process are very large and our model reproduces the
total cross section very nicely except at $p_\text{lab}\simeq 600$ MeV$/c$, where we underestimate the observable.
We obtain the general trend of the $K^- p \to \pi^0 \Lambda$ data 
but they are poorly reproduced for  $p_\text{lab}<800$ MeV$/c$ and  $p_\text{lab}>1300$ MeV$/c$ 
as a direct consequence of our difficulties in describing the $S_{11}$ 
partial wave for $s<3$ GeV$^2$ and $s>3.9$ GeV$^2$ as shown 
in Fig \ref{fig:PW_S11P11}.
From the theoretical point of view, 
the $K^- p \to \pi^0 \Lambda$ and $K^- p \to \pi^0 \Sigma^0$
processes are very interesting. The first has only isospin-1 contributions ($\Sigma^*$'s)
and the second has only isospin-0 contributions ($\Lambda^*$'s). 
This selectivity allows to decouple both sets of resonances and partial waves. 
However, in practice, both channels are difficult to separate experimentally 
 \cite{Conforto1976,Prakhov2009,Manweiler2008},
which leads to systematic uncertainties in the data analysis.
This is very well exposed if we compare total cross sections for two experimental data sets:
Armenteros \textit{et al.} \cite{Armenteros1970} and the most recent by Prakhov \textit{et al.} \cite{Prakhov2009} 
in the energy region between $p_\text{lab} = 450$ and $800$ MeV$/c$.
For $K^-p\to \bar{K}^0n$ the agreement between both data sets is excellent as shown in 
Fig.~10 in \cite{Prakhov2009}. 
This indicates that uncertainties are well under control in both experiments for this reaction. 
However, for $K^- p \to \pi^0 \Sigma^0$,
Armenteros \textit{et al.} shows a certain structure in the total cross section
that, with better statistics and better control on the systematics, disappears in
Prakhov \textit{et al.}, showing a flatter total cross section.

In Fig.~\ref{fig:xseclog2} we display the $K^- p \to \pi^\pm \Sigma^\mp$ reactions. 
The $K^- p \to \pi^+ \Sigma^-$ total cross section is very well reproduced except for the 
peak at $p_\text{lab} = 750$ MeV$/c$ and the energy region between 
1400 and 1750 MeV$/c$ where the observable is underestimated.
The shape of the $K^- p \to \pi^- \Sigma^+$ total cross section is well reproduced although the absolute
value of the observable is largely underestimated. The main source of disagreement is the inability of the
model to provide a good description of the $S_{11}$ partial wave for the $\bar{K}N\to\pi \Sigma$ channel
(Fig.~\ref{fig:PW_S11P11}) below $s=3$ GeV$^2$. 

We find a certain level of inconsistency between the total cross section data for $K^- p \to \pi^+ \Sigma^-$, 
$K^- p \to \pi^0 \Sigma^0$ and $K^- p \to \pi^- \Sigma^+$ reactions.
We reproduce $K^- p \to \pi^0 \Sigma^0$ data, which have only isospin one contributions --see Eq.~(\ref{eq:pi0sigma0})--
and we also reproduce $K^- p \to \pi^+ \Sigma^-$ data whose amplitude is obtained as the isospin one amplitude minus
the isospin zero amplitude. Hence, we should be able to predict correctly the 
$K^- p \to \pi^- \Sigma^+$ cross section, which corresponds 
to the addition of the two isospin amplitudes --see Eq.~(\ref{eq:piminussugmaplus}).
Instead, we underestimate the $K^- p \to \pi^- \Sigma^+$ total cross section.

\subsubsection{Differential Cross Sections and Polarizations} \label{sec:xsecandp}
In this section we compare with the differential cross section and polarization data. 
Almost all of the database is from experiments performed during the late 60's and the 70's except
for \cite{Prakhov2009} and \cite{Manweiler2008} published in 2009 and 2008, respectively.
These two  data sets come from the same BNL experiment and report 
measurements on the differential cross sections and polarizations
for $K^-p\to \pi^0 \Sigma^0$ \cite{Manweiler2008,Prakhov2009}
and for $K^-p\to \bar{K}^0n$ and  $K^-p\to \pi^0 \Lambda $ \cite{Prakhov2009} 
for eight anti-kaon momenta. There are some discrepancies  between these two data sets 
 that will be apparent  in the discussion of the $K^-p\to \pi^0 \Sigma^0$ polarization. 

We first compare to $K^-p\to K^- p$ and $K^-p\to \bar{K}^0 n$
because one of our main interests is the $\bar{K}N\to\bar{K}N$ amplitude due to its importance in the 
rescattering of heavy-baryon decays and $K\bar{K}$ photoproduction  experiments.
The $\bar{K}N\to\bar{K}N$ data constitute almost half of the experimental database.
However, the amount of polarization data is small with no data
below $p_\text{lab}=865$ MeV$/c$ \cite{Albrow1971}.
In Figs.~\ref{fig:k-p1} and \ref{fig:k-p2} we compare our results to a wide sample of the $K^-p\to K^- p$ database. 
It is the best known reaction under consideration in this paper and 
the general description we obtain is excellent for both differential cross sections and polarizations.
The only exceptions happen at low momenta, around $p_\text{lab} \simeq 400$ MeV$/c$, and at very high momentum,
$p_\text{lab} = 1815$ MeV$/c$. At low momentum we do not expect that a model like ours, built to 
describe the whole resonant region, provides an accurate description of the
amplitude because we lack additional constrains like chiral symmetry that drives the physics at low energies.
In the region around $p_\text{lab} \simeq 400$ MeV$/c$ ($\Lambda (1520)$ region)
we capture the main behavior of the differential cross section
although our model is not able to keep up with the rapid fall off of the cross section at forward and backward 
angles. This happens because, as shown in Section \ref{sec:fits}, the variation of the single-energy partial-wave data
is faster than the variation of the model, making difficult to capture
the full extent of the partial wave in such region. 
At very high energy ($p_\text{lab} > 1800$ MeV$/c$) our model overestimates the differential cross section,
and is no longer very accurate, although it reproduces the trend of the data.
We note the forward-angle behavior as the energy increases, 
which is the expected trend from Regge physics \cite{Mathieu15}.

We compare to $K^-p\to \bar{K}^0 n$ differential cross sections in Fig.~\ref{fig:k0n1} 
(no polarization data are available). The overall agreement is very good. 
At $p_\text{lab}=345$ MeV$/c$ we find a large discrepancy at forward angles. 
The forward peak in the amplitude is due to the constructive interference between 
$P_{01}$, $P_{03}$, and $D_{03}$ partial waves, while at backward angles the interference
between $P$ waves and $D_{03}$ is destructive. As for $K^-p\to K^- p$, the rapid variation
of the amplitude due to the presence of the $\Lambda (1520)$ is not well reproduced
and impacts the description of the data.
The same explanation applies for the $p_\text{lab}=405$ MeV$/c$ data.
Data points at $p_\text{lab}=560, 629, 687$ and $750$ MeV$/c$ and solid dots at $p_\text{lab}=514$ MeV$/c$
are from the most recent experiment in Ref.~\cite{Prakhov2009}. These data have small statistical
error bars and they were not used in the single-energy partial-wave amplitudes in \cite{Manley13b}
that we are fitting.
We systematically underestimate the $p_\text{lab} = 629$ MeV$/c$ and 687 MeV$/c$ data and we fail to reproduce 
the 750 MeV$/c$ data, where we find a larger forward contribution 
from the $P_{01}$, $P_{03}$ and specially $D_{03}$,
$\Lambda (1690)$ contribution, that the other partial waves cannot compensate.
In \cite{Prakhov2009}, the differential cross sections were fitted to Legendre polynomials 
expansion up to order five, rendering excellent fits for the $K^-p\to \bar{K}^0 n$ data except for 
$p_\text{lab}=560 , 687$ and $750$ MeV$/c$. Hence, although we do not reproduce the 750 MeV$/c$ data, 
it is not worrisome because the
data themselves might not be as good as they look according to their error bars.
As the energy increases, experimental data are very well described.

The comparison to $K^-p\to \pi^0 \Lambda $ differential cross sections is provided in Fig.~\ref{fig:p0l01}
and to polarizations in Fig.~\ref{fig:p0l02}. 
Only isospin-1 partial waves contribute to this reaction and 
as expected from the comparison to the total cross section,
the energy region above $p_\text{lab}=790$ MeV$/c$ is very well described for  differential cross sections
except at $p_\text{lab}=1740$ MeV$/c$ ($s=4.52$ GeV$^2/c^2$). 
This energy corresponds to the upper limit of the fitted energy region and neither the magnitude nor the shape
of the cross section are properly reproduced.
The shape of the the low-energy cross sections is correctly obtained but we fail to recover the right
magnitude mainly due to our poor description of the $S_{11}$ partial wave. 

We have a good description of the polarization data in Fig.~\ref{fig:p0l02}
with the exception of $p_\text{lab}=514, 960$ and $750$ MeV$/c$.
Despite the fact that we do not obtain the correct magnitude of the differential cross section
or polarization at 514 MeV$/c$  and 750 MeV$/c$ we do obtain both magnitude and shape for the
$P\frac{d\sigma}{d\Omega}$ observable. This is specially puzzling in the case of 750 MeV$/c$ where the
discrepancy between theory and experiment is very apparent.
 
Polarization data are experimentally very challenging for the $K^-p\to \pi^0 \Lambda $ and
$K^-p\to  \pi^0 \Sigma^0$. 
These difficulties become obvious when we compare to $K^-p\to \pi^0 \Sigma^0$  polarization data
from Refs. \cite{Prakhov2009,Manweiler2008}.
For  $P\frac{d\sigma}{d\Omega}$  we compare to data from \cite{Armenteros1970} 
for $p_\text{lab}=455, 514, 554, 719$ and $773$ MeV$/c$
and we construct the observable from the differential cross section and the polarization observable 
from the most recent data in \cite{Prakhov2009} for the closest possible momenta
$p_\text{lab}= 514, 560, 714$ and $750$ MeV$/c$. In this way it is possible to observe the improvement these
latest data constitute. For example, at $p_\text{lab} = 554$ (560) MeV$/c$ the forward and backward structures disappear
obtaining a flatter distribution and at  719 (714) and 773 (750) MeV$/c$ any disagreement between theory and
experiment vanishes. Hence, disagreements at 960 and 1285 MeV$/c$ for $P$ and at 455 MeV$/c$ for
$P\frac{d\sigma}{d\Omega}$ are not worrisome.
 
The measurement of the $K^-p \to \pi^0 \Sigma^0$ reaction is very challenging. 
An excellent example of the difficulties is provided by the
only two experimental papers in the last 35 years on $K^-p$ scattering, which have 
tried to tackle this reaction, 
\textit{i.e.}  Manweiler \textit{et al.} \cite{Manweiler2008} and Prakhov  \textit{et al.} \cite{Prakhov2009}.
Both analyses have been performed on the same experimental data 
at eight incident $K^-$ momenta ($p_\text{lab}= 514, 560,581,629, 659, 687, 714$, and $750$ MeV$/c$)  
reporting overall normalization uncertainties of $\pm7\%$ \cite{Prakhov2009} and $\pm10\%$ \cite{Manweiler2008} 
with serious disagreements on the systematic uncertainties treatment and their results, specially at forward angles.
Figure \ref{fig:p0s01} shows a sample of $K^-p \to \pi^0 \Sigma^0$ differential cross sections and in particular
data from these two analyses at six momenta. 
Both analyses agree for the lower energies but the discrepancies are very apparent at the higher energies.
This situation gets worse if we compare the polarization results as we do in the first column in Fig.~\ref{fig:p0s02},
where they disagree even in the sign of the polarization at every momenta
except at 750 MeV$/c$, which has very large error bars. 
The KSU single-energy partial waves that we have fitted incorporated the data from \cite{Manweiler2008}
but not the data from \cite{Prakhov2009} in their extraction. This explains why our model
has better agreement with the polarization data from \cite{Manweiler2008}.
In Fig.~\ref{fig:p0s02} we also compare to the $P\frac{d\sigma}{d\Omega}$ from \cite{Armenteros1970}
although the large error bars make difficult any meaningful comparison between theory and experiment.
At high energies, the differential cross section is not well reproduced as it is obvious from the fourth column
in Fig.~\ref{fig:p0s01}.
The KSU single-energy partial waves we have fitted do not reproduce these high-energy 
$K^-p \to \pi^0 \Sigma^0$ data, hence we do not expect to reproduce them with our model.

As expected from the results on the total cross section, the differential cross sections are systematically
underestimated at low energies for the  $K^-p\to \pi^- \Sigma^+$ reaction as it is shown in Fig.~\ref{fig:p-s+1}.
We underestimate the peak at $\cos \theta \simeq 0.7$
that shows up for $p_\text{lab}$ from $\sim790$ MeV$/c$ to $\sim1100$ MeV$/c$.
The peak shape is generated by the $F_{05}$ partial wave and its magnitude by its interference with $P_{03}$ wave.
If we compare both partial waves for the $\bar{K}N\to \pi \Sigma$ channel in 
the energy range between 2 and 3 GeV$^2$ in Figs.~\ref{fig:PW_S01P01} ($2^\text{nd}$ row, $3^\text{rd}$ column) 
and \ref{fig:PW_D05F05} ($2^\text{nd}$ row, $2^\text{nd}$ column) 
we find that there is a sizable underestimation of the single-energy partial waves by our fits that are responsible
for the result we obtain for the differential cross sections. This explains also the deviation
from the $P \frac{d\sigma}{d\Omega}$ data in Fig.~\ref{fig:p-s+2} at the same energies 
(although polarization data are fairly well reproduced). 
The rest of the polarization and $P \frac{d\sigma}{d\Omega}$ data are well reproduced considering the
large uncertainties and that many experimental data points have unphysical values of $|P|>1$.
Unfortunately, in the angular region for the polarization data  ($\cos \theta \simeq -0.25$)
where the most interesting structure shows up we lack
experimental information.
 
The comparison to $K^-p\to \pi^+ \Sigma^-$ differential cross sections is
presented in Fig.~\ref {fig:p+s-1}. No polarization data are available for this reaction.
The agreement between theory and experiment is excellent for all energies and angles
except for the forward region at low energies (see $p_\text{lab}=436, 455$, and $495$ MeV$/c$)
where the reduction of the differential cross section can be achieved through the
interference of  $D_{05}$ with higher-order partial waves, however,  $D_{03}$ contribution
compensates $D_{05}$ and  $P_{01}$, $P_{13}$ and $D_{13}$ contribute to the
large overestimation shown in the plots. 
Nevertheless, all the low-energy experimental points come from the
same experiment in Ref.~\cite{Armenteros1970} 
and further independent experimental information would be useful.
  
\section{Conclusions and Outlook} \label{sec:conclusions}
We have presented a coupled-channel model for $\bar{K}N, \pi \Sigma,  \ldots \to \bar{K}N,\pi \Sigma,  \ldots$ 
processes  in the resonant region (up to $s=4.70$ GeV$^2$) 
incorporating all the relevant channels. 
The approach presented is based on $K$-matrix formalism and Breit--Wigner parameterizations. 
The $T$ matrix is analytical,
the first Riemann sheet  has no poles  
(at least in a very wide area that envelopes the physical region of interest).
Unitarity gives the discontinuity of the $T$-matrix elements 
across the right-hand cuts and determines continuation 
to complex values of $s$ below the real axis where resonance poles are located
in the unphysical Riemann sheets.
Analytical amplitudes enable the application of dispersion relations 
to connect the resonance region with the high-energy domain 
that is dominated by Regge poles in cross-channels, \textit{e.g.}  in a similar  fashion to 
that used recently in the analysis of $\pi N$ scattering \cite{Mathieu15}. 
The construction of amplitudes valid in a wide range of energies is required
in the analysis of processes that have $\bar{K} N$ in the final states, 
\textit{e.g.} three-body decays to $\bar{K}N+$meson in pentaquark searches \cite{LHCbpentaquark}
and real and quasi-real diffractive photoproduction of $K\bar{K}$ pairs on the proton
in the search for strangeonia and exotic mesons with hidden strangeness \cite{JLAB}.

For simplicity and computational reasons 
we have fitted our model to the single-energy partial waves from \cite{Manley13a}.
We present the results of our fits in Table \ref{tab:fits} and Figs.~\ref{fig:PW_S01P01}--\ref{fig:PW_D15F15}.
Statistical errors have been estimated by means of the bootstrap technique.
In general the fits obtained are very good except for $S_{01}$, $P_{01}$ and $S_{11}$
partial waves, whose resonance extraction is less reliable than for other partial waves.
For these three partial waves we have performed 
additional analyses on systematic errors by randomly pruning and refitting the data base. 
Due to their nature, these systematic uncertainties have not been propagated to the resonances 
or observables error estimation. 

We have reported the most comprehensive analysis of the $Y^*$ spectrum to date.
All the obtained resonances are summarized in Tables \ref{tab:poles0} and \ref{tab:poles1}
together with their uncertainties  and a comparison to
previous pole extractions by Zhang \textit{et al.} \cite{Manley13b} and Kamano \textit{et al.} \cite{Kamano15}.
We provide graphical representations of the location of the resonances
in Fig.~\ref{fig:poles} ($T$-matrix pole positions in the unphysical Riemann sheets) 
and Fig.~\ref{fig:polesregge} (Regge trajectories).
The Regge trajectories provide additional insight into the nature 
 of  the hyperon spectrum. 
Gaps in the trajectories provide hints on possible missing states
and the shape of the trajectories and their (non-)linearity 
information on the quark-gluon dynamics \cite{reggetrajectories}.
We find that most of the states fit within linear trajectories, implying a three-quark state nature.
An exception is the $\Lambda (1600)$ whose mass and width are very well established
and does not fit  within the daughter natural parity linear Regge trajectory. 
Hence, it is likely that its nature is not that of a three-quark state.
We report a $3\slash2^+$ state in the $P_{03}$ partial wave with a mass of $1690$ MeV
and a narrow width of 46 MeV that fills in the gap in the parent $0^-$ Regge trajectory
(see Fig.~\ref{fig:poles0regge}). 
A similar  $P_{03}$ state was found in \cite{Kamano15} at $M_p=1671$ MeV and $\Gamma_p=10$ MeV,
although with a model that does not obtain the four-star $\Lambda (1830)$ state also present in the  $P_{03}$ partial wave. 
Neither present nor  \cite{Manley13b,Kamano15} analyses find evidence 
of the three-star $\Sigma (1940)$ state in the $D_{13}$ partial wave, however, the structure of the Regge trajectories in 
Fig.~\ref{fig:poles1regge} suggests that this state is necessary to fill in a gap in the $1^+$ daughter
Regge trajectory and further studies are mandatory.

Finally, we have compared our model predictions to the experimental observables
for $K^-p \to K^-p $, $\bar{K}^0 n$, $\pi^0 \Lambda$, $\pi^0\Sigma^0$,  $\pi^-\Sigma^+$,  $\pi^+\Sigma^-$ reactions,
namely, total and differential cross sections, polarizations and $P\frac{d\sigma}{d\Omega}$.
The $K^-p \to K^-p$ and $K^-p \to \bar{K}^0n$ data are well reproduced and our amplitudes
are an adequate input for  $\bar{K}N+$meson decays
and $\gamma p \to K\bar{K} p$  partial-wave analyses.
The model also provides a general good description for 
$K^-p\to \pi^+ \Sigma^-$ and $K^-p\to \pi^0 \Sigma^0$
processes and a not-so-good description of the 
$K^-p\to \pi^- \Sigma^+$ and $K^-p\to \pi^0 \Lambda$ reactions
depending on the energy range under consideration.
The reasons for discrepancies, database inconsistencies
and systematics have been addressed in Section \ref{sec:experiment}.

The next step in a comprehensive description and analysis
of the hyperon spectrum consists on fitting directly the experimental data as done 
in \cite{Kamano14} bypassing the single-energy partial waves from \cite{Manley13a}.
The partial waves presented in this paper can be used as starting point in the fitting process.
The examination of the experimental database shows how in dire need of new data we are
due to discrepancies encountered between different experimental
analyses. 
Considering how increasingly important
$\bar{K}N$ amplitudes are becoming in the data analysis for hadron spectroscopy research 
programs at LHCb \cite{LHCbpentaquark}
and Jefferson Lab  \cite{JLAB} an ambitious experimental
program should be seriously considered in the future experimental research 
programs at hadron beam facilities \cite{KNstatus}.

The codes employed to compute the partial waves and the observables are available
for downloading  as well as in an interactive form online at 
the Joint Physics Analysis Center (JPAC) webpage \cite{JPACwebpage}.

\begin{acknowledgments}
This work is part of the efforts of the Joint Physics Analysis Center (JPAC).
We thank Ra\'ul A.~Brice\~no, Michael U.~D\"oring,
Victor Mokeev, Emilie Passemar, Michael R.~Pennington, and Ron L.~Workman for useful discussions.
We thank Manoj Shrestha for making available the single-energy partial waves 
of Kent State University analysis as well as the
experimental database employed in the analysis.
This material is based upon work supported in part by the U.S.~Department of Energy, Office of Science, 
Office of Nuclear Physics under contract DE-AC05-06OR23177. 
This work was also supported in part by the U.S.~Department of Energy under Grant Nos. DE-FG0287ER40365
and DE-FG02-01ER41194, 
National Science Foundation under Grants PHY-1415459 and PHY-1205019, and IU Collaborative Research Grant.
\end{acknowledgments}

\appendix
\section{Solution to Eqs.~(\ref{eq:cc1}) and (\ref{eq:cc2})} \label{sec:appendix}
In this Appendix we provide the
analytic expressions of $c_{ab}(s)$ and $\mathcal{D}(s)$
for the case of six $K$ matrices that satisfy
the system of equations defined by Eqs.~(\ref{eq:cc1}) and (\ref{eq:cc2}).
Throughout this Appendix we drop the $s$ dependence in the equations.
The solution reads:
\begin{widetext}
\begin{equation}
      c_{aa} =    \mathcal{T}_a \left( 1 +  2 i\: f_{bcdef}\mathcal{T}_{bcdef}
      +\sum_{\{j,k\}} \varepsilon_{jk}\mathcal{T}_{jk}
      + 2 i \: \sum_{\{j,k,l\}} f_{jkl} \: \mathcal{T}_{jkl}
      + \sum_{\{j,k, l, m \}} f_{jklm}  \mathcal{T}_{jklm}  \right) \:, \label{eq:caa}
\end{equation}
\begin{equation}
      c_{ab} =  i \varepsilon_{ab} \mathcal{T}_a \mathcal{T}_b \left(    1 +\sum_{\{j,k\}} \varepsilon_{jk}\mathcal{T}_{jk}
      + 2 i \: \sum_{\{j,k,l\}} f_{jkl} \: \mathcal{T}_{jkl}
      + \sum_{\{j,k, l, m \}} f_{jklm}  \mathcal{T}_{jklm}  
      +  2 i\: f_{jklmn}\mathcal{T}_{jklmn} \right) \:, \quad a \neq b \: ; \label{eq:cab}
\end{equation}
\begin{equation}
      \mathcal{D} = 1 + \sum_{\{j,k\}} \varepsilon_{jk}^2 \: \mathcal{T}_{jk}
       + 2 i \: \sum_{\{j,k, l \}} f_{jkl} \: \mathcal{T}_{jkl}   
       + \sum_{\{j,k, l, m \}} f_{jklm}  \mathcal{T}_{jklm} 
       + 2 i \: \sum_{\{j,k, l, m,n \}} f_{jklmn}  \mathcal{T}_{jklmn}   
       +  f_{123456} \mathcal{T}_{123456} \:, \label{eq:ds}
\end{equation}
\end{widetext}
where we define $\{j,k\}$, $\{j, k,l\}$, $\{j, k,l,m\}$ and $\{j, k, l, m,n\}$
as the set of combinations without repetition of six elements taken in sets of 
two, three, four, and five elements at a time, respectively,
where $1$ to $6$ label each one of the $K_a$ matrices.
In Eqs.~(\ref{eq:caa}) and (\ref{eq:cab}) $a,b \neq j,k,l,m,n$.
$\varepsilon_{ij}$ is defined by Eq.~(\ref{eq:varepsilon}). 
The $\mathcal{T}$'s are defined by
\begin{eqnarray}
      \mathcal{T}_{123456} &= &\mathcal{T}_1\: \mathcal{T}_2 \:\mathcal{T}_3 \:\mathcal{T}_4\: \mathcal{T}_5\: \mathcal{T}_6 \:,\\
      \mathcal{T}_{jklmn} &=& \mathcal{T}_j\: \mathcal{T}_k \: \mathcal{T}_l\: \mathcal{T}_m\: \mathcal{T}_n \:,\\
       \mathcal{T}_{jklm} &= & \mathcal{T}_j\: \mathcal{T}_k \:\mathcal{T}_l \:\mathcal{T}_m \:,\\
       \mathcal{T}_{jkl} &= & \mathcal{T}_j\: \mathcal{T}_k\: \mathcal{T}_l \:,\\
       \mathcal{T}_{jk} &= & \mathcal{T}_j\: \mathcal{T}_k  \:, \\
\end{eqnarray}
where $\mathcal{T}_j$ is defined by Eq.~(\ref{eq:Ta}) if $j$ denotes 
a pole $K$ matrix and by Eq.~(\ref{eq:Tb}) if $j$ denotes a background $K$ matrix.

The $f$ functions are defined as follows:
\begin{equation}
f_{jkl} = \varepsilon_{jk} \varepsilon_{kl} \varepsilon_{lj}   \:,
\end{equation}

\begin{equation}
\begin{split}
      f_{ijkl} = &\: \varepsilon_{il}^2\:  \varepsilon_{jk}^2 
      + \varepsilon_{ij}^2 \: \varepsilon_{kl}^2 
      + \varepsilon_{ik}^2\: \varepsilon_{jl}^2- 2\: \varepsilon_{ij} \: \varepsilon_{ik}\: \varepsilon_{jl}\: \varepsilon_{kl} \\
      &- 2\: \varepsilon_{ik}  \:\varepsilon_{il} \: \varepsilon_{jk} \: \varepsilon_{jl} 
      - 2\: \varepsilon_{ij}\: \varepsilon_{il}\: \varepsilon_{jk} \: \varepsilon_{kl} \:,
\end{split}
\end{equation}

\begin{equation}
\begin{split}
f_{ijklm} =& \: \varepsilon_{im}^2\varepsilon_{jk}\varepsilon_{jl}\varepsilon_{kl} 
+ \varepsilon_{ik}\varepsilon_{il}\varepsilon_{jm}^2\varepsilon_{kl} \\
+& \: \varepsilon_{ij}\varepsilon_{im}\varepsilon_{jm}\varepsilon_{kl}^2
+ \varepsilon_{ik}\varepsilon_{im}\varepsilon_{jl}^2\varepsilon_{km} \\
+& \:  \varepsilon_{il}^2\varepsilon_{jk}\varepsilon_{jm}\varepsilon_{km}
+ \varepsilon_{ij}\varepsilon_{il}\varepsilon_{jl}\varepsilon_{km}^2\\
+&  \: \varepsilon_{il}\varepsilon_{im}\varepsilon_{jk}^2\varepsilon_{lm} 
+ \varepsilon_{ik}^2\varepsilon_{jl}\varepsilon_{jm}\varepsilon_{lm}\\
+& \:  \varepsilon_{ij}^2\varepsilon_{kl}\varepsilon_{km}\varepsilon_{lm}
+ \varepsilon_{ij}\varepsilon_{ik}\varepsilon_{jk}\varepsilon_{lm}^2\\
-&  \: \varepsilon_{il}\varepsilon_{im}\varepsilon_{jk}\varepsilon_{jm}\varepsilon_{kl} 
- \varepsilon_{ik}\varepsilon_{im}\varepsilon_{jl}\varepsilon_{jm}\varepsilon_{kl} \\
- & \: \varepsilon_{il}\varepsilon_{im}\varepsilon_{jk}\varepsilon_{jl}\varepsilon_{km} 
- \varepsilon_{ik}\varepsilon_{il}\varepsilon_{jl}\varepsilon_{jm}\varepsilon_{km}\\
- & \: \varepsilon_{ij}\varepsilon_{im}\varepsilon_{jl}\varepsilon_{kl}\varepsilon_{km} 
- \varepsilon_{ij}\varepsilon_{il}\varepsilon_{jm}\varepsilon_{kl}\varepsilon_{km} \\
- & \: \varepsilon_{ik}\varepsilon_{im}\varepsilon_{jk}\varepsilon_{jl}\varepsilon_{lm}
-\varepsilon_{ik}\varepsilon_{il}\varepsilon_{jk}\varepsilon_{jm}\varepsilon_{lm} \\
-& \:  \varepsilon_{ij}\varepsilon_{im}\varepsilon_{jk}\varepsilon_{kl}\varepsilon_{lm} 
- \varepsilon_{ij}\varepsilon_{ik}\varepsilon_{jm}\varepsilon_{kl}\varepsilon_{lm} \\
- & \: \varepsilon_{ij}\varepsilon_{il}\varepsilon_{jk}\varepsilon_{km}\varepsilon_{lm}
- \varepsilon_{ij}\varepsilon_{ik}\varepsilon_{jl}\varepsilon_{km}\varepsilon_{lm} \:,
\end{split}
\end{equation}

\begin{equation}
f_{123456} =  \: \alpha - 4\beta + 2\left( \gamma_1+\gamma_2 -  \delta_1-\delta_2 \right) \:,
\end{equation}
where
\begin{widetext}
\begin{equation}
\begin{split}
      \alpha = 
      &\: \varepsilon_{16}^2\varepsilon_{25}^2\varepsilon_{34}^2 
      + \varepsilon_{15}^2\varepsilon_{26}^2\varepsilon_{34}^2 
      +\varepsilon_{16}^2\varepsilon_{24}^2\varepsilon_{35}^2
      + \varepsilon_{14}^2\varepsilon_{26}^2\varepsilon_{35}^2 
      +  \varepsilon_{15}^2\varepsilon_{24}^2\varepsilon_{36}^2 
      + \varepsilon_{14}^2\varepsilon_{25}^2\varepsilon_{36}^2 \\
      +&\: \varepsilon_{12}^2\varepsilon_{35}^2\varepsilon_{46}^2 
      +\varepsilon_{13}^2\varepsilon_{24}^2\varepsilon_{56}^2
       +  \varepsilon_{16}^2\varepsilon_{23}^2\varepsilon_{45}^2
       +\varepsilon_{12}^2\varepsilon_{36}^2\varepsilon_{45}^2 
       + \varepsilon_{13}^2\varepsilon_{26}^2\varepsilon_{45}^2
        + \varepsilon_{12}^2\varepsilon_{34}^2\varepsilon_{56}^2\\
       + &\:  \varepsilon_{14}^2\varepsilon_{23}^2\varepsilon_{56}^2 
       +\varepsilon_{13}^2\varepsilon_{25}^2\varepsilon_{46}^2 
       + \varepsilon_{15}^2\varepsilon_{23}^2\varepsilon_{46}^2   \:,
\end{split}
\end{equation}

\begin{equation}
\begin{split}
      \beta =&\: \varepsilon_{12}\varepsilon_{13}\varepsilon_{23}\varepsilon_{45}\varepsilon_{46}\varepsilon_{56} 
      + \varepsilon_{12}\varepsilon_{14}\varepsilon_{24}\varepsilon_{35}\varepsilon_{36}\varepsilon_{56}
      + \varepsilon_{12}\varepsilon_{16}\varepsilon_{26}\varepsilon_{34}\varepsilon_{35}\varepsilon_{45} 
          + \varepsilon_{13}\varepsilon_{16}\varepsilon_{24}\varepsilon_{25}\varepsilon_{36}\varepsilon_{45} \\
          + &\: \varepsilon_{14}\varepsilon_{15}\varepsilon_{23}\varepsilon_{26}\varepsilon_{36}\varepsilon_{45} 
          + \varepsilon_{14}\varepsilon_{16}\varepsilon_{23}\varepsilon_{25}\varepsilon_{35}\varepsilon_{46} 
          +\varepsilon_{13}\varepsilon_{15}\varepsilon_{24}\varepsilon_{26}\varepsilon_{35}\varepsilon_{46} 
          + \varepsilon_{12}\varepsilon_{15}\varepsilon_{25}\varepsilon_{34}\varepsilon_{36}\varepsilon_{46} \\
          + &\: \varepsilon_{15}\varepsilon_{16}\varepsilon_{23}\varepsilon_{24}\varepsilon_{34}\varepsilon_{56} 
          + \varepsilon_{13}\varepsilon_{14}\varepsilon_{25}\varepsilon_{26}\varepsilon_{34}\varepsilon_{56} \:,
\end{split}
\end{equation}

\begin{equation}
\begin{split}
\gamma_1 =&\: \varepsilon_{15}\varepsilon_{16}\varepsilon_{24}\varepsilon_{26}\varepsilon_{34}\varepsilon_{35} 
+ \varepsilon_{14}\varepsilon_{16}\varepsilon_{25}\varepsilon_{26}\varepsilon_{34}\varepsilon_{35} 
+  \varepsilon_{15}\varepsilon_{16}\varepsilon_{24}\varepsilon_{25}\varepsilon_{34}\varepsilon_{36} 
 +  \varepsilon_{14}\varepsilon_{15}\varepsilon_{25}\varepsilon_{26}\varepsilon_{34}\varepsilon_{36} \\
+ &\: \varepsilon_{14}\varepsilon_{16}\varepsilon_{24}\varepsilon_{25}\varepsilon_{35}\varepsilon_{36} 
+ \varepsilon_{14}\varepsilon_{15}\varepsilon_{24}\varepsilon_{26}\varepsilon_{35}\varepsilon_{36} 
+ \varepsilon_{15}\varepsilon_{16}\varepsilon_{23}\varepsilon_{26}\varepsilon_{34}\varepsilon_{45}
+   \varepsilon_{13}\varepsilon_{16}\varepsilon_{25}\varepsilon_{26}\varepsilon_{34}\varepsilon_{45} \\
 +&\: \varepsilon_{15}\varepsilon_{16}\varepsilon_{23}\varepsilon_{24}\varepsilon_{36}\varepsilon_{45} 
 + \varepsilon_{14}\varepsilon_{16}\varepsilon_{23}\varepsilon_{25}\varepsilon_{36}\varepsilon_{45} 
 + \varepsilon_{13}\varepsilon_{15}\varepsilon_{24}\varepsilon_{26}\varepsilon_{36}\varepsilon_{45} 
 + \varepsilon_{13}\varepsilon_{14}\varepsilon_{25}\varepsilon_{26}\varepsilon_{36}\varepsilon_{45} \\
 +&\:  \varepsilon_{12}\varepsilon_{16}\varepsilon_{25}\varepsilon_{34}\varepsilon_{36}\varepsilon_{45} 
 + \varepsilon_{12}\varepsilon_{15}\varepsilon_{26}\varepsilon_{34}\varepsilon_{36}\varepsilon_{45} 
  + \varepsilon_{12}\varepsilon_{16}\varepsilon_{24}\varepsilon_{35}\varepsilon_{36}\varepsilon_{45} 
  + \varepsilon_{12}\varepsilon_{14}\varepsilon_{26}\varepsilon_{35}\varepsilon_{36}\varepsilon_{45} \\
+&\:  \varepsilon_{13}\varepsilon_{15}\varepsilon_{25}\varepsilon_{26}\varepsilon_{34}\varepsilon_{46} 
+ \varepsilon_{15}\varepsilon_{16}\varepsilon_{23}\varepsilon_{24}\varepsilon_{35}\varepsilon_{46} 
+   \varepsilon_{13}\varepsilon_{16}\varepsilon_{24}\varepsilon_{25}\varepsilon_{35}\varepsilon_{46} 
+ \varepsilon_{14}\varepsilon_{15}\varepsilon_{23}\varepsilon_{26}\varepsilon_{35}\varepsilon_{46}\\
 + &\: \varepsilon_{13}\varepsilon_{14}\varepsilon_{25}\varepsilon_{26}\varepsilon_{35}\varepsilon_{46} 
 + \varepsilon_{12}\varepsilon_{16}\varepsilon_{25}\varepsilon_{34}\varepsilon_{35}\varepsilon_{46} 
 + \varepsilon_{15}\varepsilon_{16}\varepsilon_{23}\varepsilon_{25}\varepsilon_{34}\varepsilon_{46}
 +  \varepsilon_{12}\varepsilon_{15}\varepsilon_{26}\varepsilon_{34}\varepsilon_{35}\varepsilon_{46}  \\
  + &\: \varepsilon_{14}\varepsilon_{15}\varepsilon_{23}\varepsilon_{25}\varepsilon_{36}\varepsilon_{46} 
  + \varepsilon_{13}\varepsilon_{15}\varepsilon_{24}\varepsilon_{25}\varepsilon_{36}\varepsilon_{46}
  + \varepsilon_{14}\varepsilon_{16}\varepsilon_{23}\varepsilon_{26}\varepsilon_{35}\varepsilon_{45}
+  \varepsilon_{13}\varepsilon_{16}\varepsilon_{24}\varepsilon_{26}\varepsilon_{35}\varepsilon_{45} \\
 +&\: \varepsilon_{12}\varepsilon_{15}\varepsilon_{24}\varepsilon_{35}\varepsilon_{36}\varepsilon_{46} 
 + \varepsilon_{12}\varepsilon_{14}\varepsilon_{25}\varepsilon_{35}\varepsilon_{36}\varepsilon_{46}\:,
\end{split}
\end{equation}

\begin{equation}
\begin{split}
\gamma_2 =  &\:\varepsilon_{13}\varepsilon_{15}\varepsilon_{23}\varepsilon_{26}\varepsilon_{45}\varepsilon_{46} 
 + \varepsilon_{13}\varepsilon_{16}\varepsilon_{23}\varepsilon_{25}\varepsilon_{45}\varepsilon_{46} 
+  \varepsilon_{12}\varepsilon_{16}\varepsilon_{23}\varepsilon_{35}\varepsilon_{45}\varepsilon_{46} 
+ \varepsilon_{12}\varepsilon_{13}\varepsilon_{26}\varepsilon_{35}\varepsilon_{45}\varepsilon_{46} \\
 +  &\: \varepsilon_{12}\varepsilon_{15}\varepsilon_{23}\varepsilon_{36}\varepsilon_{45}\varepsilon_{46} 
 + \varepsilon_{12}\varepsilon_{13}\varepsilon_{25}\varepsilon_{36}\varepsilon_{45}\varepsilon_{46} 
 +  \varepsilon_{14}\varepsilon_{16}\varepsilon_{23}\varepsilon_{25}\varepsilon_{34}\varepsilon_{56} 
 + \varepsilon_{13}\varepsilon_{16}\varepsilon_{24}\varepsilon_{25}\varepsilon_{34}\varepsilon_{56}\\
  + &\: \varepsilon_{14}\varepsilon_{15}\varepsilon_{23}\varepsilon_{26}\varepsilon_{34}\varepsilon_{56} 
  + \varepsilon_{13}\varepsilon_{15}\varepsilon_{24}\varepsilon_{26}\varepsilon_{34}\varepsilon_{56}  
  +    \varepsilon_{14}\varepsilon_{16}\varepsilon_{23}\varepsilon_{24}\varepsilon_{35}\varepsilon_{56}
 + \varepsilon_{13}\varepsilon_{14}\varepsilon_{24}\varepsilon_{26}\varepsilon_{35}\varepsilon_{56} \\
  +  &\:  \varepsilon_{12}\varepsilon_{16}\varepsilon_{24}\varepsilon_{34}\varepsilon_{35}\varepsilon_{56} 
  + \varepsilon_{12}\varepsilon_{14}\varepsilon_{26}\varepsilon_{34}\varepsilon_{35}\varepsilon_{56} 
 + \varepsilon_{14}\varepsilon_{15}\varepsilon_{23}\varepsilon_{24}\varepsilon_{36}\varepsilon_{56} 
+ \varepsilon_{13}\varepsilon_{14}\varepsilon_{24}\varepsilon_{25}\varepsilon_{36}\varepsilon_{56}\\
+   &\:   \varepsilon_{12}\varepsilon_{14}\varepsilon_{23}\varepsilon_{36}\varepsilon_{45}\varepsilon_{56} 
+ \varepsilon_{12}\varepsilon_{13}\varepsilon_{24}\varepsilon_{36}\varepsilon_{45}\varepsilon_{56} 
+   \varepsilon_{12}\varepsilon_{15}\varepsilon_{24}\varepsilon_{34}\varepsilon_{36}\varepsilon_{56} 
+ \varepsilon_{12}\varepsilon_{14}\varepsilon_{25}\varepsilon_{34}\varepsilon_{36}\varepsilon_{56}\\
+ &\: \varepsilon_{12}\varepsilon_{13}\varepsilon_{25}\varepsilon_{34}\varepsilon_{46}\varepsilon_{56} 
 + \varepsilon_{13}\varepsilon_{16}\varepsilon_{23}\varepsilon_{24}\varepsilon_{45}\varepsilon_{56} 
  +   \varepsilon_{13}\varepsilon_{14}\varepsilon_{23}\varepsilon_{26}\varepsilon_{45}\varepsilon_{56}
+  \varepsilon_{12}\varepsilon_{16}\varepsilon_{23}\varepsilon_{34}\varepsilon_{45}\varepsilon_{56} \\
+ &\: \varepsilon_{12}\varepsilon_{14}\varepsilon_{23}\varepsilon_{35}\varepsilon_{46}\varepsilon_{56}  
+ \varepsilon_{12}\varepsilon_{13}\varepsilon_{26}\varepsilon_{34}\varepsilon_{45}\varepsilon_{56}
+  \varepsilon_{13}\varepsilon_{15}\varepsilon_{23}\varepsilon_{24}\varepsilon_{46}\varepsilon_{56} 
+ \varepsilon_{13}\varepsilon_{14}\varepsilon_{23}\varepsilon_{25}\varepsilon_{46}\varepsilon_{56} \\
 + &\: \varepsilon_{12}\varepsilon_{13}\varepsilon_{24}\varepsilon_{35}\varepsilon_{46}\varepsilon_{56} 
+ \varepsilon_{12}\varepsilon_{15}\varepsilon_{23}\varepsilon_{34}\varepsilon_{46}\varepsilon_{56} \:,
\end{split}
\end{equation}

\begin{equation}
\begin{split}
\delta_1 =&\:  \varepsilon_{12}\varepsilon_{13}\varepsilon_{25}\varepsilon_{35}\varepsilon_{46}^2 
+  \varepsilon_{12}\varepsilon_{15}\varepsilon_{23}\varepsilon_{35}\varepsilon_{46}^2 
+ \varepsilon_{13}^2\varepsilon_{24}\varepsilon_{25}\varepsilon_{46}\varepsilon_{56}   
+  \varepsilon_{12}\varepsilon_{13}\varepsilon_{24}\varepsilon_{34}\varepsilon_{56}^2 \\
  + &\: \varepsilon_{13}\varepsilon_{14}\varepsilon_{25}^2\varepsilon_{36}\varepsilon_{46}
+ \varepsilon_{12}\varepsilon_{14}\varepsilon_{23}\varepsilon_{34}\varepsilon_{56}^2 
+ \varepsilon_{14}\varepsilon_{15}\varepsilon_{23}^2\varepsilon_{46}\varepsilon_{56}
 +\varepsilon_{15}\varepsilon_{16}\varepsilon_{25}\varepsilon_{26}\varepsilon_{34}^2 \\
  +&\:\varepsilon_{15}\varepsilon_{16}\varepsilon_{24}^2\varepsilon_{35}\varepsilon_{36} 
 +\varepsilon_{16}^2\varepsilon_{24}\varepsilon_{25}\varepsilon_{34}\varepsilon_{35} 
 +  \varepsilon_{14}^2\varepsilon_{23}\varepsilon_{26}\varepsilon_{35}\varepsilon_{56} 
+  \varepsilon_{13}\varepsilon_{14}\varepsilon_{23}\varepsilon_{24}\varepsilon_{56}^2 \\
 +&\: \varepsilon_{12}^2\varepsilon_{34}\varepsilon_{35}\varepsilon_{46}\varepsilon_{56}
 + \varepsilon_{14}^2\varepsilon_{25}\varepsilon_{26}\varepsilon_{35}\varepsilon_{36} 
 + \varepsilon_{13}^2\varepsilon_{25}\varepsilon_{26}\varepsilon_{45}\varepsilon_{46} 
 + \varepsilon_{13}\varepsilon_{16}\varepsilon_{24}^2\varepsilon_{35}\varepsilon_{56} \\
  + &\: \varepsilon_{15}\varepsilon_{16}\varepsilon_{23}^2\varepsilon_{45}\varepsilon_{46}
   + \varepsilon_{13}^2\varepsilon_{24}\varepsilon_{26}\varepsilon_{45}\varepsilon_{56}
   + \varepsilon_{14}\varepsilon_{15}\varepsilon_{24}\varepsilon_{25}\varepsilon_{36}^2
   + \varepsilon_{12}\varepsilon_{16}\varepsilon_{25}\varepsilon_{34}^2\varepsilon_{56}\\
+&\: \varepsilon_{12}\varepsilon_{15}\varepsilon_{26}\varepsilon_{34}^2\varepsilon_{56}
 +  \varepsilon_{12}\varepsilon_{14}\varepsilon_{26}\varepsilon_{35}^2\varepsilon_{46} 
 + \varepsilon_{15}^2\varepsilon_{23}\varepsilon_{24}\varepsilon_{36}\varepsilon_{46}
  + \varepsilon_{14}\varepsilon_{16}\varepsilon_{25}^2\varepsilon_{34}\varepsilon_{36}\:,
 \end{split}
\end{equation}

\begin{equation}
\begin{split}
  \delta_2 =&\: \varepsilon_{15}^2\varepsilon_{23}\varepsilon_{26}\varepsilon_{34}\varepsilon_{46} 
+  \varepsilon_{12}\varepsilon_{15}\varepsilon_{24}\varepsilon_{36}^2\varepsilon_{45} 
+\varepsilon_{12}\varepsilon_{14}\varepsilon_{25}\varepsilon_{36}^2\varepsilon_{45}
+ \varepsilon_{13}\varepsilon_{14}\varepsilon_{26}^2\varepsilon_{35}\varepsilon_{45}\\
+ &\:\varepsilon_{13}\varepsilon_{16}\varepsilon_{23}\varepsilon_{26}\varepsilon_{45}^2 
 + \varepsilon_{12}\varepsilon_{16}\varepsilon_{23}\varepsilon_{36}\varepsilon_{45}^2
+ \varepsilon_{13}\varepsilon_{15}\varepsilon_{24}^2\varepsilon_{36}\varepsilon_{56} 
+\varepsilon_{12}\varepsilon_{16}\varepsilon_{24}\varepsilon_{35}^2\varepsilon_{46}\\
 +&\:\varepsilon_{13}\varepsilon_{15}\varepsilon_{23}\varepsilon_{25}\varepsilon_{46}^2
 +  \varepsilon_{14}^2\varepsilon_{23}\varepsilon_{25}\varepsilon_{36}\varepsilon_{56}
  + \varepsilon_{12}^2\varepsilon_{34}\varepsilon_{36}\varepsilon_{45}\varepsilon_{56}
 +  \varepsilon_{14}\varepsilon_{16}\varepsilon_{23}^2\varepsilon_{45}\varepsilon_{56}\\
 + &\:\varepsilon_{13}\varepsilon_{15}\varepsilon_{26}^2\varepsilon_{34}\varepsilon_{45}
  + \varepsilon_{12}^2\varepsilon_{35}\varepsilon_{36}\varepsilon_{45}\varepsilon_{46} 
 + \varepsilon_{16}^2\varepsilon_{23}\varepsilon_{24}\varepsilon_{35}\varepsilon_{45} 
 +  \varepsilon_{12}\varepsilon_{13}\varepsilon_{26}\varepsilon_{36}\varepsilon_{45}^2 \\
 +&\: \varepsilon_{16}^2\varepsilon_{23}\varepsilon_{25}\varepsilon_{34}\varepsilon_{45}
  + \varepsilon_{14}\varepsilon_{15}\varepsilon_{26}^2\varepsilon_{34}\varepsilon_{35} 
   + \varepsilon_{14}\varepsilon_{16}\varepsilon_{24}\varepsilon_{26}\varepsilon_{35}^2
 + \varepsilon_{12}^2\varepsilon_{35}\varepsilon_{36}\varepsilon_{45}\varepsilon_{46} \\
 + &\: \varepsilon_{13}\varepsilon_{15}\varepsilon_{23}\varepsilon_{25}\varepsilon_{46}^2
  + \varepsilon_{13}\varepsilon_{16}\varepsilon_{25}^2\varepsilon_{34}\varepsilon_{46} \:.
\end{split}
\end{equation}
\end{widetext}


\end{document}